\documentclass[pre,floatfix,reprint,superscriptaddres,amsmath,amssymb]{revtex4-2}

\usepackage{color,graphics}
\usepackage{graphicx}
\usepackage[sort&compress]{natbib}
\usepackage{hhline}
\usepackage{soul}
\usepackage{url}
\usepackage{hyperref}

\usepackage{graphicx}
\usepackage{dcolumn}
\usepackage{bm}
\usepackage[utf8]{inputenc}
\usepackage[T1]{fontenc}
\usepackage{etoolbox}
\usepackage{mathptmx}
\usepackage[usenames,dvipsnames]{xcolor}

\newcommand{\mycomment}[1]{}

\makeatletter
\def\@email#1#2{%
 \endgroup
 \patchcmd{\titleblock@produce}
  {\frontmatter@RRAPformat}
  {\frontmatter@RRAPformat{\produce@RRAP{#1\href{mailto:#2}{#2}}}
  \frontmatter@RRAPformat}
  {}{}
}%
\makeatother
\begin{document}

\title{Structural and Dynamical Crossovers in Dense Electrolytes}
\author{Daehyeok Kim$^{a}$, Taejin Kwon$^{b,\dagger}$, and Jeongmin Kim$^{c,*}$}
\affiliation
{$^a$Department of Energy Engineering, Korea Institute of Energy Technology (KENTECH), Naju 58330, Republic of Korea \\
$^b$Department of Chemistry and Cosmetics, Jeju National University, Jeju 63243, Republic of Korea \\
$^c$Department of Chemistry Education and Graduate Department of Chemical Materials, Pusan National University, Busan 46241, Republic of Korea}
\email{$^*$jeongmin@pusan.ac.kr}
\email{$^\dagger$tjkwon@jejunu.ac.kr}

\date{\today}

\begin{abstract}
Electrostatic interactions fundamentally govern the structure and transport of electrolytes. In concentrated electrolytes, however, electrostatic and steric correlations, together with ion–solvent coupling, give rise to complex behavior, such as underscreening, that remains challenging to explain despite extensive theoretical effort. Using molecular dynamics simulations of primitive electrolytes with and without space-filling solvent particles, we elucidate the structural and dynamical crossovers and their connection that emerge with increasing salt concentration. Explicit-solvent electrolytes exhibit a screening transition from a charge-dominated dilute regime to a density-dominated concentrated regime, accompanied by dynamical crossovers in ion self-diffusion and ion-pair lifetimes. These dynamical crossovers display a marked discontinuity, unlike the smoother variation of the screening crossover, which originates from short-range ion–counterion structures. The pronounced growth of ionic clusters leads to a percolation transition only at higher concentrations, whereas the onset of the structural and dynamical crossovers is associated primarily with local ion pairing or small aggregates. Both structural and dynamical behaviors are found to depend sensitively on ion–solvent coupling: implicit-solvent electrolytes exhibit a screening transition between two charge-dominated regimes, accompanied by qualitatively distinct dynamical behavior. Finally, we demonstrate that the diffusion-corrected ion-pair lifetime provides a consistent descriptor linking ionic structure and dynamics across electrolyte systems.
\end{abstract}

\maketitle

\section{Introduction}
Electrolytes are indispensable in diverse systems spanning energy storage and conversion~\cite{yamada2014unusual,suo2015water,wang2015development,long2016polymer,wang2016superconcentrated,xia2017electrolytes,yamada2019advances,ye2020recent,kwon2024borate}, environmental processes~\cite{kim2021selective,hand2021electrochemical,alkhadra2022electrochemical,dang2023bias}, and biological functions~\cite{oliot2016ion,kefauver2020discoveries,chen2020fundamentals}. Understanding of their physicochemical properties at the molecular level is therefore essential for the development of advanced and next-generation electrolytes. At sufficiently low salt concentrations (typically below  $\sim$0.1 mol$\cdot$L$^{-1}$ for aqueous monovalent electrolytes), the Debye-H\"uckel (DH) theory~\cite{huckel1923theorie,kontogeorgis2018debye} provides a successful description based mainly on electrostatic interactions among point ions embedded in a uniform dielectric-continuum solvent. However, as the salt concentration increases, deviations from DH behavior emerge due to ion-ion correlations, excluded-volume effects, and ion-solvent coupling; such complex interplay challenges a molecular-level understanding of concentrated electrolytes~\cite{gebbie2013ionic,adar2019screening,dinpajooh2024beyond}.

Underscreening~\cite{Gebbie2015,gebbie2017long,Lee2017a,Jager2023screening} represents one of the most intriguing anomalies in concentrated electrolytes, which emerges as a result of competition between electrostatic and steric interactions. Recent surface force balance (SFB) experiments~\cite{Gebbie2015,Lee2017a,Jager2023screening} have investigated electrostatic screening across a wide range of electrolytes and concentrations~\cite{Avni2020,Goodwin2017,Lee2017,Krucker-Velasquez2021,Rotenberg2018,zhang2024long}. These studies revealed that concentrated electrolytes exhibit a longer electrostatic decay length ($\lambda_s$) with higher concentration, in contrast to the prediction of the DH theory, and in some cases $\lambda_s$ reaches up to $\sim$10 nm, significantly exceeding the Debye screening length $\lambda_D$. A clear crossover between dilute to concentrated regimes was found when $\lambda_D$ becomes comparable to the mean ion size $a$, with underscreening primarily observed in the concentrated regime. A phenomenon was previously predicted as the Kirkwood crossover~\cite{kirkwood1936statistical}. Remarkably, in the concentrated regime, a universal scaling relation was identified across chemically diverse electrolytes when normalized appropriately:
\begin{equation}\label{eq:scaling}
\bigg(\frac{\lambda_s}{\lambda_D}\bigg)\sim\bigg(\frac{a}{\lambda_D}\bigg)^{\alpha},
\end{equation}
where the scaling exponent $\alpha = 3$ in the SFB experiments. The intriguing scaling relation still lacks a clear physical origin, as many bulk simulations~\cite{coles2020correlation,zeman2021ionic,Krucker-Velasquez2021} and theoretical studies~\cite{Adar2019,Cats2021,Rotenberg2018} have reproduced similar screening crossovers, albeit with consistently underestimated exponents ($\alpha\approx1-2$).

In addition to the structural features, understanding and predicting the dynamic behavior of concentrated electrolytes remain equally challenging.
Theoretical approaches, such as  mean-spherical approximation (MSA)~\cite{waisman1972mean,simonin2021full} and stochastic density functional theory (sDFT)~\cite{demery2016conductivity,avni2022conductance,illien2024stochastic}, have extended the classical Debye-H\"uckel-Onsager theory~\cite{onsager1927report,PhysRevE.55.2814} by incorporating steric ion-ion interactions. Recent applications of sDFT~\cite{avni2022conductivity,illien2024stochastic,robin2024correlation} have reported encouraging agreement with experimental trends in the transport properties of moderately concentrated aqueous electrolytes. Nevertheless, a subsequent study~\cite{bernard2023analytical} indicated that the quantitative accuracy of these predictions may be limited by compensating effects among different theoretical contributions. However, a unified microscopic understanding that consistently connects ion dynamics and electrostatic screening remains elusive, seeking to establish a direct connection between structure and dynamics. That is, how microscopic ion and solvent organization governs macroscopic transport properties, such as electrical conductivity and diffusion.

The formation of ionic clusters, as a result of structural correlations, plays a crucial role in determining ion transport in concentrated electrolytes by governing correlations in ion motion~\cite{feng2019free,molinari2019general,france2019correlations, bi2024cluster}. A striking example is water-in-salt electrolytes~\cite{shen2021water,suo2015water}, where nanoscale phase separation emerges into a salt-rich domain that enhances interfacial stability and a water-rich domain that facilitates lithium-ion transport. The emergence of nanoscale ionic organization has also been observed in model electrolytes~\cite{krucker2021underscreening,wang2024structure,liu2024minimal,markiewitz2025ionic}. Such nonideality becomes particularly significant in nonaqueous electrolytes with low dielectric permittivity, where electrostatic correlations dominate over the thermal fluctuations. These ionic clusters contribute differently to the total ionic conductivity, depending on their diffusivity~\cite{feng2011supercapacitor,lee2023anomalous,bi2024cluster}. With increasing salt concentration, ion association proceeds through the progressive linking of ion pairs into extended networks~\cite{borodin2006litfsi,borodin2017liquid,mceldrew2020theory}. When the connectivity of these networks spans the entire system, a system-spanning ionic structure emerges, often referred to as percolation or gelation in the context of polymer physics~\cite{rubinstein2003polymer,mceldrew2020theory,mceldrew2021ion,mceldrew2021salt,goodwin2022gelation}. The percolation of ionic domains was also suggested to be closely related to the cubic scaling (Eq.~\ref{eq:scaling} with $\alpha=3$) observed in SFB experiments~\cite{pincus2023soft,skarmoutsos2025length}. These findings motivate the need to further investigate how percolation in ionic structures, and its onset, relate to potential structural and dynamical crossovers in concentrated systems.
 
In this work, we employ molecular dynamics simulations of primitive electrolytes with both explicit and implicit solvents to elucidate the relationships between structural and dynamical correlations in dense ionic fluids across a wide range of salt concentrations. Primitive models are chosen because, despite their simplicity, they have proven successful in capturing essential features of electrostatic screening and ionic organization~\cite{giera2015electric,Cats2021,hartel2023anomalous,yang2023solvent,bi2024cluster,ribar2024cluster}. While explaining the microscopic origin of the cubic scaling is beyond the scope of this study, our primary focus is to clarify the mechanistic connections among electrostatic screening, ionic clustering, and charge transport in dense ionic environments.

 We identify dynamic crossovers in ion self-diffusion and ion-pair dynamics that accompany the screening crossover and can be unified by scaling with the ratio of the mean ion size to the Debye screening length, as used in the SFB measurements. The nature of the screening crossover depends sensitively on explicit-solvent packing: explicit-solvent systems transition from a charge-dominated dilute regime to a density-dominated concentrated regime, whereas implicit-solvent systems undergo a transition between two charge-dominated regimes. Although enhanced ion–ion correlations promote a percolation transition in ionic clustering, we find that this percolation is not directly coupled to either screening or dynamical crossovers. Instead, it correlates more strongly with ion density, reflecting that percolation requires sufficiently extensive aggregation to form a system-spanning network and therefore typically occurs only at higher salt concentrations, whereas the onset of underscreening is associated with the emergence of local ion pairing or small aggregates. The dynamical crossovers exhibit a marked discontinuity, linked to qualitative changes in short-range ion–counterion structure. Finally, we identify the diffusion-corrected ion-pair lifetime as a unified microscopic descriptor that consistently bridges ionic structure and transport across electrolyte models.

The remainder of this paper is organized as follows. In Section~\ref{sec:methods}, we describe the simulation models and methods, detailing the procedures for calculating structural and dynamical properties as well as identifying ionic clusters. Section~\ref{sec:results} presents the results and discussion of the coupled crossovers of electrostatic screening and dynamic properties, including ion self diffusion and ion-pair lifetime, highlighting the diffusion-corrected ion-pair lifetime as a unified basis for both explicit-solvent electrolytes over a wide range of salt concentration. We also examine the decoupling of the percolation transition from both structural and dynamical crossovers. Finally, Section~\ref{sec:conclusions} provides concluding remarks.

\section{Simulation Methods and model systems}\label{sec:methods}

\subsection{Model electrolytes}\label{subsec:model}
We consider model electrolytes consisting of neutral LJ solvent particles and ion particles, all of the same size and of the same mass ($m$)~\cite{joly2006liquid}. The particles interact via the LJ and Coulomb interactions. All LJ interactions ($U_{LJ}$) for solvent and ion particles were truncated and shifted at a cutoff distance $r_c$. 
\begin{equation}\label{eq:lj}
\begin{split}
U_{LJ}(r)=&4\epsilon\bigg[\bigg(\frac{\sigma}{r}\bigg)^{12}-
\bigg(\frac{\sigma}{r}\bigg)^{6}-\bigg(\frac{\sigma}{r_c}\bigg)^{12}+\bigg(\frac{\sigma}{r_c}\bigg)^{6}\bigg],
\end{split}
\end{equation}
assuming the same LJ energy $\epsilon$, and diameter $\sigma$ for interactions between all types of particles.
Two cutoff distances are chosen: $r_c=2.5~\sigma$ and $1.222~\sigma$. In the former case, both attractive and repulsive LJ interactions are considered, while only the repulsive interactions (WCA) are considered in the latter case. In this study, electrolytes employing cutoff distances of $r_c=2.5~\sigma$ and $1.122~\sigma$ are referred as "LJ" and "WCA", respectively. In contrast, the Coulomb interaction ($U_C$) between the ions is identical across all electrolyte systems, and is given by:
\begin{equation}
U_{C}(r)=\frac{1}{4\pi\varepsilon_0\varepsilon_s}\frac{q_iq_j}{r}=k_BTZ_iZ_j\frac{l_B}{r} 
\end{equation}
where $\varepsilon_0$ is the vacuum permittivity, and $l_B$ denotes the Bjerrum length with $k_B$ being the Boltzmann constant and $T$ being the temperature. $l_B$ varies with the uniform background dielectric constant $\varepsilon_s$, ranging between 0.2 and 5 in this work. The LJ ions are monovalent, \emph{i.e.}, valency $Z_\alpha=1$ for all ionic species $\alpha$, carrying either $q^*_\alpha=eZ_\alpha/\sqrt{4\pi\varepsilon_0\sigma\epsilon}=+1$ or -1. Here, the asterisk represents a quantity in reduced LJ units. Regardless of the LJ cutoff $r_c$, $U_C(r)$ is split into two contributions: The short-range contribution is cut at $r=3.5~\sigma$, and the long-range contribution is calculated using the particle-particle and particle-mesh (PPPM) method with the Ewald screening parameter chosen for a fixed cutoff distance in order to achieve a given relative error in forces ($10^{-4}$ in this work). All computations were performed in the $N_{solv}N_{salt}p^*T^*$ ensemble at $p^*=p\sigma^3/\epsilon=1$ and $T^*=Tk_B/\epsilon=1$, which corresponds to a liquid phase \cite{scalfi2021gibbs,kim2024JCP}. The number of solvent particles, $N_{\mathrm{solv}}$, was fixed at 5000, while the number of salt pairs, $N_{\mathrm{salt}}$, varied according to solution concentration unless otherwise noted. 

During equilibrium MD, the equations of motion were integrated using the velocity Verlet integrator with a timestep $\delta t^*=\delta t\sqrt{\epsilon(m\sigma^2)^{-1}}=0.001$.
The desired pressure $p^*=1$ and temperature $T^*=1$ were maintained using the Nos\'e-Hoover barostat and thermostat, with time constants of 1000 and 100 LJ units, respectively. For each condition, we obtained at least ten independent trajectories starting from different initial states, from which statistical averages and errors were computed.

\textbf{Implicit-solvent models.} As a limiting case to examine explicit space-filling solvent effects, we performed Langevin dynamics (LD) simulations, implemented in LAMMPS~\cite{brunger1984stochastic,dunweg1991brownian}, for electrolyte systems without explicit solvent particles. The ions interact via the same interaction potentials and were assigned identical LJ parameters, charges, and masses as in the explicit-solvent model. Long-range electrostatic interactions and all other simulation settings were kept identical to those used in the explicit-solvent systems. The friction coefficient was determined from the diffusion coefficient obtained in the explicit-solvent simulations at the smallest salt-to-solvent ratio ($r_{salt} = 0.01$) and was kept constant throughout all LD simulations. All LD simulations were carried out in the canonical ensemble, with the box size chosen to match the average volume of the equilibrated explicit-solvent systems at each condition, and with the identical temperature maintained at $T^* = 1.0$. We note that the ions in the explicit and implicit solvent models do not follow the same equation of state, which, however, is not the primary focus of the comparison between the two solvent models. We also note that, in concentrated LJ electrolytes, the ions undergo phase segregation~\cite{ribar2024cluster} when explicit solvent particles are absent. Therefore, only low-concentration cases that show no phase segregation are included in our analysis (see Fig.~S7 in the SM~\cite{SI}).

\subsection{Computation of observables}\label{subsec:method_comp}

\textbf{Structural correlation functions.}
We investigated the short- and long-range ion structures by computing the radial distribution function (RDF). 
\begin{eqnarray}
g_{\alpha\beta}(r) = \frac{\langle V\rangle}{4 \pi r^2 N_{\alpha} N_{\beta}} \sum_{i,j\neq i} \langle \delta(r-|\vec{r}^{\alpha}_i - \vec{r}^{\beta}_j|) \rangle,
\label{eq:rdf}
\end{eqnarray}
where $\langle V\rangle$, and $N_{\alpha}$ denote the average volume of a simulation cell, the number of ionic species $\alpha$, respectively: the subindices $\alpha$ and $\beta\in\{+,-,s\}$. $\vec{r}^{\alpha}_i$ represents the position vectors of $i^{th}$ particle of species $\alpha$. Here, $\langle \cdots \rangle$ denotes the ensemble average.

Using the ion-ion RDFs, we also calculated the density-density ($g_{NN}(r)$) and charge-charge ($g_{ZZ}(r)$) correlation function as follows,
\begin{eqnarray}\label{eq:gnn}
g_{NN}(r) = \sum_{\alpha} \sum_{\beta}  \chi_{\alpha} \chi_{\beta} g_{\alpha\beta}(r),
\end{eqnarray}\label{eq:gzz}
\begin{eqnarray}
g_{ZZ}(r) = \sum_{\alpha} \sum_{\beta}  q_{\alpha} q_{\beta} \chi_{\alpha} \chi_{\beta} g_{\alpha\beta}(r), 
\end{eqnarray}
where $\chi_{\alpha}$ denotes the number fraction of the ionic species $\alpha$, respectively. For the 1:1 symmetric electrolytes in the present study, $\chi_\alpha=1/2$ for any species $\alpha$. We note that $g_{NN}(r\rightarrow \infty)\rightarrow 1$, while $g_{ZZ}(r\rightarrow \infty)\rightarrow 0$. These systems exhibit the charge inversion symmetry, and therefore the charge-density coupling is strictly zero: $g_{ZN}(r)=\sum_{\alpha} \sum_{\beta}  q_{\alpha}\chi_{\alpha} \chi_{\beta} g_{\alpha\beta}(r)=\frac{e}{4}[g_{++}(r)+g_{+-}(r)-g_{-+}(r)-g_{--}(r)]=0$. 

In a range of salt concentrations, an oscillatory exponential function describes the decay of the total correlation functions $h(r)$ at long distances~\cite{coles2020correlation, zeman2021ionic}:
\begin{eqnarray}\label{eq:hfit}
h(r)\approx A\cdot r^{-1}\exp\bigg(-\frac{r}{\lambda}\bigg) 
\cdot \cos\left(\frac{2\pi r}{l} + \phi \right).
\end{eqnarray}
Here, the first term represents the Yukawa-type exponential decay characterized by the decay length $\lambda$, while the second term captures the oscillatory modulation with amplitude $A$, phase $\phi$, and period $l$. The total correlation function $h(r)$ corresponds to $h_{ZZ}(r)=g_{ZZ}(r)$, $h_{NN}(r)=g_{NN}(r)-1$, $h_{s+}(r)=g_{s+}(r)-1$, or $h_{ss}(r) g_{ss}(r)-1$, ensuring $h(r\to\infty)\to0$. In our analysis, each structural correlation function was approximated by a single Yukawa-type decay. In principle, however, multiple decay modes may coexist according to the dressed-ion theory~\cite{kjellander1994dressed,leote1994decay,ennis1995dressed,ulander2001decay,forsberg2005dressed}.

Regarding electrostatic interactions, the exponential term dominates $h_{ZZ}(r)$ in sufficiently dilute electrolytes ($\lesssim 0.1$ mol$\cdot$L$^{-1}$), consistent with Debye–Hückel theory, where the decay length satisfies $\lambda_Z \approx \lambda_D$, with $\lambda_D$ being the Debye screening length:
\begin{equation}\label{eq:debye}
\lambda_D = \sqrt{\frac{1}{8 \pi l_{\mathrm{B}} c_{\mathrm{salt}}}}.
\end{equation}
In contrast, the oscillatory component of $h_{ZZ}(r)$ becomes prominent at high salt concentrations owing to the strong ion–ion correlations~\cite{coles2020correlation}. In this work, we refer to the transition between these two regimes as the electrostatic screening crossover.

In a similar manner, the density-density total correlation $h_{NN}(r)$ was analyzed. The dominant length scale is determined by the longest decay length, not by the presence of the oscillation~\cite{hartel2023anomalous,coles2020correlation}

\textbf{Ionic clusters and gelation.} We analyzed ionic clusters by constructing an undirected graph $G = (V, E)$ from the instantaneous ionic configuration at a given time using networkX~\cite{Hagberg2008}. Each ion $i \in V$ is represented as a node located at position $\vec{r}_i$, and an edge $(i, j) \in E$ is assigned between ions $i$ and $j$ if their separation is less than a cutoff distance $r_c$. In other words, we define ionic clusters using a distance-based connectivity. In this work, $r_c=1.5\sigma$, taken from the first minimum of $g_{+-}(r)$.  The ionic clusters are then identified as the connected components of the graph $G$, where each cluster consists of ions linked through the edges. The size $s$ of a cluster is defined as the sum of the numbers of constituent cations $n_+$ and anions $n_-$: $s = n_+ + n_-$. The charge inversion symmetry in the model electrolytes imposes $n_+ = n_-$.

We analyzed the size distribution of ionic clusters as a function of salt concentration, modeled by an exponential distribution with a power-law tail~\cite{rubinstein2003polymer}:
\begin{equation}\label{eq:cluster}
P(s) \approx \frac{1}{N_s}s^{-\gamma_s}\cdot\exp\left(-\frac{s}{\xi_s}\right) .
\end{equation}
The normalization constant $N_s$ ensures $\sum_s P(s)=1$. This analysis is analogous to the scaling approaches used in critical phenomena~\cite{de1979scaling,stauffer2018introduction,mceldrew2020theory}. Two characteristic length scales, $\xi_s$ and $\gamma_s$, describe different parts of the distribution: the former governs the exponential decay, while the latter characterizes the power-law tail. Their relative contributions change across the percolation (or \textit{gelation}) transition. In the pre-gel regime, the exponential part dominates, corresponding to small, localized ion clusters. At the gelation point, $P(s)$ reduces to a pure power-law distribution, implying the emergence of a scale-free ionic network spanning the simulation box. In the post-gel regime, the percolating ionic network coexists with small clusters. Additional analysis of ionic clusters, including association patterns and comparison to the Cayley-tree-based ionic gelation theory~\cite{mceldrew2020theory,mceldrew2021correlated,mceldrew2021salt, goodwin2023theory,zhang2024long}, is provided in the SM~\cite{SI}.

\textbf{Ion-pair lifetime.} To quantify the lifetime of ion pairs, we computed the ion-pair survival function $H_{\alpha\beta}(t)$ using MD analysis package~\cite{michaud2011mdanalysis,Gowers2016}:
\begin{equation}\label{eq:pairdyn}
    H_{\alpha\beta}(t) = \langle h^{\alpha\beta}_{ij}(t) h^{\alpha\beta}_{ij}(0) \rangle,
\end{equation}
where $h^{\alpha\beta}_{ij}(t)$ is an indicator function defined such that $h^{\alpha\beta}_{ij}(t) = 1$ if ions $i$ of species  and $j$ reside within a certain distance $r_h$ at time $t$, and 0 otherwise. Here, ions $i$ and $j$ belong to species $\alpha$ and $\beta$, respectively. In this work, we used the same distance cutoff  ($1.5\sigma$) to define neighboring ion pairs at each time. $H_{\alpha\beta}(t)$ follows a stretched exponential function: $H_{\alpha\beta}(t)\approx \exp[-(t/\tau_H)^{\beta_H}]$, where $\tau_H$ is a characteristic time constant, and $\beta_H$ is a stretching exponent. The mean ion-pair lifetime was then estimated as $\tau_{\text{pair}}=\int_0^\infty tH_{\alpha\beta}(t)dt=\tau_{\text{H}}\Gamma(1/\beta_{\text{H}})/\beta_{\text{H}}$ with $\Gamma$ denoting the Gamma function.

\textbf{Diffusion.} We computed diffusion of ion and solvent particle using the mean-squared displacement (MSD) $\Delta\vec{r}_{\alpha\beta}(t)$ ($\alpha,\beta\in\{+,-,s\}$): $\Delta\vec{r}_{\alpha\beta}(t) \equiv \langle \Delta \vec{r}_i^{\alpha}(t) \cdot \Delta \vec{r}_j^{\beta}(t) \rangle$,
where $\Delta \vec{r}_i^{\alpha}(t)~(=\vec{r}_i^{\alpha}(t)- \vec{r}_i^{\alpha}(0))$ denotes the displacement of the $i^{th}$ ion of species $\alpha$ during time $t$~\cite{fong2021ion, fong2020onsager}. The corresponding diffusion coefficient $D_{\alpha\beta}$ was obtained in the long-time limit where the MSD grows linearly with time.
\begin{equation}
    D_{\alpha\beta}= \lim_{t\to\infty} \frac{\Delta\vec{r}_{\alpha\beta}(t)}{6t}.
\label{eq:diff}
\end{equation}
In this work, we focused on the self-diffusion ($\alpha = \beta$), where the MSD of each species grows linearly with time within the simulation window (see Sec. S7 in the Supplemental Material (SM)~\cite{SI}). Thanks to the charge inversion symmetry, we used $D_{\mathrm{ion}}=(D_{++}+D_{--})/2=D_{++}=D_{--}$. Using $D_{\alpha\beta}$, we also estimated the characteristic diffusion timescale over which an ion diffuses, on average, a distance comparable to its diameter: $\tau_{\mathrm{diff}} = \frac{\sigma^2}{6D_{\mathrm{ion}}}$~\cite{Kim2023Farad}. 

\begin{figure*}[htbp]\centering
\includegraphics[width=\textwidth]{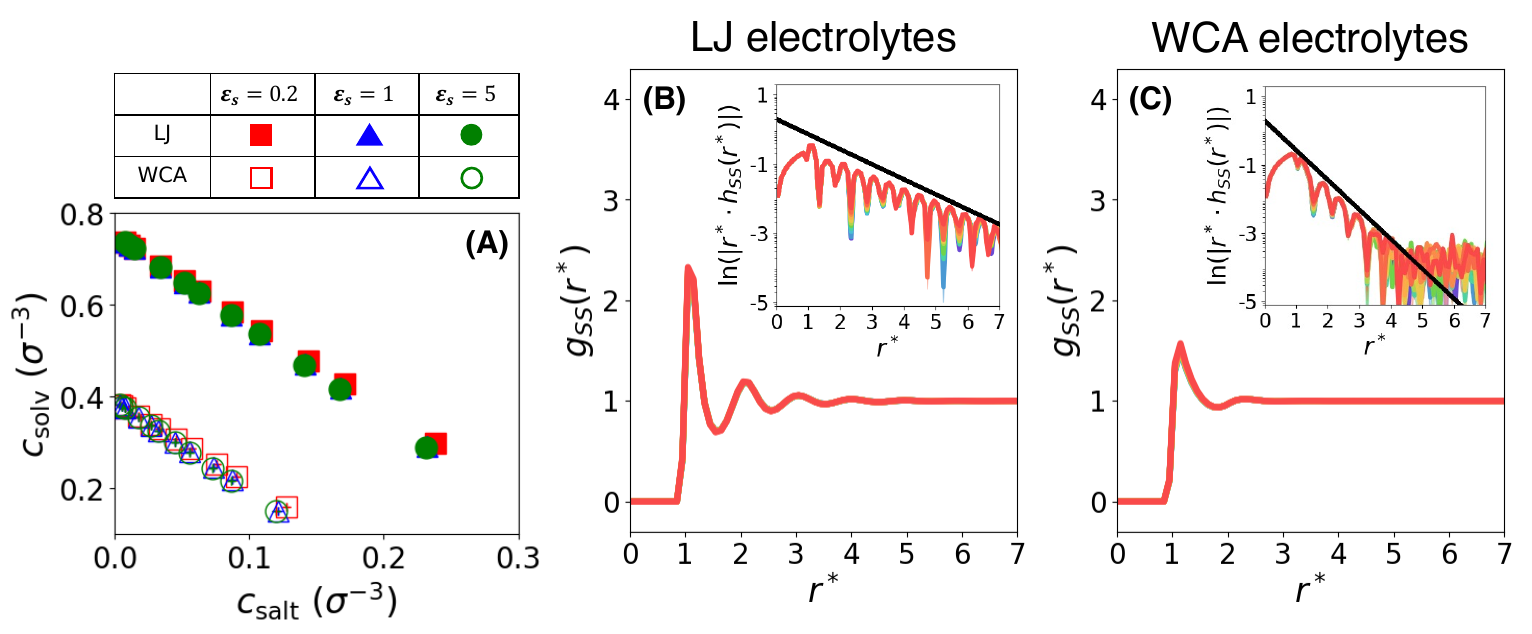}
\caption{Space-filling solvent effects. (A) Dependence of solvent concentration ($c_{\mathrm{solv}}$) and salt concentration ($c_{\mathrm{salt}}$) on the number of salt pairs $N_{\mathrm{salt}}$, with the number of solvent particles fixed at $N_{\mathrm{solv}} = 5000$. (B) Solvent–solvent radial distribution function $g_{SS}(r)$ for LJ electrolytes and (C) for WCA electrolytes. In both cases, the results collapse onto a single curve across all $c_{\mathrm{salt}}$. Insets show the distinct decay behavior of $g_{SS}(r)$, with black solid lines indicating decay lengths $\lambda_s=1~\sigma$ and $\sigma/2$, respectively. In this work, “LJ” and “WCA” refer to electrolytes with and without Lennard–Jones attractive interactions (Eq.~\ref{eq:lj}), corresponding to cutoff distances of $r_c = 2.5~\sigma$ and $1.122~\sigma$, respectively.}
\label{fig:solvent}
\end{figure*}

\section{Results and Discussion}\label{sec:results}
In this section, we primarily present the simulation results for the explicit-solvent electrolytes up to Sec.~\ref{subsec:implicit}, where they are compared with the implicit-solvent models. Results for all other models are provided in the SM~\cite{SI}.

\textbf{Solvent structure.} 
Figure~\ref{fig:solvent}(A) shows the inverse relationship between salt concentration, $c_{\mathrm{salt}}$, and the solvent concentration, $c_{\mathrm{solv}}$, as a function of $N_{\mathrm{salt}}$, clearly illustrating the steric effects in dense electrolytes. Here, $c_{\mathrm{salt}} = N_{\mathrm{salt}} / \langle V \rangle$ and $c_{\mathrm{solv}} = N_{\mathrm{solv}} / \langle V \rangle$, where $\langle V \rangle$ denotes the average system volume with $\langle \cdots \rangle$ denoting the ensemble average.

In the model systems studied, the non-electrostatic interaction strongly modifies the solvent organization: LJ systems exhibit more pronounced peaks in $g_{ss}(r)$ than WCA systems, indicative of stronger short-range correlations among solvent particles (Figs.~\ref{fig:solvent}(B) and~\ref{fig:solvent}(C)). At long distances, the decay length $\lambda_{s}$ of $\ln(|r\cdot h_{ss}(r)|)$ also differs significantly: $\lambda_{s}\approx\sigma$ for LJ electrolytes, while it is halved ($\lambda_{s}\approx\sigma/2$) for WCA ones. Nevertheless, the solvent structures in both systems are largely insensitive to variations in $\varepsilon_s$ and $c_{\text{salt}}$. We note that the ion-solvent interaction is also different in LJ and WCA systems.

\subsection{Structural crossover from a charge-dominated dilute regime to a density-dominated concentrated regime with $c_{\text{salt}}$}

\begin {figure*}[htbp]\centering
\includegraphics [width=\textwidth] {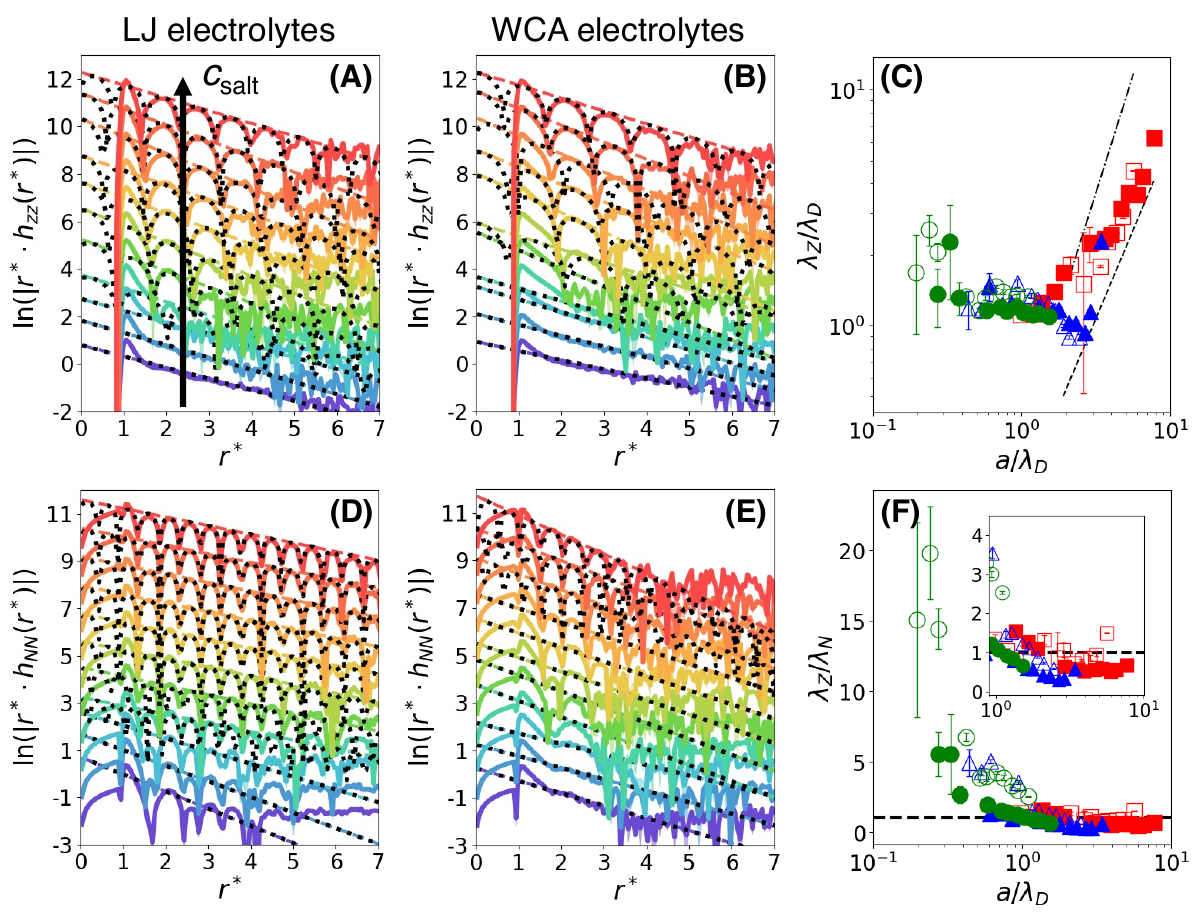}
\caption{Crossover in structural correlations of explicit-solvent electrolytes.
(A) Charge–charge correlation functions, $\ln(|r \cdot h_{ZZ}(r)|)$, for LJ electrolytes with $\varepsilon_s = 0.2$, and (B) the corresponding results for WCA electrolytes.
(C) Decay length $\lambda_Z$ of $h_{ZZ}(r)$ as a function of $a/\lambda_D$, where $a$ is the ion size and $\lambda_D$ is the Debye screening length (Eq.~\ref{eq:debye}). Black dash–dot ($\alpha = 2$) and dashed ($\alpha = 1.5$) lines serve as visual guides to the scaling behavior described in Eq.~\ref{eq:scaling}.
(D) Density–density correlation functions, $\ln(|r \cdot h_{NN}(r)|)$, for LJ electrolytes with $\varepsilon_s = 0.2$, and (E) the corresponding results for WCA electrolytes.
(F) Ratio of decay lengths, $\lambda_Z/\lambda_N$, as a function of $a/\lambda_D$. The black dashed horizontal line indicates $\lambda_Z/\lambda_N = 1$. The inset shows a magnified view of the transition region.
In panels (A–B, D–E), $c_{\mathrm{salt}}$ increases with the color changing from purple to green to red.
Panels (C) and (F) present results across all explicit-solvent systems examined in this study, with the colors and markers consistent with those in Fig.~\ref{fig:solvent}. Structural correlation functions and their corresponding fitting results for the other systems are provided in the SM~\cite{SI}.}\label{fig:gzzgnn}
\end{figure*}

The decay of charge–charge correlations (Figs.~\ref{fig:gzzgnn}(A) and~\ref{fig:gzzgnn}(B)) reveals that two distinct regimes emerge for both LJ and WCA electrolytes with varying $c_{\text{salt}}$, similar to those reported in previous studies~\cite{Cats2021,hartel2023anomalous,park2025bridging}. $\ln(|r\cdot h_{ZZ}(r)|)$ follows a simple exponential decay in the dilute regime with low $c_{\text{salt}}$, and its decay length $\lambda_Z$ decreases with $c_{\text{salt}}$, being $\lambda_Z\approx\lambda_D$, consistent with Debye–H\"uckel theory~\cite{smith2016electrostatic}. Beyond a threshold salt concentration ($a\lambda^{-1}_D\gtrsim1$), however, the oscillatory contribution to $\ln(|r\cdot h_{ZZ}(r)|)$ emerges. This threshold marks a screening crossover between the dilute and concentrated regimes. While the mean-field theories such as Debye–H\"uckel theory describes the dilute regime well, it breaks down in the concentrated regime, where strong electrostatic and steric correlations become significant. In the concentrated regime, the so-called underscreening takes place: $\lambda_Z$ increases with $c_{\text{salt}}$, exceeding $\lambda_D$~\cite{Jager2023screening,Krucker-Velasquez2021}. We note that most simulation results, including ours, exhibits the concurrent emergence of underscreening and oscillations in $\ln(|r\cdot h_{ZZ}(r)|)$, in clear contrast to the purely exponential decay of surface forces measured in SFB experiments~\cite{Lee2017}; the oscillatory features are often regarded as signatures of \textit{over}screening~\cite{dean2021overscreening,lee2021ion}.

Figure~\ref{fig:gzzgnn}(C) summarizes our results for both LJ and WCA electrolytes, which reproduce the electrostatic screening crossovers observed in SFB experiments, yet yield underestimated scaling exponents ($\alpha \approx 1.5$–$2$). Notably, the range of measured $\lambda_Z$ extends only up to $\sim10~\lambda_D$, whereas anomalous underscreening in SFB experiments has been reported to reach $\sim100~\lambda_D$. Consistent with previous bulk molecular simulations~\cite{coles2020correlation,zeman2021ionic,Krucker-Velasquez2021} and classical DFT studies~\cite{Adar2019,Cats2021,Rotenberg2018}, this underestimation leaves the physical basis of the experimental scaling unresolved, a discrepancy that persists across both chemically detailed models and coarse-grained systems such as primitive electrolytes. Recent studies~\cite{hartel2023anomalous,Kumar2022,wang2024structure} have further suggested that confinement- or interface-induced effects, including enhanced ion association, may contribute to the anomalous cubic scaling observed in SFB experiments, but investigating these effects is beyond the scope of the present study.

The LJ electrolytes exhibit strong density--density correlations across all $c_{\mathrm{salt}}$, evidenced by the oscillatory exponential decay of $\ln\left(\lvert r\, h_{NN}(r)\rvert\right)$ (Fig.~\ref{fig:gzzgnn}(D)), in marked contrast to the Debye--Hückel assumption that $g_{NN}(r)=1$ at all distances. For $c_{\mathrm{salt}} \gtrsim 0.0343~\sigma^{-3}$ ($r_{\mathrm{salt}}=0.05$), both the decay length and the oscillation period remain nearly constant, with $\lambda_N \approx \sigma$ and $l_N \approx \sigma/2$, largely independent of $c_{\mathrm{salt}}$. The WCA electrolytes also exhibit oscillatory behavior (Fig.~\ref{fig:gzzgnn}(E)), though with comparatively weaker oscillations. Non-electrostatic attractions strongly influence long-range density packing, as reflected in the decay lengths: at all $c_{\mathrm{salt}}$, the WCA systems show $\lambda_N \approx \sigma/2$, approximately half of that in the LJ systems. This trend is consistent with the long-range decay of the solvent structure $h_{SS}(r)$ (Fig.~\ref{fig:solvent}). In contrast, the oscillation period of $h_{NN}(r)$ remains largely unchanged regardless of attractive interactions, consistent with that of $h_{SS}(r)$: $l_N \approx l_s\approx\sigma/2$. Overall, these results underscore the crucial role of solvent packing and, more broadly, ion-solvent coupling in determining the structural organization of ions.

The comparison of the decay lengths ($\lambda_N$ versus $\lambda_Z$, Fig.~\ref{fig:gzzgnn}(F)) shows that both explicit-solvent electrolytes exhibit a structural crossover at $a/\lambda_D \approx 1$, from a charge-dominated dilute regime ($\lambda_Z \gtrsim \lambda_N$) to a density-dominated concentrated regime ($\lambda_Z \lesssim \lambda_N$). This transition is consistent with previous predictions from a hard-sphere ion–solvent mixture (HISM) at high solvent density~\cite{PRL2018hartel} and from simulations of primitive models with a variable solvent dielectric function~\cite{ribar2024cluster}. Our results further confirm these predictions: in the density-dominated regime, most decay lengths collapse onto a common value, except for $\lambda_Z$. Supporting this, the oscillatory decays of $h_{s+}(r)$ give $\lambda_{s+} \approx \sigma$ in the LJ systems and $\lambda_{s+} \approx \sigma/2$ in the WCA systems (see Fig.~S6 in the SM~\cite{SI}). Taken together, these structural signatures highlight the significant influence of ion–solvent coupling across concentration regimes and its strong dependence on the non-electrostatic interactions.

\subsection{Formation of a percolating ionic network is not directly coupled to the structural crossover}

\begin{figure*}[htbp]\centering
\includegraphics[width=\textwidth]{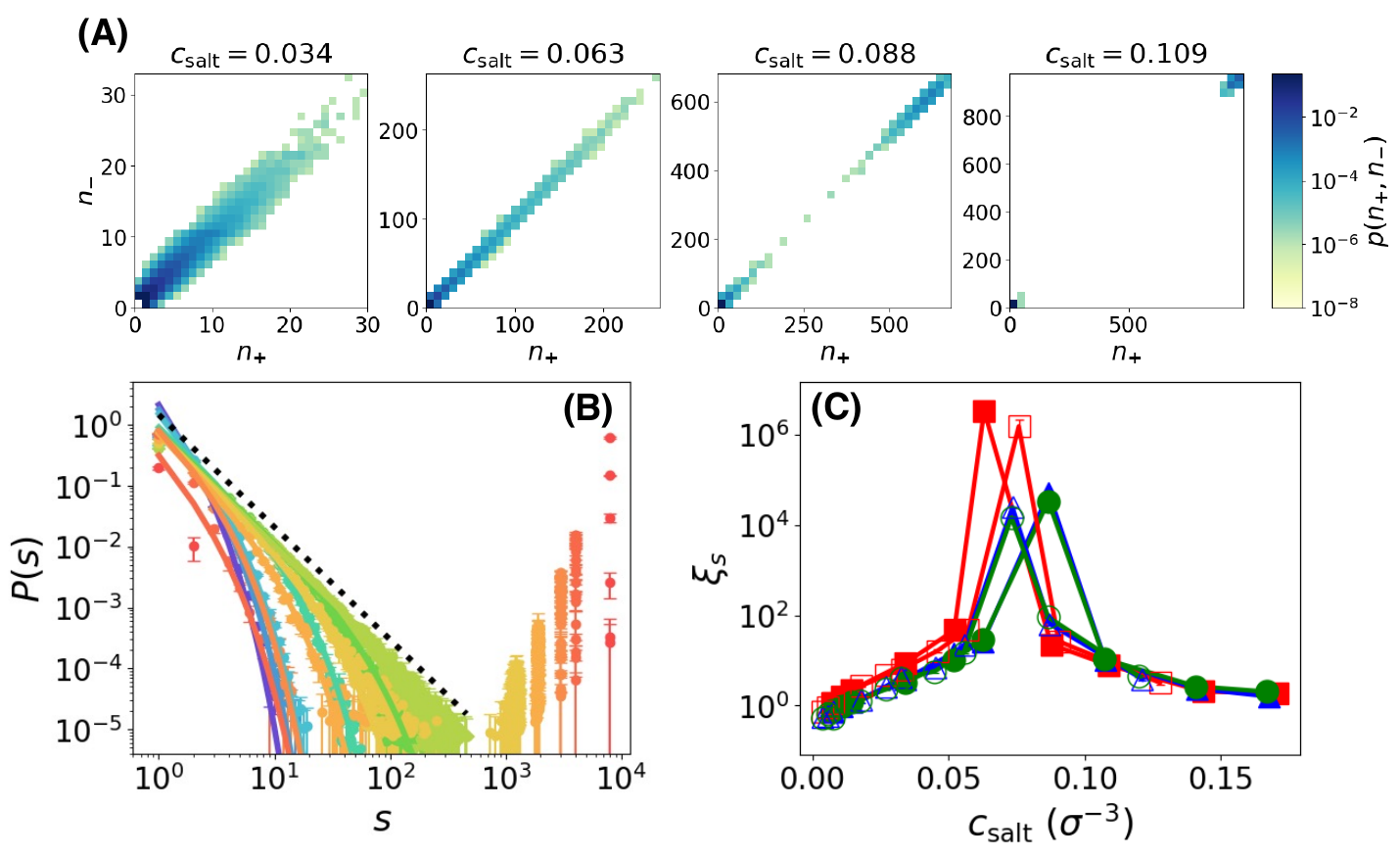}
\caption{Percolation transition of ionic clusters in explicit-solvent electrolytes. (A) Evolution of ionic clusters with salt concentration $c_{\mathrm{salt}}$. The composition of each cluster is defined by the number pair ($n_+$, $n_-$). (B) Size distribution $P(s)$ (Eq.~\ref{eq:cluster}) of ionic clusters with $s = n_+ + n_-$. $c_{\mathrm{salt}}$ increases with the color changing from purple to green to red. The black dotted line indicates the power-law decay at the percolation threshold, $c_{\mathrm{gel}}\approx0.063~\sigma^{-3}$, $P(s)\sim s^{-1.85}$. Panels (A) and (B) show results for explicit-solvent LJ electrolytes with $\varepsilon_s = 0.2$. (C) Dependence of the exponential decay length $\xi_s$, extracted from $P(s)$, on salt concentration $c_{\mathrm{salt}}$ across all explicit-solvent electrolytes. In all subpanels of panel (A), the probability at the point $(n_+,n_-)=(0,0)$ is zero; the apparent non-zero value indicated by the dark blue color arises solely from the finite binning used in constructing the heat map. In panel (C), the colors and markers are consistent with those in Fig.~\ref{fig:solvent}.} \label{fig:cluster}
\end{figure*}

In an attempt to examine a connection between the screening crossover and the formation of a percolating ionic network, we quantify the growing size of ionic clusters with increasing $c_{\mathrm{salt}}$ due to increased ion-ion correlations (Figs.~\ref{fig:cluster}). In defining an ionic cluster, a pair of a cation and an anion belongs to the same cluster if their interionic distance is within the cutoff distance $r_{cut}=1.5~\sigma$ taken from the first minimum of $g_{+-}(r)$ (see Figs.~S10 and S14-S15 in the SM~\cite{SI}). Due to the charge symmetry in the systems of interest, the numbers of constituent cations $n_+$ and anions $n_-$ of an ionic cluster are the same. We characterize the ionic clusters by their size distribution $P(s)$ (Fig.~\ref{fig:cluster} and Tab.~\ref{tab:cluster}). Two characteristic length scales ($\gamma_s$ and $\xi_s$ in Eq.~\ref{eq:cluster}) describe different parts of the distribution: the former governs the exponential decay, while the latter characterizes the power-law tail. At sufficiently low $c_{\mathrm{salt}}$, $P(s)$ is well described by the exponential part alone with its decay length $\xi_s$. Far below the threshold salt concentration (the pre-gel regime), $\gamma_s$ increases with $c_{\text{salt}}$ as more ionic clusters form, while the contribution of the power-law tail grows with $\gamma_s$, approaching the threshold. At the threshold concentration $c_{\mathrm{gel}}$, $P(s)$ follows a pure power-law distribution, indicating the emergence of a scale-free ionic \textit{network} spanning the simulation box. This threshold corresponds to the gelation or percolation point ($c_{\text{salt}}\approx0.063~\sigma^{-3}$ for LJ electrolytes with $\varepsilon_s=0.2$). In the post-gel regime, the power-law tail diminishes with increasing salt concentration, as a larger fraction of ions participates in the spanning network, thereby reducing the relative contribution of small ionic clusters.

\begin{table}[h]
\centering
\begin{tabular}{l|c||cc}
\hline
System & $\varepsilon_s$ & $c_{\text{gel}}$ ($\sigma^{-3}$) & $\gamma_s(c_{\text{gel}})$ ($\sigma$) \\
\hline
LJ  & 5   & 0.087 & 1.84$\pm$0.03  \\
LJ  & 1   & 0.087 & 1.84$\pm$0.05  \\
LJ  & 0.2 & 0.063 & 1.85$\pm$0.01 \\
\hline
WCA & 5   & 0.073 & 1.85$\pm$0.06  \\
WCA & 1   & 0.074 & 1.89$\pm$0.07  \\
WCA & 0.2 & 0.076 & 1.97$\pm$0.01 \\
\hline
\end{tabular}
\caption{Formation of a percolating ionic network near the gelation point, $c_{\text{gel}}$, in the explicit-solvent electrolytes. The exponent $\gamma_s(c_{\text{gel}})$ quantifies the power-law tail of the size distribution, $P(s)$ (see Eq.~\ref{eq:cluster} and Fig.~\ref{fig:cluster}(B)).}\label{tab:cluster}
\end{table}

Despite the intriguing $c_{\text{salt}}$ dependence of $P(s)$, the formation of an ionic network does not appear to be strongly correlated with the screening crossover (Fig.~\ref{fig:cluster}(B)). Instead, the emergence of a percolating network seems to be primarily governed by the spatial proximity of ions, as it occurs even in relatively dilute electrolytes (e.g., $\varepsilon_s = 5$), with gelation taking place at a similar transition concentration, $c_{\mathrm{gel}}$ (Tab.~\ref{tab:cluster}). Therefore, the screening crossover appears to be decoupled from the ionic network formation in these solvent-dense concentrated electrolytes. This decoupling can be rationalized by the distinct physical requirements of the two phenomena: percolation requires sufficiently extensive aggregation to form a system-spanning network and thus typically occurs only at higher salt concentrations, whereas the onset of underscreening is associated with the emergence of local ion pairing or small aggregates.
We note that the emergence of a percolating network, in general, provides a structural basis for strong correlations in ionic motion~\cite{mceldrew2020theory,mceldrew2021ion,mceldrew2021salt,goodwin2022gelation}. Recent studies~\cite{pincus2023soft,skarmoutsos2025length} have further suggested its relation to the cubic scaling observed in SFB experiments. In connection with the growing ionic domains, two competing length scales were introduced—the screening length, defined in terms of renormalized charges, and the average cluster size, which increases up to the percolation threshold—whose larger value is proposed to govern the scaling behavior. 

The Cayley-tree-based ionic gelation theory~\cite{mceldrew2020theory,mceldrew2021correlated,mceldrew2021salt,goodwin2023theory,zhang2024long} provides a theoretical framework for describing ionic clusters in concentrated electrolytes at the mean-field level, thereby linking structural changes—particularly gelation—to other physical properties. This framework has been successfully applied to a range of concentrated electrolytes, including (diluted) ionic liquids and their electrochemical interfaces. A key assumption of the theory is that ions form Cayley-tree clusters among other possible morphologies, such as chains, rings, and branched ionic aggregates. In contrast, owing to the simplicity of our electrolyte models, which employ isotropic interaction potentials between all ions and solvent particles, the resulting association patterns are more complex than ideal Cayley-tree-like structures (see Figs.~S30-S31 in the SM~\cite{SI}), particularly owing to the presence of co-ion associations and intra-cluster loops. These deviations, together with ambiguities in defining the ion-association functionality - a crucial input parameter of the theory - hinder the theory from accurately locating the gelation threshold $c_{\text{gel}}$ of our simple models over the estimated counterion-association functionality range of 2-5.

Similar limitations of the ionic gelation theory have been reported previously; for example, theoretical predictions of gelation points were found to be systematically lower than those estimated from simulations for nonaqueous LiPF$_6$ and LiBF$_4$ electrolytes, a discrepancy likewise attributed to deviations from ideal Cayley-tree clusters~\cite{mceldrew2021salt}. Nevertheless, we find that below $c_{\text{gel}}$ the ionic gelation theory captures the counterion association constant reasonably well, yielding values of comparable order of magnitude to those observed in simulations when an appropriate ion-association functionality is chosen (see Fig.~S32 in the SM~\cite{SI}). A detailed analysis and discussion of ionic association, together with its comparison with the ionic gelation theory, are provided in Sec.~S10 of the SM~\cite{SI}.

\subsection{Dynamic crossover of ion self-diffusion and ion-pair lifetime}\label{sec:dynamic}
\begin{figure*}[htbp]\centering
\includegraphics[width=\textwidth]{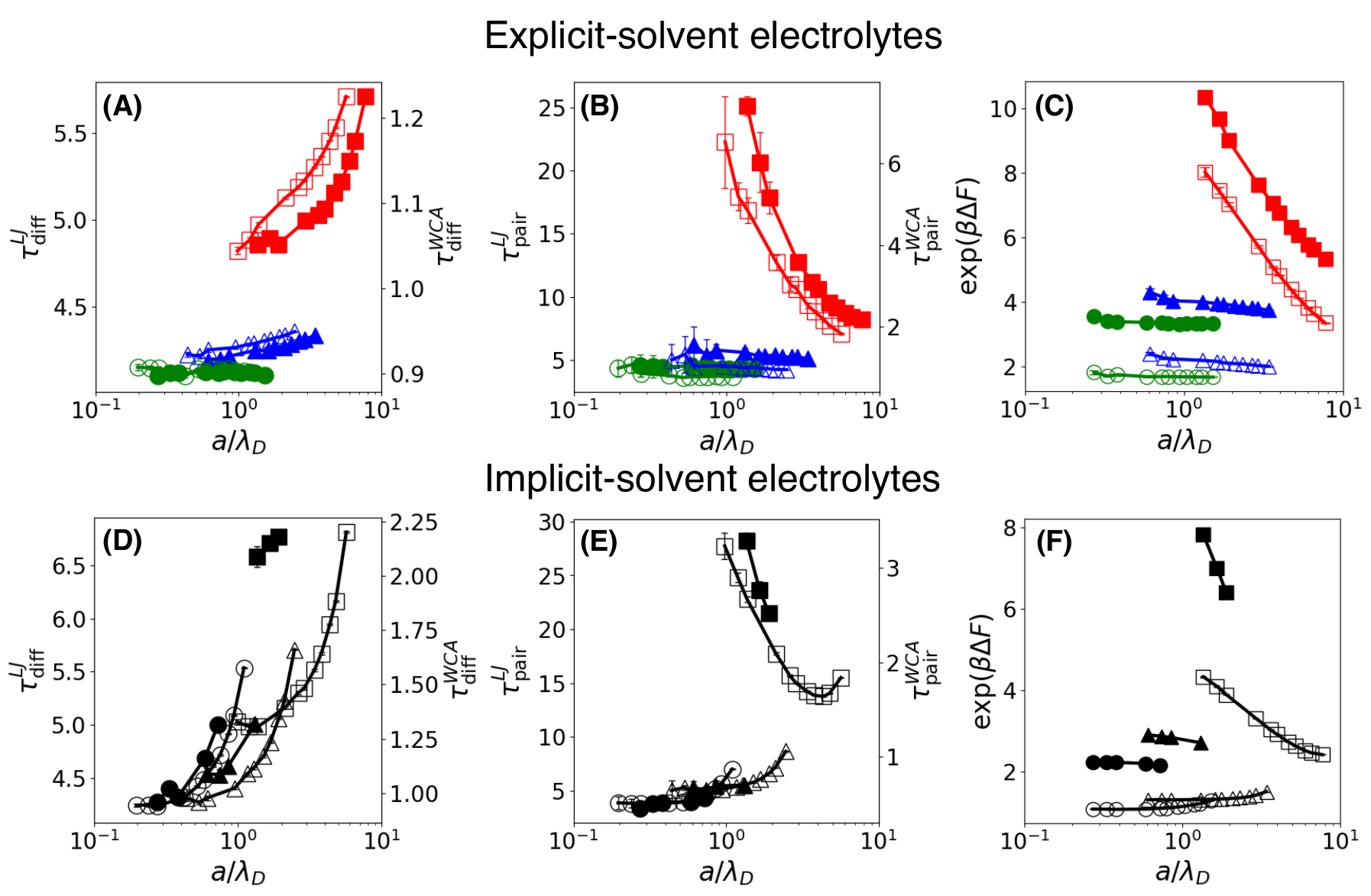}
\caption{Crossovers in ion dynamics and their structural origins as a function of the ratio $a/\lambda_D$ across all electrolytes examined. 
(A, D) Diffusion relaxation time, $\tau_{\mathrm{Diff}}$. 
(B, E) Mean ion-pair lifetime, $\tau_{\mathrm{pair}}$. 
(C, F) Free-energy barrier for ion-counterion dissociation, $\beta\Delta F$. 
Panels (A–C) on the top row correspond to explicit-solvent models, while panels (D–F) on the bottom row correspond to implicit-solvent models.
In panels (A-B) and (D-E), the left and right axes represent results for LJ (filled markers) and WCA (open markers), respectively. 
For the LJ electrolytes, the colors and markers are consistent with those in Fig.~\ref{fig:solvent}, and the black counterparts correspond to the WCA electrolytes.}\label{fig:msd}
\end{figure*}

We examine $c_{\text{salt}}$-dependent dynamic properties, including ion self-diffusion and ion-pair lifetime across the structural crossover from the charge-dominated dilute regime to the density-dominated concentrated regime. In particular, the dynamic properties are organized with respect to the ratio $a/\lambda_D$, in the same manner as for the structural correlations, enabling a systematic investigation across a wide range of $c_{\text{salt}}$ (top row of Fig.~\ref{fig:msd}). This representation is also consistent with the Debye–H\"uckel–Onsager (DHO) and DFT theories~\cite{demery2016conductivity,avni2022conductivity,illien2024stochastic}, where $a/\lambda_D$ serves as a physically meaningful variable for describing ionic transport.

Figure~\ref{fig:msd}(A) displays the diffusion relaxation time defined by ion self-diffusion, $\tau_{\mathrm{diff}}=\sigma^2/6D_{\mathrm{ion}}$, which exhibits crossover behavior (see Fig.~S19(A,D) for ion self-diffusion coefficients $D_{\mathrm{ion}}$ in the SM~\cite{SI}). It is clear that plotting the results against $a/\lambda_D$ is better than $c_{\text{salt}}$ for systematic comparison (see Fig.~S19 in the SM~\cite{SI}). Ions in the LJ electrolytes diffuse approximately four times more slowly than those in the WCA electrolytes, as expected from the increased viscosity arising from non-electrostatic attractive interactions (see Fig.~S18 for solvent diffusion coefficients in the SM~\cite{SI}). The relaxation time $\tau_{\mathrm{diff}}$ exhibits a discontinuous change across the crossover. When $a\lambda^{-1}_D\lesssim1$, $\tau_{\mathrm{diff}}$ depends only weakly on $c_{\text{salt}}$ (green and blue markers in Fig.~\ref{fig:msd}), whereas in the concentrated regime, strong ion–ion electrostatic interactions further suppress ion diffusion, as clearly demonstrated for $\varepsilon_s=0.2$ (red markers in Fig.~\ref{fig:msd}). In the concentrated WCA electrolytes with $\varepsilon_s = 0.2$, the ratio $D_{\text{ion}}/D_{\text{solv}}$ decreases noticeably, implying a breakdown of the Stokes–Einstein relation (see Fig.~S19(C, F) in the SM~\cite{SI}).

A clear crossover is also observed in the ion-pair lifetime $\tau_{\mathrm{pair}}$ when plotted against $a/\lambda_D$ (Fig.~\ref{fig:msd}(B)). The lifetime was estimated using the ion-pair survival function $H(t)$ (Eq.~\ref{eq:pairdyn}, see Figs.~S23 and~S24 in the SM~\cite{SI}). Both dilute LJ and WCA electrolytes exhibits the nearly constant lifetime of cation–anion pairs, independent of both $c_{\text{salt}}$ and $\varepsilon_s$. Upon entering the concentrated regime, however, this behavior changes markedly with a notable discontinuity. For both LJ and WCA electrolytes with $\varepsilon_s=0.2$, $\tau_{\mathrm{pair}}$ decreases at a fast rate, i.e., a rapid decay of $H(t)$ with increasing $c_{\text{salt}}$ despite the enhanced ion–ion correlations and increased viscosity. The short-lived nature of ion pairs is often presumed to manifest the formation of an ionic network that enables frequent partner exchange~\cite{mceldrew2021salt,kazmierczak2021dynamics,bi2024cluster}. Whereas ion pairs in dilute electrolytes can persist until they encounter other pairs or free ions, the dense ionic environment in concentrated systems can promote facile partner exchange without compromising the integrity of the network~\cite{bi2024cluster}. 

Instead of being indirectly linked to the percolating ion network, these dynamic crossovers are more directly encoded in the underlying ion structures at short distances, particularly in the potential of mean force between cations and anions: $\beta w_{+-}(r)=-\ln[g_{+-}(r)]$ (see Fig.~S14, S15 in the SM~\cite{SI}). Each minimum in $w_{+-}(r)$ corresponds to a distinct solvation state of the ion pair, such as contact, solvent-shared, solvent-separated, or separated ones~\cite{guardia1991potential,guardia1991na,pratt1994ion}. Thus, the profile of $\beta w_{+-}(r)$ represents the free-energy landscape governing ion-pair association and dissociation, which can be viewed as a barrier-crossing process~\cite{geissler1999kinetic}. The non-electrostatic attractive interactions result in a clear distinction in $w_{+-}(r)$ between LJ and WCA electrolytes, similar to the differences observed in $g_{NN}(r)$ and $g_{ss}(r)$: $w_{+-}(r)$ is more structured in LJ systems at short distances. Despite this difference, both LJ and WCA electrolytes exhibit $w_{+-}(r)$ that remains largely insensitive to $c_{\text{salt}}$ in the dilute regime ($\varepsilon_s=1$ and 5). 

In contrast, in the density-dominated concentrated regime ($\varepsilon_s=0.2$), $w_{+-}(r)$ becomes strongly dependent on $c_{\text{salt}}$. This dependence is most clearly reflected in the free energy barrier $\Delta F$ between the first and second solvation states of the ion pair  (Fig.~\ref{fig:msd}(C)). While $\Delta F$ remains nearly constant in dilute electrolytes, it progressively decreases with increasing $c_{\text{salt}}$ in concentrated electrolytes, driven by pronounced ion–ion correlations. The reduction in $\beta\Delta F$ accelerates the cation-anion dissociation, leading to a shorter $\tau_{\mathrm{pair}}$. Notably, the barrier itself also exhibits a discontinuity between the dilute and concentrated systems, as observed in $\tau_{\mathrm{pair}}$ and $\tau_{\mathrm{diff}}$. These trends are consistently observed for both LJ and WCA electrolytes.

In the case of ionic conductivity $\sigma$ (see Sec.~S9 in the SM~\cite{SI} for the results and computational details), it is more appropriately represented as a function of $c_{\mathrm{salt}}$ because of the differences in $l_B$, as conventionally employed following the DHO theory and its extensions~\cite{avni2022conductivity,kalikin2024modified}. Across all explicit-solvent systems, $\sigma$ exhibits no turnover; instead, it increases monotonically with increasing $c_{\mathrm{ion}}$ within the range examined. This observation indicates that the screening crossover does not necessarily result in a decrease in ionic conductivity. The monotonic behavior of $\sigma$ primarily reflects the relatively modest changes in $D_{\mathrm{ion}}$ with $c_{\mathrm{salt}}$ across the crossover (Fig.~\ref{fig:msd}).

The ionicity $\sigma/\sigma_{\mathrm{NE}}$ exhibits an intriguing $c_{\mathrm{salt}}$ dependence, quantifying the degree of correlated ion motion (see Fig.~S29 in the SM~\cite{SI}). Unlike the ionic conductivity, $\sigma/\sigma_{\mathrm{NE}}$ is better represented with $a/\lambda_D$ than with $c_{\mathrm{salt}}$, clearly showing that it gradually decreases from unity to $\sim 0.5$ as the screening crossover is approached. This decrease indicates the breakdown of the Nernst-Einstein approximation due to dynamic correlations, particularly through the ion-counterion association. The ionicity appears to remain nearly constant in the concentrated regime, with no further decrease or upturn observed within the $c_{\mathrm{salt}}$ window examined. Overall, this behavior reflects enhanced dynamic correlations and collective ion motion with increasing $c_{\mathrm{salt}}$, arising from the growth of underlying structural correlations.

\subsection{Screening crossover between two charge-dominated regimes in implicit-solvent electrolytes}\label{subsec:implicit}

\begin {figure}[htbp]\centering
\includegraphics [width=3in] {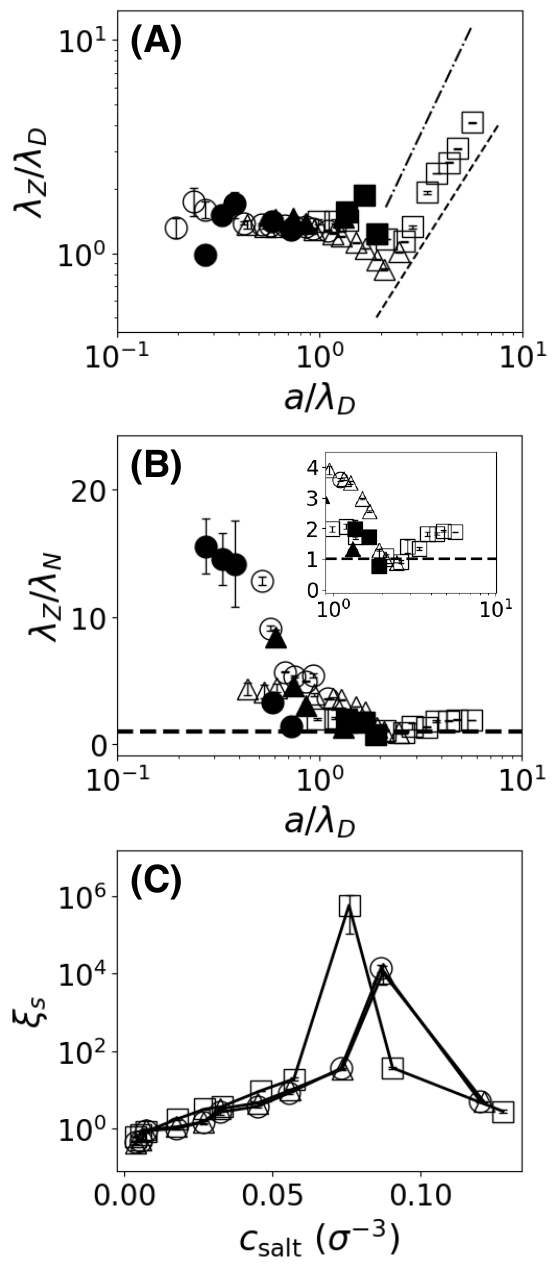}
\caption{Crossover in structural correlations of implicit-solvent electrolytes. (A) Decay length $\lambda_Z$ of $h_{ZZ}(r)$ as a function of $a/\lambda_D$, where $a$ is the ion size and $\lambda_D$ is the Debye screening length (Eq.~\ref{eq:debye}). Black dash–dot ($\alpha=2$) and dashed ($\alpha=1.5$) lines serve as visual guides for the scaling behavior described in Eq.~\ref{eq:scaling}. (B) Ratio of decay lengths $\lambda_Z/\lambda_N$ as a function of $a/\lambda_D$. Black dashed horizontal line indicates $\lambda_Z/\lambda_N=1$. The inset shows a magnified view of the transition region. (C) The growth and percolation of ionic clusters with salt concentration $c_{\mathrm{salt}}$, characterized by the exponential decay length $\xi_s$ extracted from $P(s)$ (Eq.~\ref{eq:cluster}). The colors and markers are consistent with those in Fig.~\ref{fig:msd}(D-F). Structural correlation functions and their corresponding fitting results for the implicit-solvent systems are provided in the SM~\cite{SI}.}\label{fig:implicit}
\end{figure}

To examine the effects of strong solvent-ion coupling, we performed Langevin dynamics (LD) simulations without explicit, space-filling solvent particles as a limiting case. The implicit-solvent models successfully reproduce the essential ion–ion structural features observed in the explicit-solvent systems at both short and long distances (see Sec.~S4 in the SM~\cite{SI}). As in the explicit-solvent models, differences in non-electrostatic attractions account for the variations in short-range ion structure: WCA ions display minimal short-range density correlations in $g_{NN}(r)$, whereas LJ ions exhibit pronounced short-range structuring, indicative of the onset of phase segregation~\cite{ribar2024cluster}. Consequently, because $h_{NN}(r)$ does not exhibit a well-defined exponential or oscillatory exponential decay in these cases, $\lambda_N$ could not be extracted for some implicit-solvent systems.

In the absence of explicit solvent, a similar screening crossover also emerges (Fig.~\ref{fig:implicit}(A)), with a slight difference in the scaling exponent in Eq.~\ref{eq:scaling}, consistent with predictions from classical density functional theory~\cite{Rotenberg2018}. However, the screening transition occurs between two charge-dominated regimes—from a dilute to a concentrated one in the implicit-solvent WCA systems with $\varepsilon_s = 0.2$. This is evidenced by the ratio $\lambda_Z/\lambda_N > 1$ (Fig.~\ref{fig:implicit}(B)), in clear contrast to the behavior observed in the explicit-solvent systems (Fig.~\ref{fig:gzzgnn}(F)). This difference underscores the role of ion–solvent coupling in the screening transition, consistent with previous theoretical predictions~\cite{PRL2018hartel}.

Ion dynamics also show notable differences depending on the presence of explicit solvent. Ion self-diffusion in the implicit-solvent models (bottom row in Fig.~\ref{fig:msd}) slows down with increasing $c_{\mathrm{salt}}$ at a markedly faster rate than in the explicit-solvent systems. This slowdown is gradual in both the dilute and concentrated regimes, leading to a smooth increase in $\tau_{\mathrm{diff}}$. Ion-pair lifetimes exhibit a similar trend in the dilute systems ($\varepsilon_s = 1$ and 5), where $\tau_{\mathrm{pair}}$ increases slowly with salt concentration.

Notably, $\tau_{\mathrm{pair}}$ exhibits a turnover in the concentrated WCA electrolytes ($\varepsilon_s = 0.2$), a behavior not observed in the other systems: it initially decreases—consistent with the explicit-solvent counterparts—but increases again at the highest salt concentrations. This increase cannot be attributed to changes in the ion–counterion local structure $w_{+-}(r)$, as $\Delta F$ decreases monotonically over the entire range of $c_{\mathrm{salt}}$. Instead, the turnover may reflect the emergence of a transition to a charge-dominated concentrated regime, rather than to a density-dominated one, or it may be associated with the percolation transition. Further systematic studies are warranted to resolve this intriguing behavior, which is beyond the scope of the present work.

\subsection{Diffusion-corrected ion-pair lifetime combines ion structure and dynamics}\label{sec:corrected}

\begin {figure}[htbp]\centering
\includegraphics [width=3.2in] {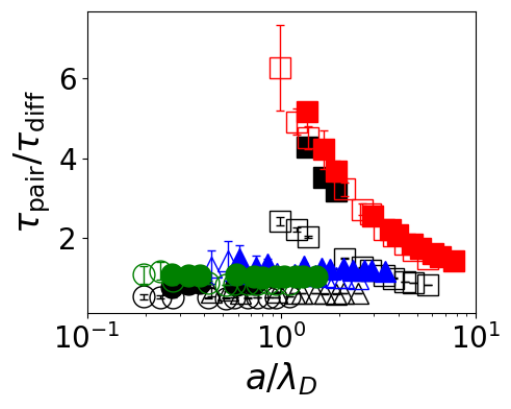}
\caption{Diffusion-corrected ion-pair lifetime $\tau_{\mathrm{pair}}/\tau_{\mathrm{Diff}}$ as a function of the ratio $a/\lambda_D$ across all electrolytes examined in this work. For the LJ electrolytes, the colors and markers are consistent with those in Fig.~\ref{fig:solvent}, and the black counterparts correspond to the WCA electrolytes.}\label{fig:relax}
\end{figure}

The distinct $c_{\text{salt}}$-dependence of ion-pair dynamics and ion self-diffusion, as shown in Fig.~\ref{fig:msd}, gives rise to a \textit{bifurcation} in the relaxation times across the crossover. In the charge-dominated dilute regime, both ion diffusion and ion-pair dynamics in explicit-solvent systems depend relatively weakly on $c_{\text{salt}}$. By contrast, in the concentrated regime, either density- or charge-dominated, ion diffusion slows down while the ion-pair lifetime decreases with increasing $c_{\text{salt}}$. This bifurcation behavior is consistently observed across all electrolytes systems studied in this work.

To capture this behavior consistently, we propose the diffusion-corrected ion-pair lifetime, $\tau_{\text{pair}}/\tau_{\text{diff}}$, which provides a unified basis for interpreting ion dynamics in relation to the underlying ion structure (Fig.~\ref{fig:relax}). The ratio effectively compensates for viscosity-related effects that primarily arise from differences in non-electrostatic interactions in explicit-solvent systems. Remarkably, the dependence of $\tau_{\text{pair}}/\tau_{\text{diff}}$ on the structural parameter $a/\lambda_D$ parallels that of the barrier-crossing factor $\exp(\beta\Delta F)$, underscoring the close connection between ion-pair dynamics and the short-range potential of mean force, $\beta w_{+-}(r)$.

Consistent with the behavior of $\exp(\beta\Delta F)$, the ratio $\tau_{\text{pair}}/\tau_{\text{diff}}$ exhibits a clear discontinuity across different $\varepsilon_s$: it remains nearly constant in the dilute regime ($\varepsilon_s = 1$ and 5), but decreases systematically in the concentrated regime ($\varepsilon_s = 0.2$). Notably, $\tau_{\text{pair}}/\tau_{\text{diff}}$ further enables qualitative comparison across models with and without explicit solvents, despite quantitative deviations that are more pronounced in the charge-dominated concentrated regime (black open squares in Fig.~\ref{fig:relax}). The comparatively small decrease observed for the implicit-solvent WCA electrolytes can be attributed to their relatively small free-energy barrier $\Delta F$ (Fig.~\ref{fig:msd}(F)). These quantitatively distinct dynamical behavior between charge-dominated and density-dominated concentrated regimes warrants further detailed investigation.

\section{Conclusions}\label{sec:conclusions}
Using molecular dynamics simulations of model electrolytes with and without explicit solvent, we systematically investigated the relationships between structural and dynamical correlations in dense ionic fluids across a wide range of salt concentrations. We find that the electrostatic screening crossover depends sensitively on ion–solvent coupling. In explicit-solvent electrolytes, the crossover occurs between a charge-dominated dilute regime and a density-dominated concentrated regime, whereas in implicit-solvent electrolytes, it occurs between two charge-dominated regimes—from dilute to concentrated. Both explicit- and implicit-solvent systems display dynamical crossovers in ion self-diffusion and ion-pair lifetimes that are coupled to the screening transition, although the detailed dynamical behaviors differ depending on the presence of solvent. The notable discontinuity across the dynamical transition is well encoded in the ion–counterion short-range structures. Ionic clusters grow with increasing salt concentration and eventually form a percolating network beyond a threshold concentration. However, the percolation transition is not strongly coupled to the onset of either the structural or dynamical crossovers and appears instead to be governed primarily by salt concentration. This decoupling can be rationalized by the distinct physical requirements of the two phenomena: percolation requires sufficiently extensive aggregation to form a system-spanning network and thus typically occurs only at higher salt concentrations, whereas the onset of underscreening reflects the emergence of local ion pairing or small aggregates. These observations also suggest that spatial proximity alone is insufficient to explain the connection between ionic clustering and transport in these model systems.

For consistent comparison across systems, we demonstrate that the diffusion-corrected ion-pair lifetime, $\tau_{\text{pair}}/\tau_{\text{diff}}$, serves as a unified descriptor that bridges ionic structure and dynamics. The ratio effectively compensates for viscosity-related effects arising from non-electrostatic attractions, while its dependence on the structural parameter $a/\lambda_D$ mirrors the behavior of the barrier-crossing factor $\exp(\beta \Delta F)$ associated with short-range ion–counterion structure, particularly the discontinuity across the dynamical crossover. Moreover, $\tau_{\text{pair}}/\tau_{\text{diff}}$ enables consistent qualitative comparison between explicit- and implicit-solvent electrolytes, despite quantitative differences that become more pronounced at high salt concentrations, which is worth future systematic investigation.

Our simple models inevitably carry limitations, primarily due to the assumption of a continuum dielectric medium, which can lead to excessive screening of electrostatic interactions at unphysically short distances. Future studies may further explore the coupling and decoupling among electrostatic screening, ionic clustering, and charge transport using chemically more detailed models.

\section*{Supplemental material}\label{sec:SI}
Supplemental Material provides detailed summaries and fitting results of structural, dynamical, and transport properties across all electrolyte conditions examined in this study. These include charge–charge, density–density, solvent–solvent, cation–anion, cation-solvent correlation functions, cluster-size distributions, mean-squared displacements of ions and solvent, ion-pair survival functions, ionic conductivity, and a detailed analysis of ionic associations.

\section*{Acknowledgment}
This work was supported by the National Supercomputing Center with supercomputing resources including technical support(KSC-2025-CRE-0142). This research was supported by the Regional Innovation System \& Education(RISE) program through the Jeju RISE center, funded by the Ministry of Education(MOE) and the Jeju Special Self-Governing Province, Republic of Korea(2025-RISE-17-001). 

\section*{Data Availability Statement}\label{sec:DataAvail}
The data that support the findings of this article are openly available~\cite{github}.

\end{document}


\preprint{
}

\title[]{
Supplemental Material for\\
Structural and Dynamical Crossovers in Dense Electrolytes
}

\author{Daehyeok Kim}
\affiliation
{Department of Energy Engineering, Korea Institute of Energy Technology (KENTECH), Naju 58330, Republic of Korea}
\author{Taejin Kwon$^\dagger$}
\email{$^\dagger$tjkwon@jejunu.ac.kr}
\affiliation
{Department of Chemistry and Cosmetics, Jeju National University, Jeju 63243, Republic of Korea}
\author{Jeongmin Kim$^*$}
\email{$^*$jeongmin@pusan.ac.kr}
\affiliation
{Department of Chemistry Education and Graduate Department of Chemical Materials, Pusan National University, Busan 46241, Republic of Korea}

\maketitle

\newpage

\renewcommand{\thesection}{S\arabic{section}}
\setcounter{section}{0}
\renewcommand{\thefigure}{S\arabic{figure}}
\setcounter{figure}{0}
\renewcommand{\thetable}{S\arabic{table}}
\setcounter{table}{0}
\renewcommand{\theequation}{S\arabic{equation}}
\setcounter{equation}{0}

\tableofcontents
\newpage

\section{Salt concentration as a function of salt-to-solvent ratio}
Table~\ref{si:tab:c_saltprop} lists the salt concentration $c_{\text{salt}}$ as a function of the salt-to-solvent ratio $r_{\text{salt}}$ for all explicit-solvent electrolytes examined.
\begin{table}[h]
\centering
\begin{tabular}{|c||c|c|c|}
\hline
$r_{\text{salt}}$ &$c_{\text{salt}}~(\varepsilon_s=0.2)$ & $c_{\text{salt}}~(\varepsilon_s=1)$ & $c_{\text{salt}} ~(\varepsilon_s=5)$\\
\hline
0.01&0.007$\pm$ 0.001 &0.007$\pm$ 0.001 &0.007$\pm$ 0.001 \\
0.015&0.011$\pm$ 0.001 &0.011$\pm$ 0.001 &0.011$\pm$ 0.001 \\
0.02&0.014$\pm$ 0.001 &0.014$\pm$ 0.001 &0.014$\pm$ 0.001 \\
0.05&0.034$\pm$ 0.001 &0.034$\pm$ 0.001 &0.034$\pm$ 0.001 \\
0.08&0.052$\pm$ 0.001 &0.052$\pm$ 0.001 &0.052$\pm$ 0.001 \\
0.1&0.063$\pm$ 0.001 &0.063$\pm$ 0.001 &0.063$\pm$ 0.001 \\
0.15&0.088$\pm$ 0.001 &0.087$\pm$ 0.001 &0.087$\pm$ 0.001 \\
0.2&0.109$\pm$ 0.001 &0.108$\pm$ 0.001 &0.107$\pm$ 0.001 \\
0.3&0.144$\pm$ 0.001 &0.141$\pm$ 0.001 &0.141$\pm$ 0.001 \\
0.4&0.171$\pm$ 0.001 &0.168$\pm$ 0.001 &0.167$\pm$ 0.001 \\
0.8&0.239$\pm$ 0.001 &0.233$\pm$ 0.001 &0.231$\pm$ 0.001 \\
\hline
\hline
0.01&0.004$\pm$ 0.001 &0.004$\pm$ 0.001 &0.004$\pm$ 0.001 \\
0.015&0.006$\pm$ 0.001 &0.006$\pm$ 0.001 &0.006$\pm$ 0.001 \\
0.02&0.008$\pm$ 0.001 &0.007$\pm$ 0.001 &0.007$\pm$ 0.001 \\
0.05&0.018$\pm$ 0.001 &0.018$\pm$ 0.001 &0.018$\pm$ 0.001 \\
0.08&0.027$\pm$ 0.001 &0.027$\pm$ 0.001 &0.027$\pm$ 0.001 \\
0.1&0.033$\pm$ 0.001 &0.033$\pm$ 0.001 &0.032$\pm$ 0.001 \\
0.15&0.046$\pm$ 0.001 &0.045$\pm$ 0.001 &0.045$\pm$ 0.001 \\
0.2&0.057$\pm$ 0.001 &0.056$\pm$ 0.001 &0.056$\pm$ 0.001 \\
0.3&0.076$\pm$ 0.001 &0.074$\pm$ 0.001 &0.073$\pm$ 0.001 \\
0.4&0.091$\pm$ 0.001 &0.087$\pm$ 0.001 &0.087$\pm$ 0.001 \\
0.8&0.128$\pm$ 0.001 &0.121$\pm$ 0.001 &0.120$\pm$ 0.001 \\
\hline
\end{tabular}
\caption{Variation of salt concentration $c_{\mathrm{salt}}$ as a function of the salt-to-solvent ratio $r_{\mathrm{salt}}$ for various $\varepsilon_s$ values: LJ (top) and WCA (bottom) electrolytes.}\label{si:tab:c_saltprop}
\end{table}
\newpage

\section{Decay length and oscillation period of structural correlation functions}
\begin {figure}[htbp]\centering
\includegraphics [width=6in] {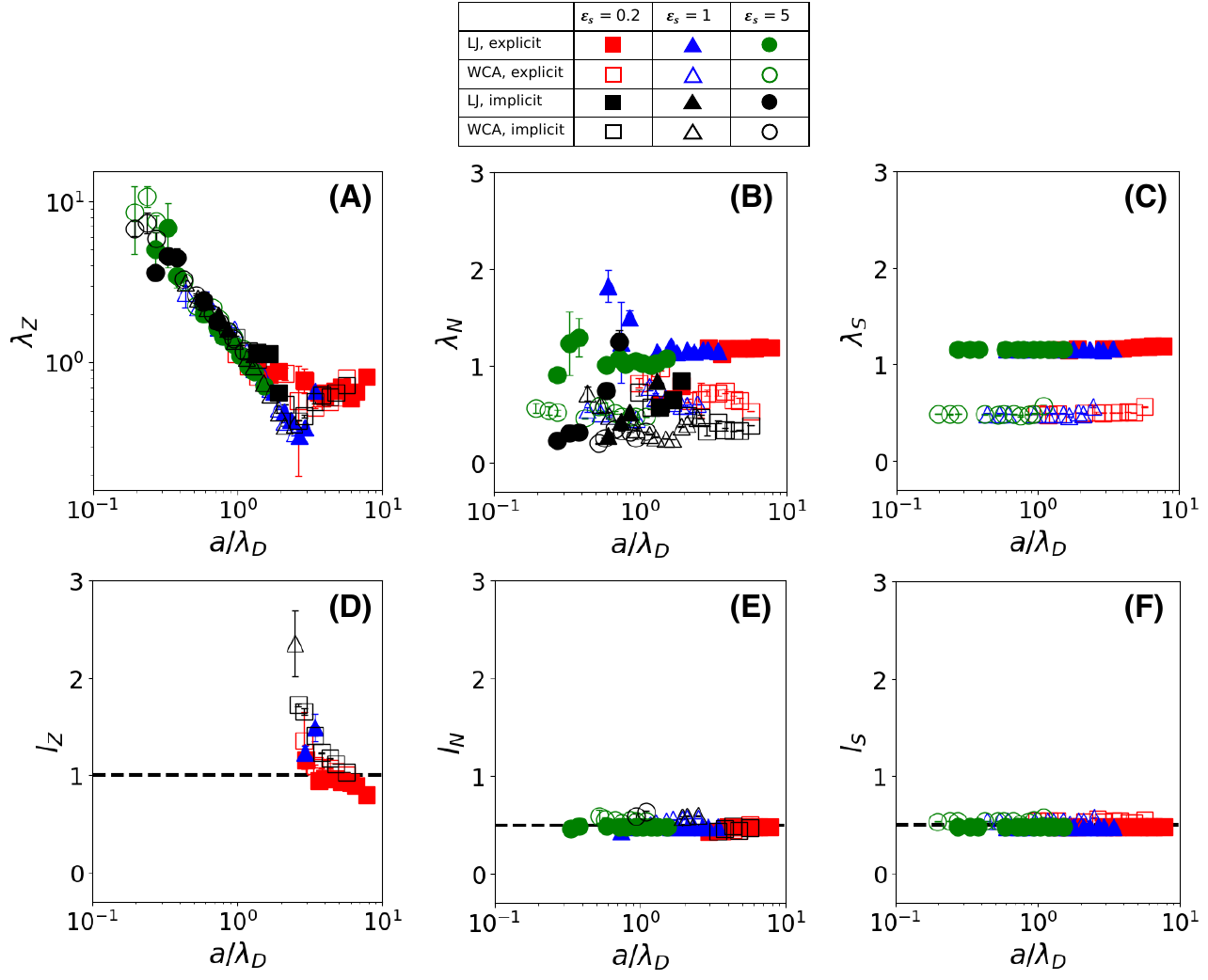}
\caption{Decay lengths and oscillation periods across all electrolytes examined. (A) Decay length $\lambda_Z$ and (D) oscillation period $l_Z$ of $h_{ZZ}(r)$. (B) Decay length $\lambda_N$ and (E) oscillation period $l_N$ of $h_{NN}(r)$. (C) Decay length $\lambda_S$ and (F) oscillation period $l_S$ of $h_{SS}(r)$. Black dashed lines serve as guides to the eye.}\label{si:fig:oscillationperiod}
\end{figure}

Figure~\ref{si:fig:oscillationperiod} displays the decay lengths and oscillation periods across all electrolytes examined. The charge–charge correlation function $h_{ZZ}(r)$ exhibits a simple Yukawa decay in the dilute regime, with the decay length decreasing as $\lambda_Z\sim1/\sqrt{c_{\mathrm{salt}}}$. At higher concentrations, however, an oscillatory contribution emerges, characterized by the oscillation period $l_Z$. In contrast, the density–density correlation function $h_{NN}(r)$ exhibits an oscillatory exponential decay in both dilute and concentrated regimes. The decay length $\lambda_N$ and oscillation period $l_N$ remain nearly constant over the range of $c_{\mathrm{salt}}$ examined.
\newpage
\clearpage

\section{Fitting results of structural correlation functions for explicit-solvent electrolytes}\label{si:sec:corrMD}
\begin{figure}[htbp]
\centering
\includegraphics[width=\textwidth]{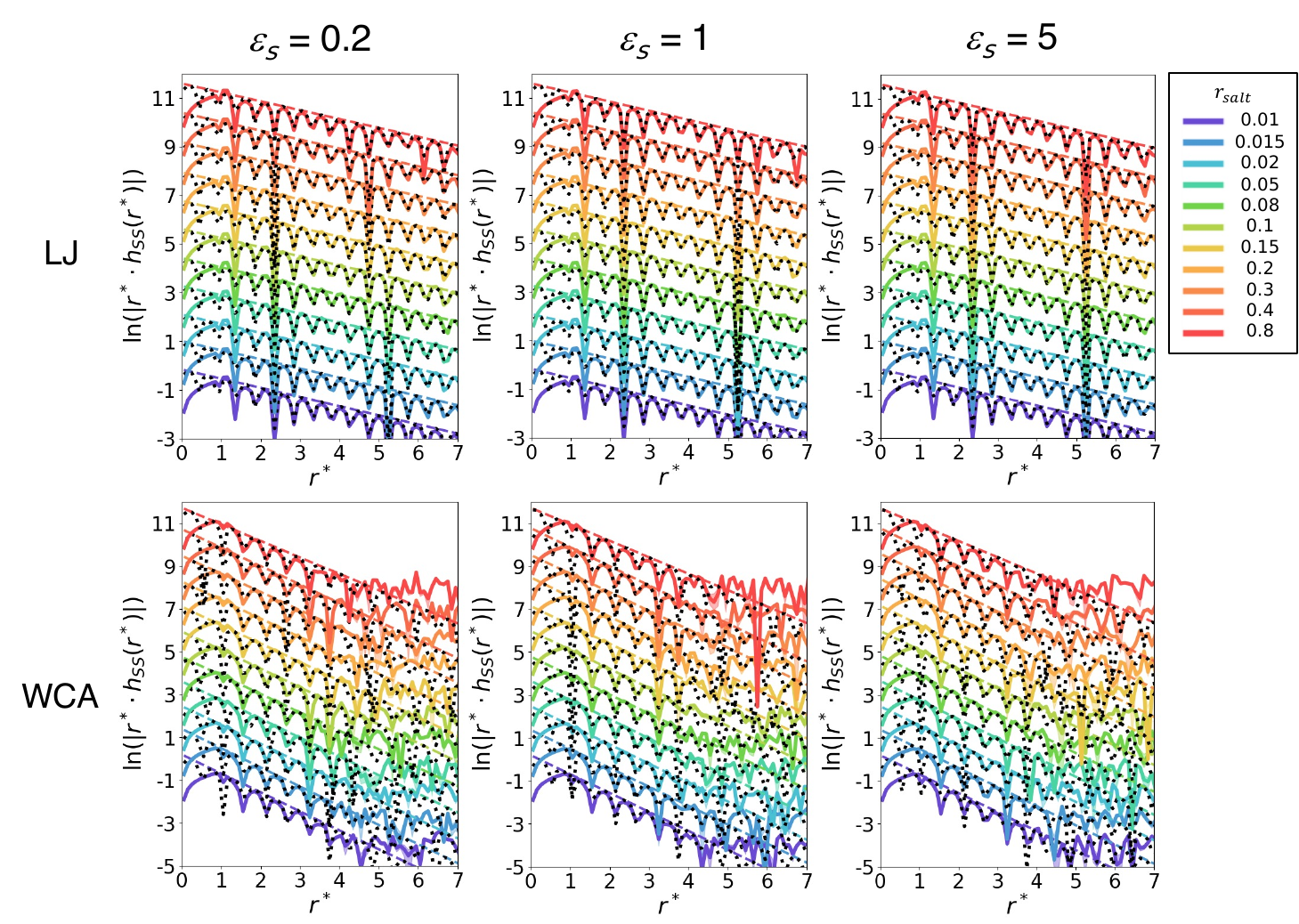} \label{si:fig:gssMD}
\caption{Solvent–solvent structural correlation functions, $|r\cdot h_{SS}(r)|$, of explicit-solvent LJ and WCA electrolytes at various $\varepsilon_s$.}
\end{figure}

\begin {figure}[htbp]\centering
\includegraphics [width=\textwidth] {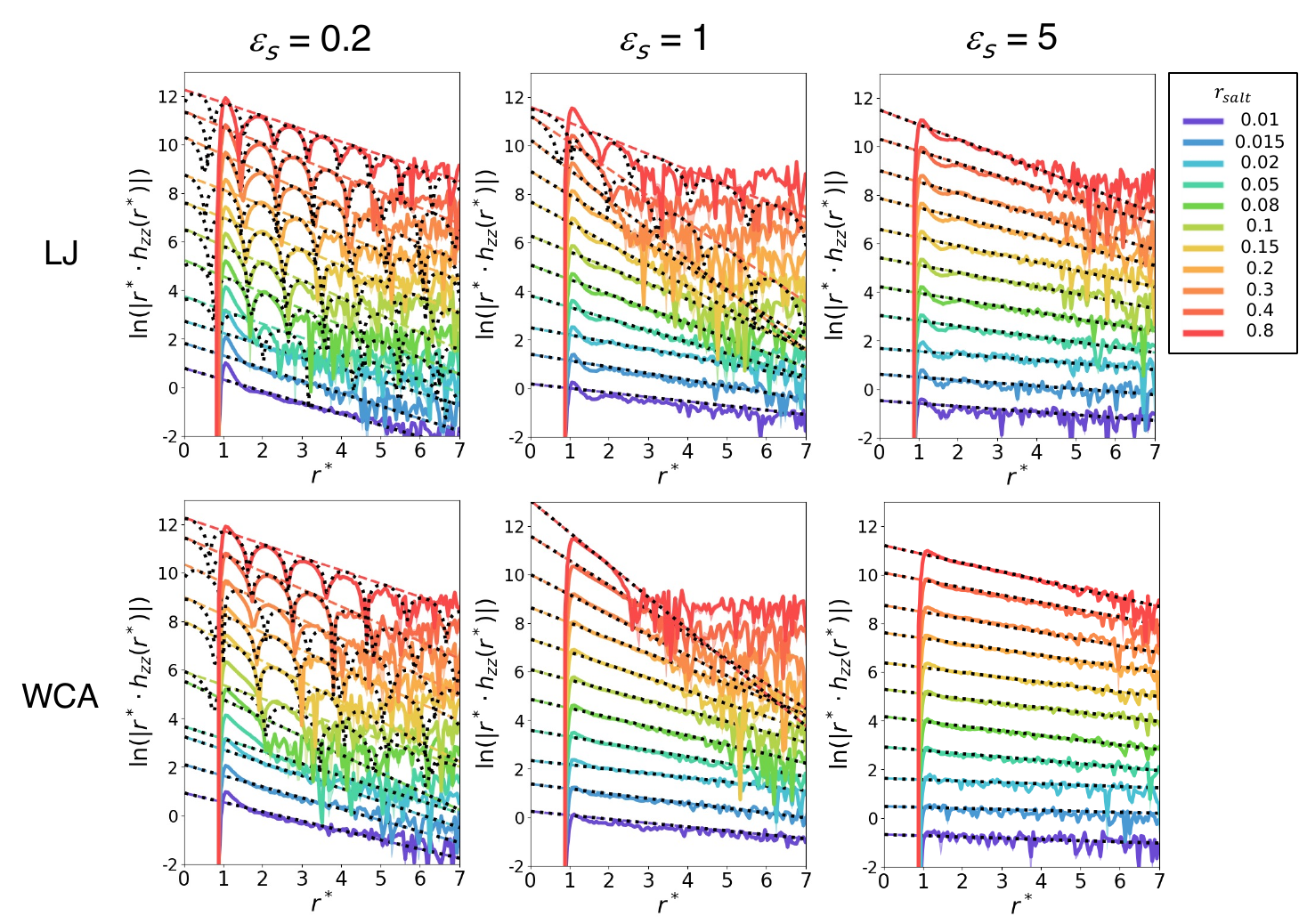} \label{si:fig:gzzMD}
\caption{Charge-charge correlation functions, $|r\cdot h_{ZZ}(r)|$, of explicit-solvent LJ and WCA electrolytes at various $\varepsilon_s$.}
\end{figure}

\begin {figure}[htbp]\centering
\includegraphics [width=\textwidth] {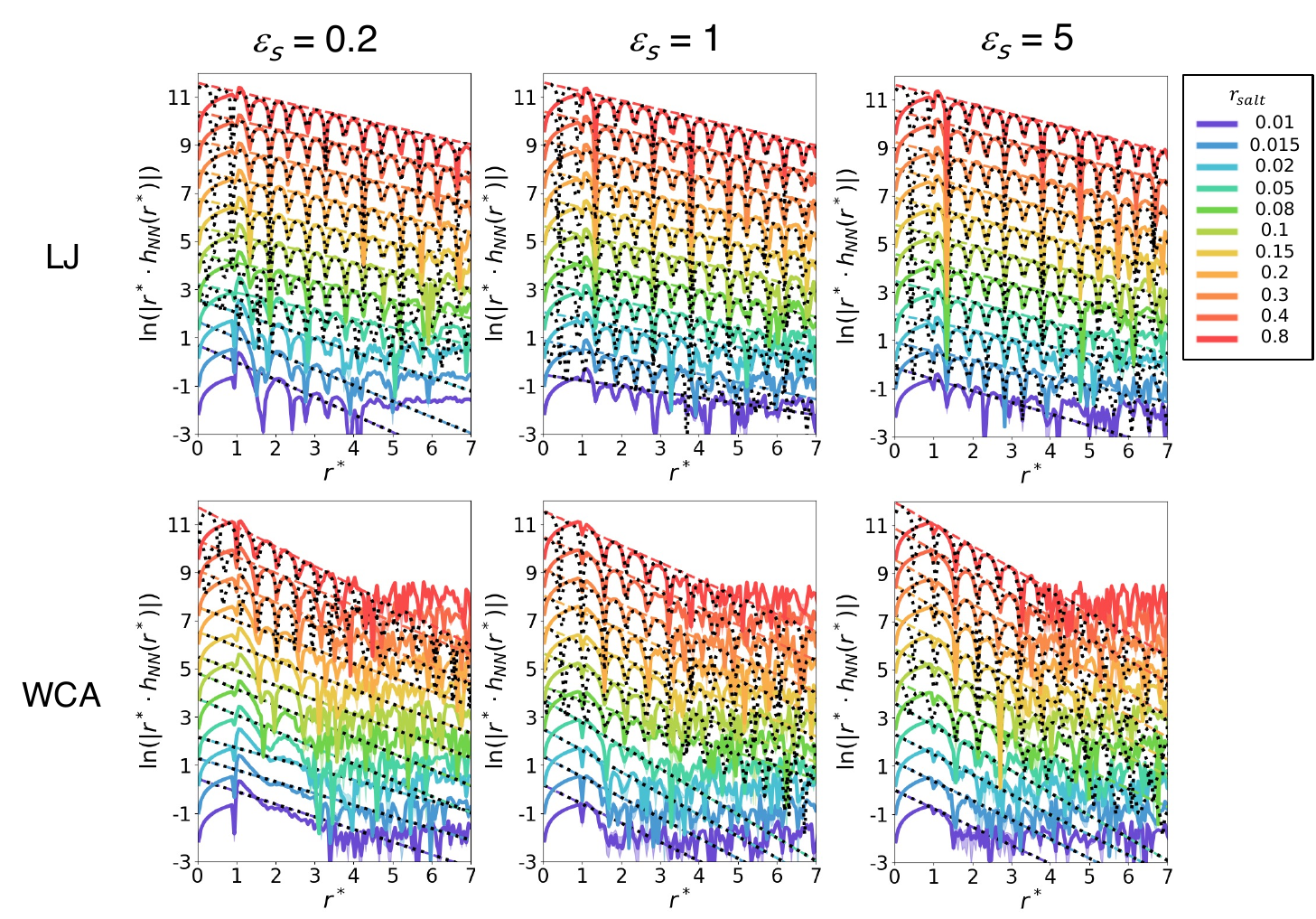}
\caption{Density-density correlation functions, $|r\cdot h_{NN}(r)|$, of explicit-solvent LJ and WCA electrolytes at various $\varepsilon_s$.}
\label{si:fig:gnnMD}
\end{figure}

\begin {figure}[htbp]\centering
\includegraphics [width=\textwidth] {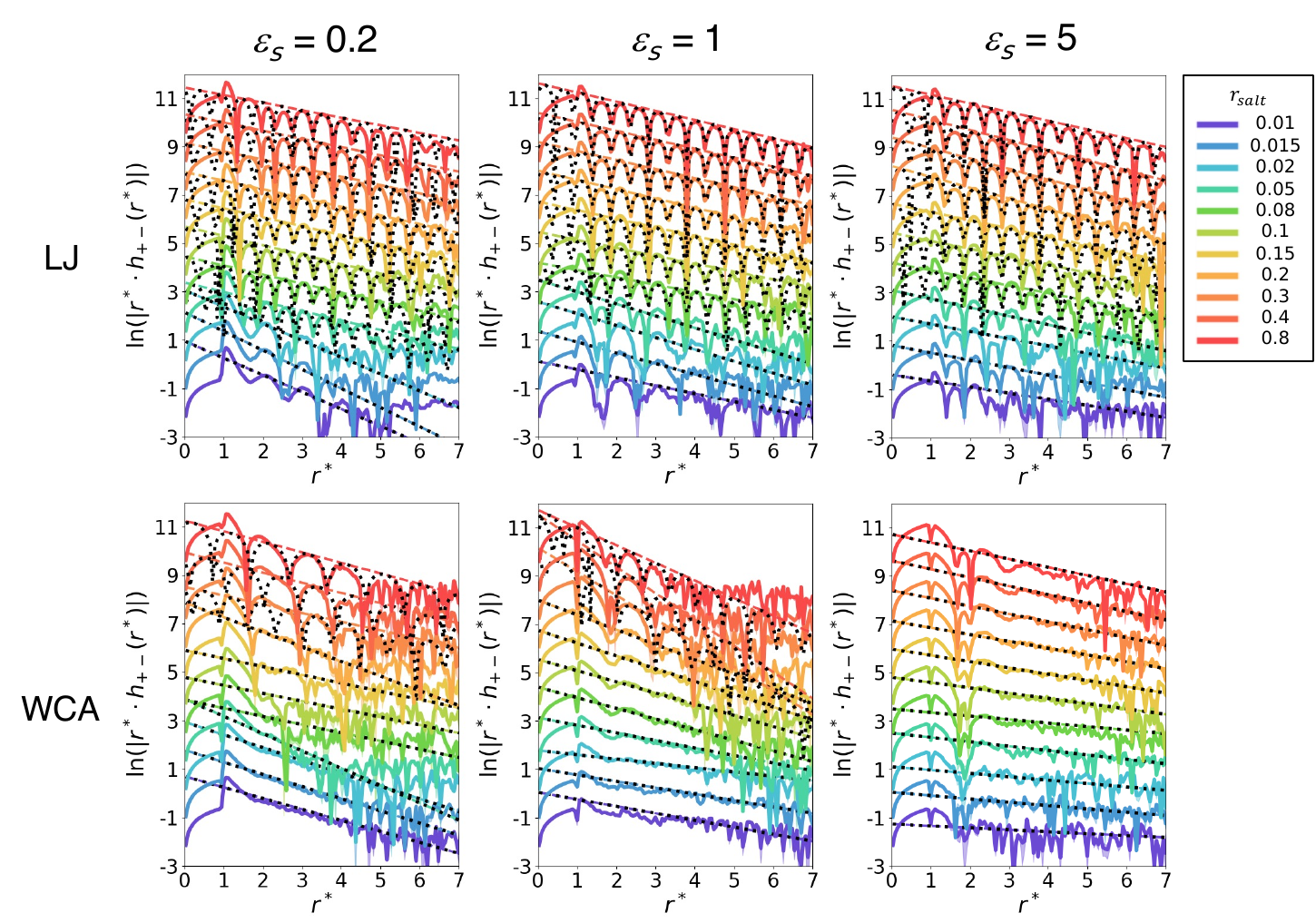}
\caption{Cation-anion correlation functions, $|r\cdot h_{+-}(r)|$, of explicit-solvent LJ and WCA electrolytes at various $\varepsilon_s$.}
\label{si:fig:gcaMD}
\end{figure}

\begin {figure}[htbp]\centering
\includegraphics [width=\textwidth] {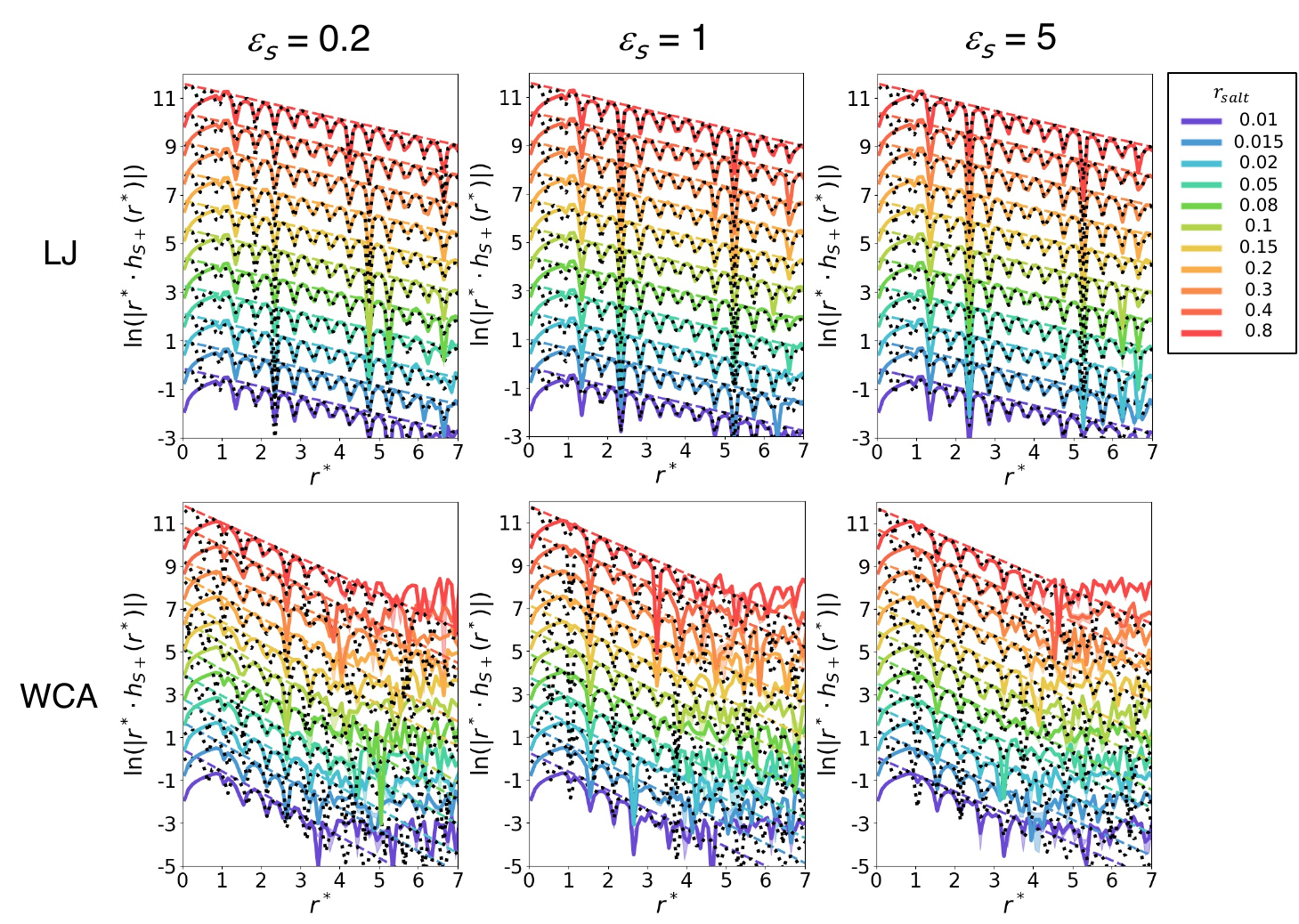}
\caption{Cation-solvent correlation functions, $|r\cdot h_{S+}(r)|$, of explicit-solvent LJ and WCA electrolytes at various $\varepsilon_s$.}\label{si:fig:gcsMD}
\end{figure}

This section presents the simulation results of structural correlation functions for explicit-solvent LJ and WCA electrolytes, together with the fitting results obtained using a single Yukawa function (Eq.~7 in the main text).  The results include:
\begin{itemize}\setlength\itemsep{0pt}
  \item the solvent–solvent structural correlation functions, $r\,|h_{SS}(r)|$ (Fig.~\ref{si:fig:gssMD}),
  \item the charge–charge correlation functions, $r\,|h_{ZZ}(r)|$ (Fig.~\ref{si:fig:gzzMD}),
  \item the density–density correlation functions, $r\,|h_{NN}(r)|$ (Fig.~\ref{si:fig:gnnMD}),
  \item the cation–anion correlation functions, $r\,|h_{+-}(r)|$ (Fig.~\ref{si:fig:gcaMD}), and
    \item the cation–solvent correlation functions, $r\,|h_{S+}(r)|$ (Fig.~\ref{si:fig:gcsMD}).
\end{itemize}
These results collectively illustrate the systematic evolution of structural correlations with $\varepsilon_s$ and concentration, demonstrating the transition from monotonic to oscillatory decay across electrolyte systems, from charge-dominated dilute to density-dominated concentrated regimes.
\newpage
\newpage
\newpage
\newpage
\newpage
\clearpage

\section{Fitting results of structural correlation functions for implicit-solvent electrolytes}\label{si:sec:corrLD}
\begin{figure}[htbp]\centering
\includegraphics[width=\textwidth]{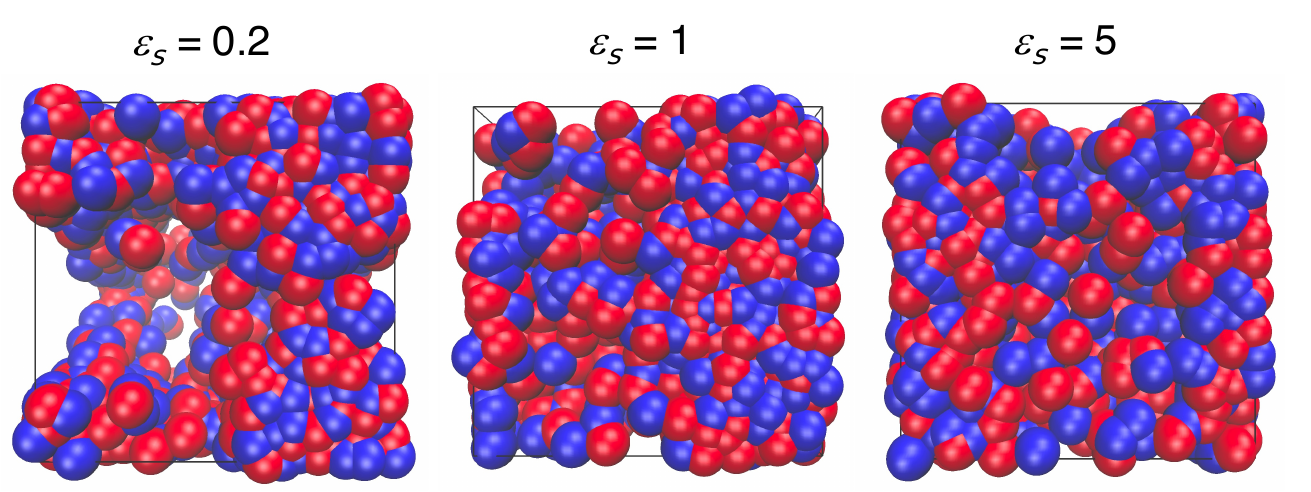}
\caption{Simulation snapshots representing phase segregation for implicit-solvent LJ electrolytes at $c_{\mathrm{salt}} = 0.171~\sigma^{-3}$ with diverse $\varepsilon_s$.}
\label{si:fig:segLD}
\end{figure}

\begin {figure}[htbp]\centering
\includegraphics [width=\textwidth] {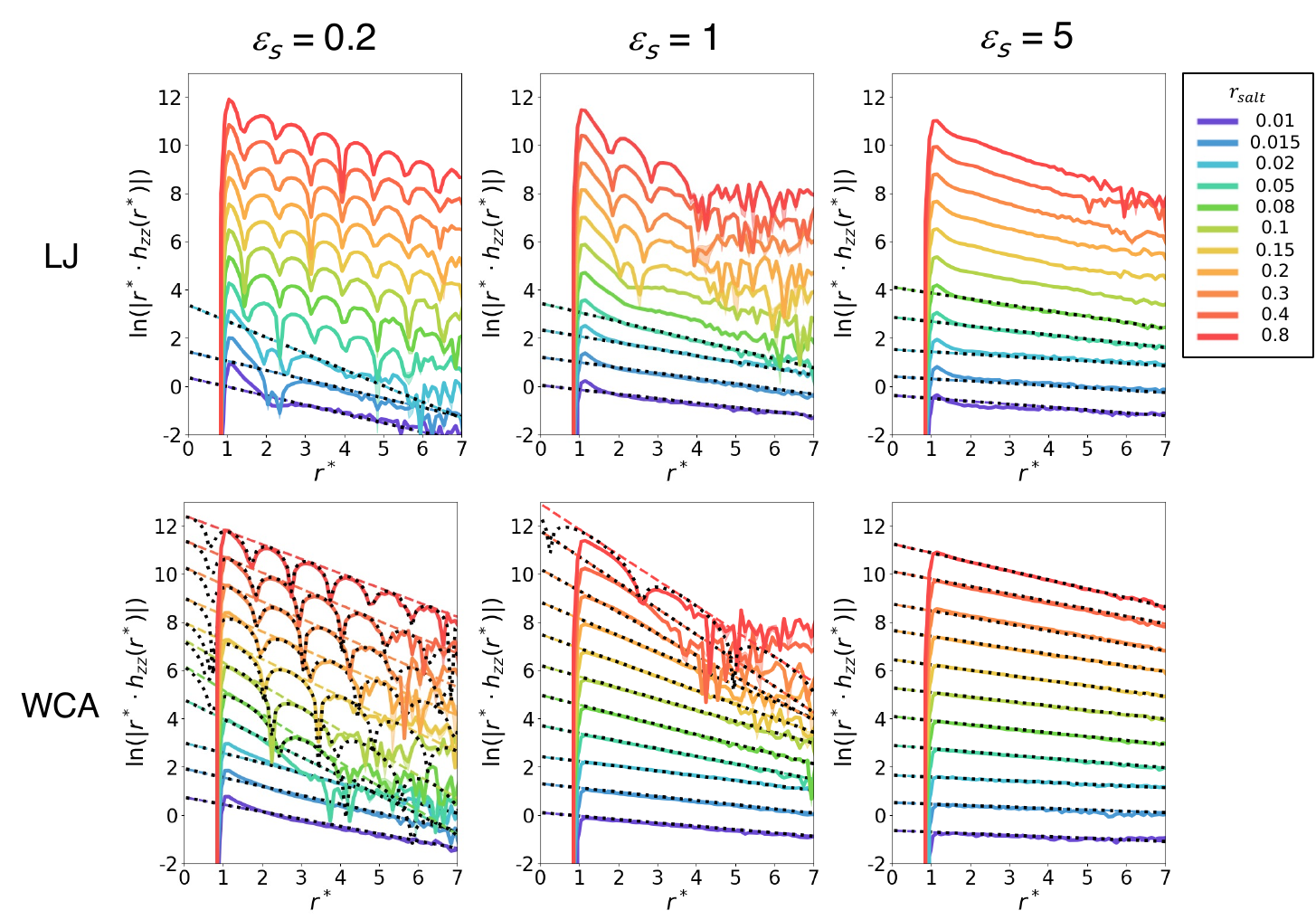}
\caption{Charge-charge correlation functions, $|r\cdot h_{ZZ}(r)|$, of implicit-solvent LJ and WCA electrolytes at various $\varepsilon_s$. The fit curves are not shown for the cases that exhibit phase segregation.}\label{si:fig:gzzLD}
\end{figure}

\begin {figure}[htbp]\centering
\includegraphics [width=\textwidth] {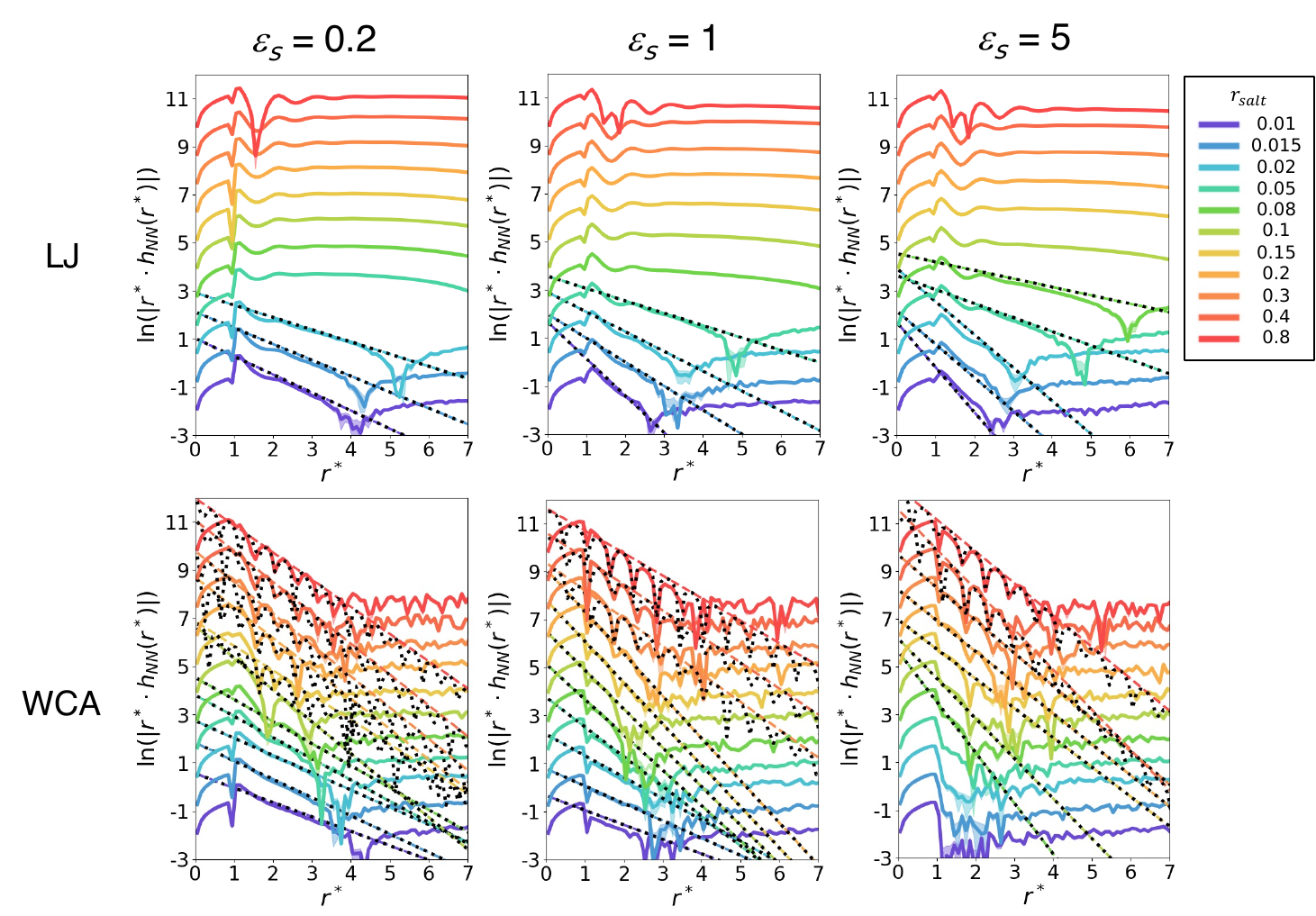}
\includegraphics [width=\textwidth] {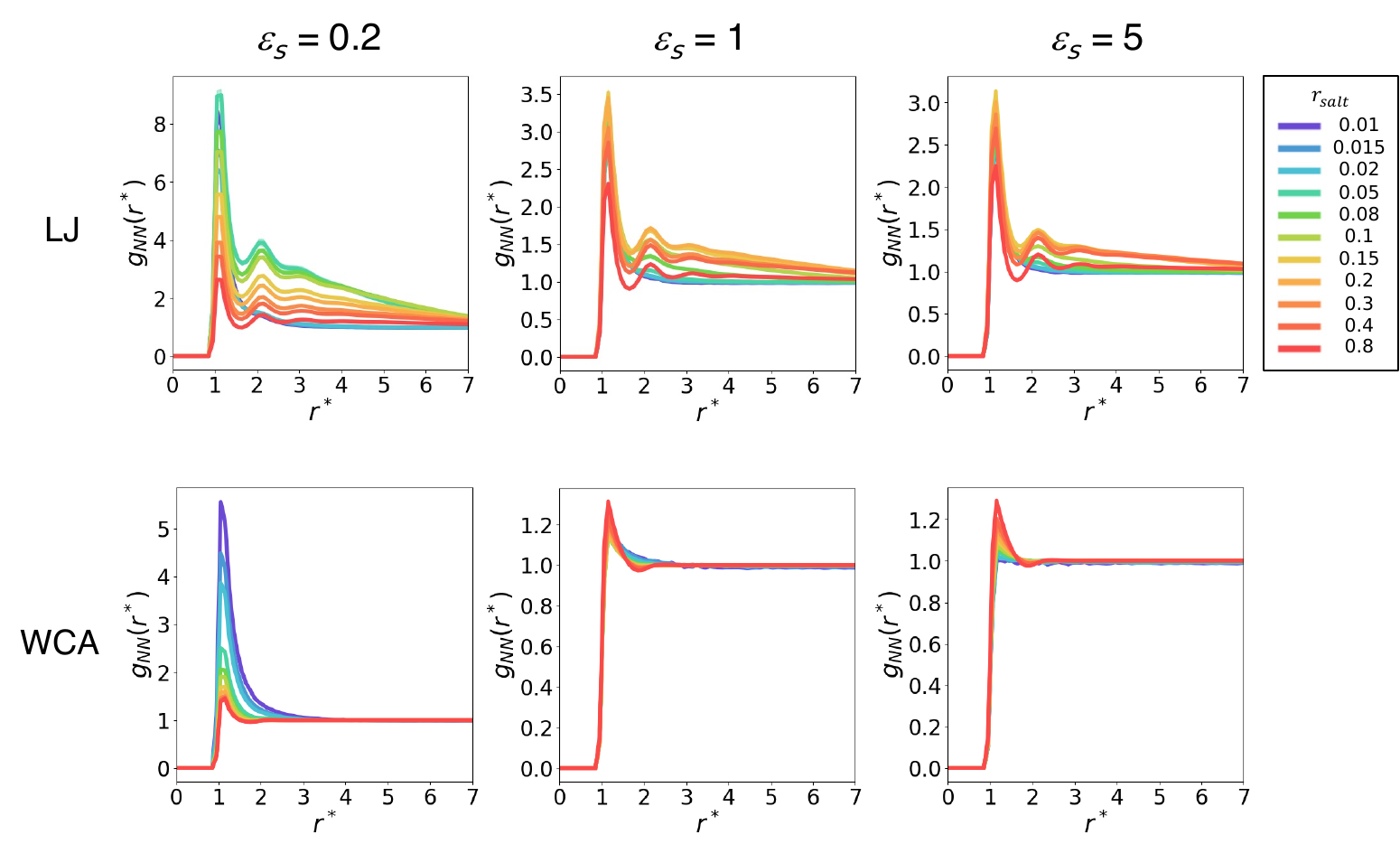}
\caption{Density-density correlation functions, $|r\cdot h_{NN}(r)|$, and $g_{NN}(r)$, of implicit-solvent LJ and WCA electrolytes at various $\varepsilon_s$. The fit curves are not shown for the cases that exhibit phase segregation.}\label{si:fig:gnnLD}
\end{figure}

\begin {figure}[htbp]\centering
\includegraphics [width=\textwidth] {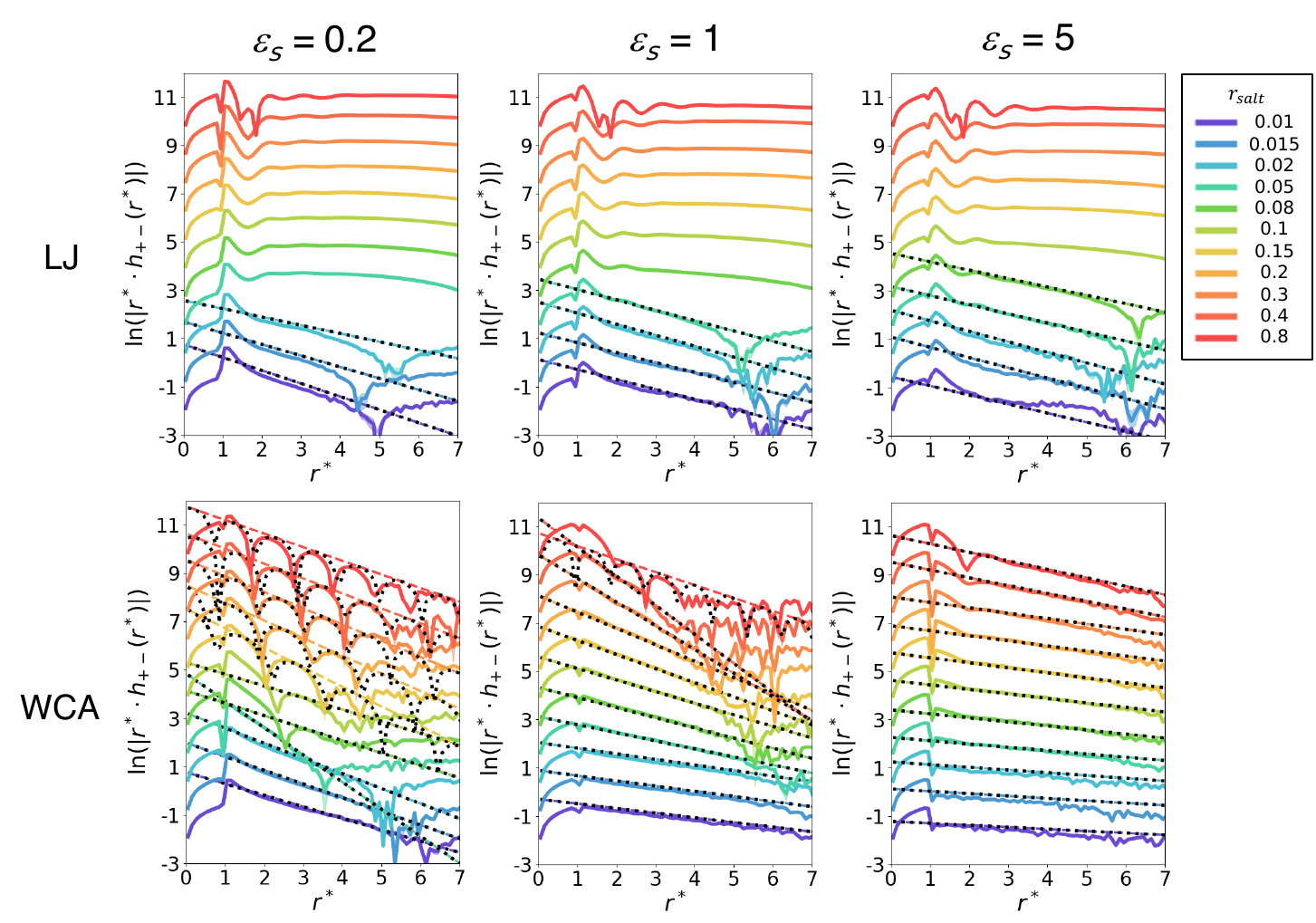}
\includegraphics [width=\textwidth] {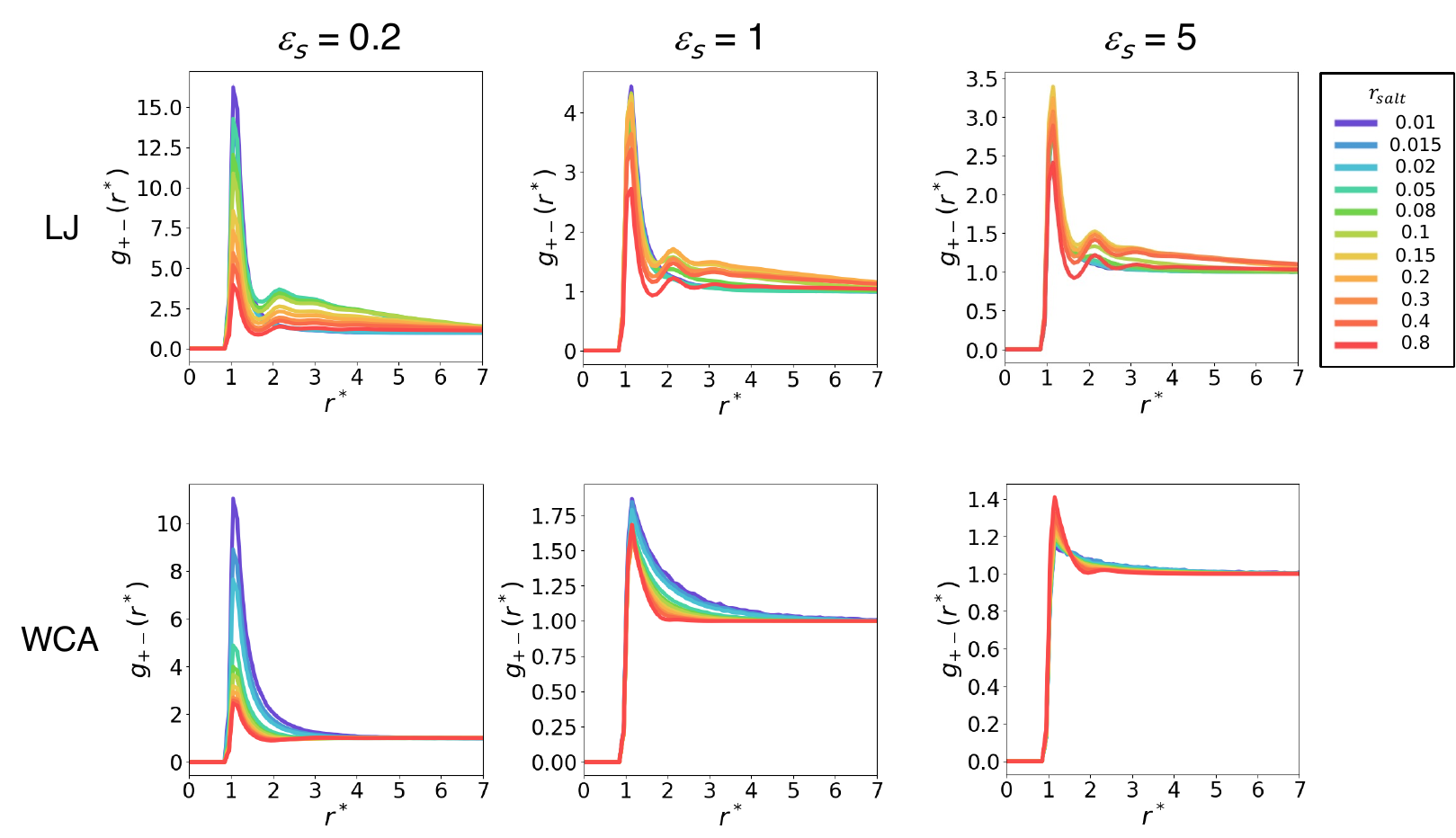}
\caption{Cation-anion correlation functions, $|r\cdot h_{+-}(r)|$, and $g_{+-}(r)$, of implicit-solvent LJ and WCA electrolytes at various $\varepsilon_s$. The fit curves are not shown for the cases that exhibit phase segregation.}\label{si:fig:gcaLD}
\end{figure}

This section, similar to Sec.~\ref{si:sec:corrMD}, presents the simulation results of structural correlation functions for implicit-solvent LJ and WCA electrolytes, together with the fitting results obtained using a single Yukawa function (Eq.~7 in the main text).  The results include:
\begin{itemize}\setlength\itemsep{0pt}
  \item the charge–charge correlation functions, $r\,|h_{ZZ}(r)|$ (Fig.~\ref{si:fig:gzzLD}),
  \item the density–density correlation functions, $r\,|h_{NN}(r)|$ (Fig.~\ref{si:fig:gnnLD}), and
  \item the cation–anion correlation functions, $r\,|h_{+-}(r)|$ (Fig.~\ref{si:fig:gcaLD}).
\end{itemize}
As discussed in the main text and shown in Fig.~\ref{si:fig:segLD}, some implicit-solvent electrolytes exhibits phase segregation.

\clearpage
\newpage

\newpage
\newpage
\clearpage

\section{Evolution of ionic clusters and their fitting results}\label{si:sec:cluster}
This section presents the results showing the evolution of ionic clusters across various electrolytes over a wide range of salt concentrations.

\begin{figure}[htbp]\centering
\includegraphics[width=\textwidth]{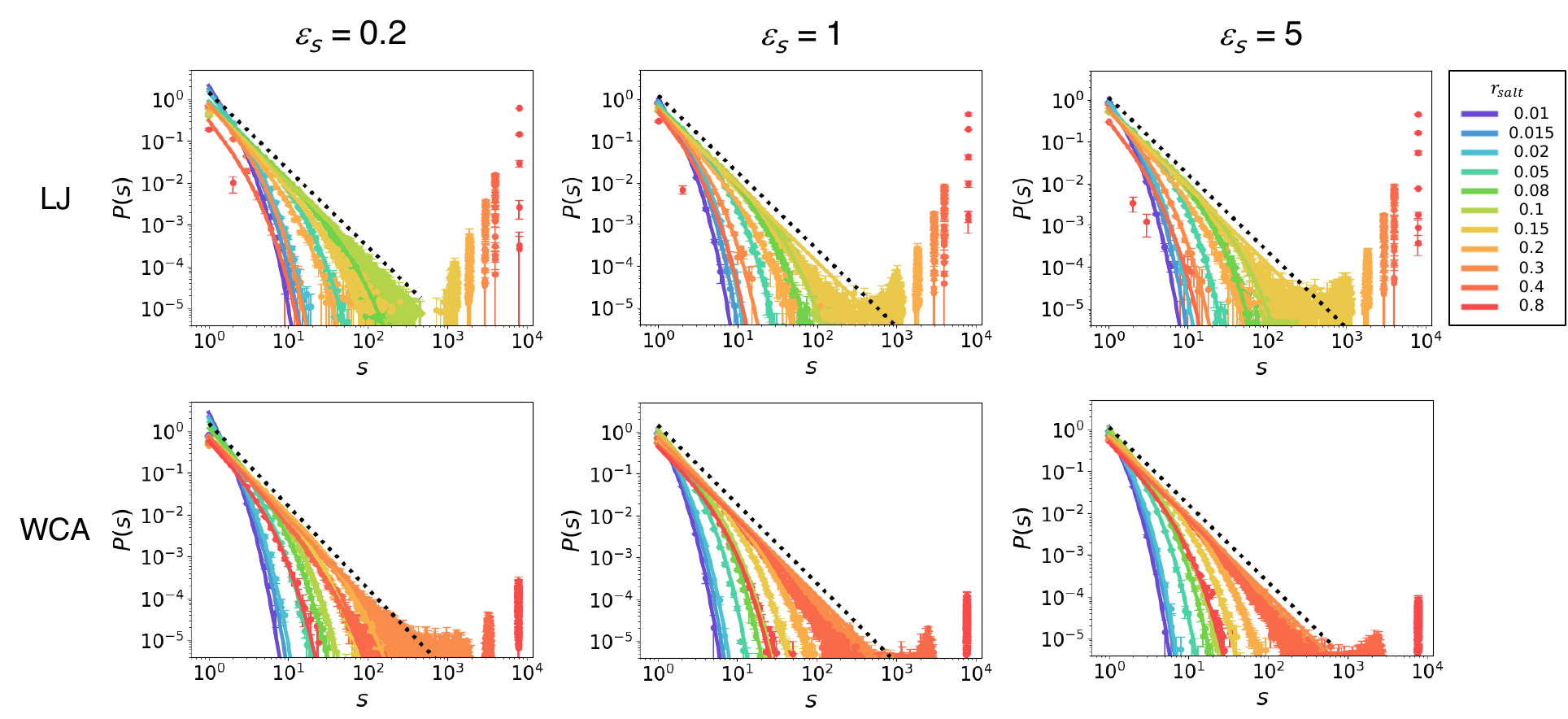}
\caption{Size distribution $P(s)$ (Eq.~9 in the main text) of ionic clusters for explicit-solvent LJ and WCA electrolytes at various $\varepsilon_s$.}
\label{si:fig:psMD}
\end{figure}

\begin{figure}[htbp]\centering
\includegraphics[width=\textwidth]{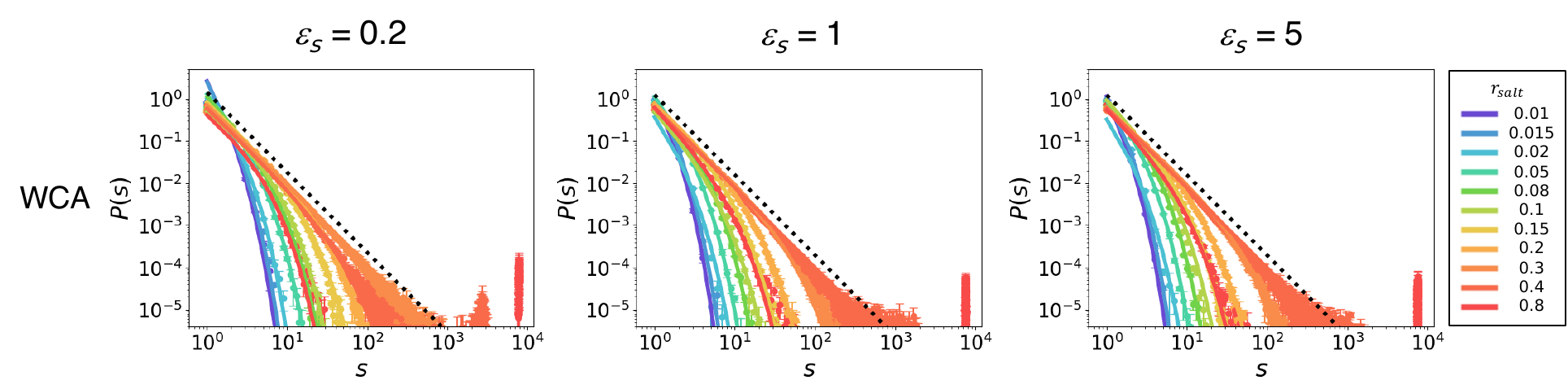}
\caption{Size distribution $P(s)$ (Eq.~9 in the main text) of ionic clusters for implicit-solvent WCA electrolytes at various $\varepsilon_s$.}
\label{si:fig:psLD}
\end{figure}
\newpage

\begin{figure}[htbp]\centering
\includegraphics[width=\textwidth]{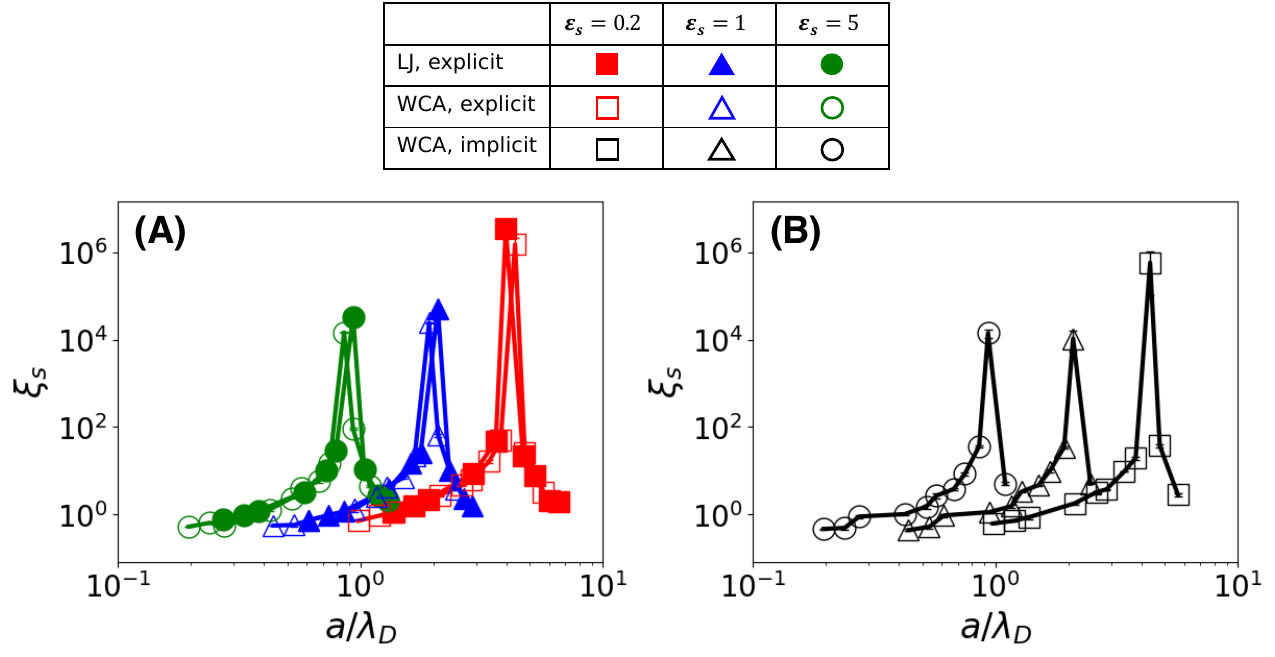}
\caption{Exponential decay length $\xi_s$, extracted from $P(s)$, as a function of $a/\lambda_D$. Each panels (A) and (B) show results for explicit-solvent LJ, WCA electrolytes, and implicit-solvent WCA electrolytes.}
\label{si:fig:xis}
\end{figure}
\newpage

\section{Potential of mean force between cations and anions}\label{si:sec:pmf}
This section presents the potential of mean force (PMF) between cations and anions, defined as $\beta w_{+-}(r) = -\ln[g_{+-}(r)]$, for various electrolytes over a wide range of salt concentrations.

\begin{figure}[htbp]\centering
\includegraphics[width=\textwidth]{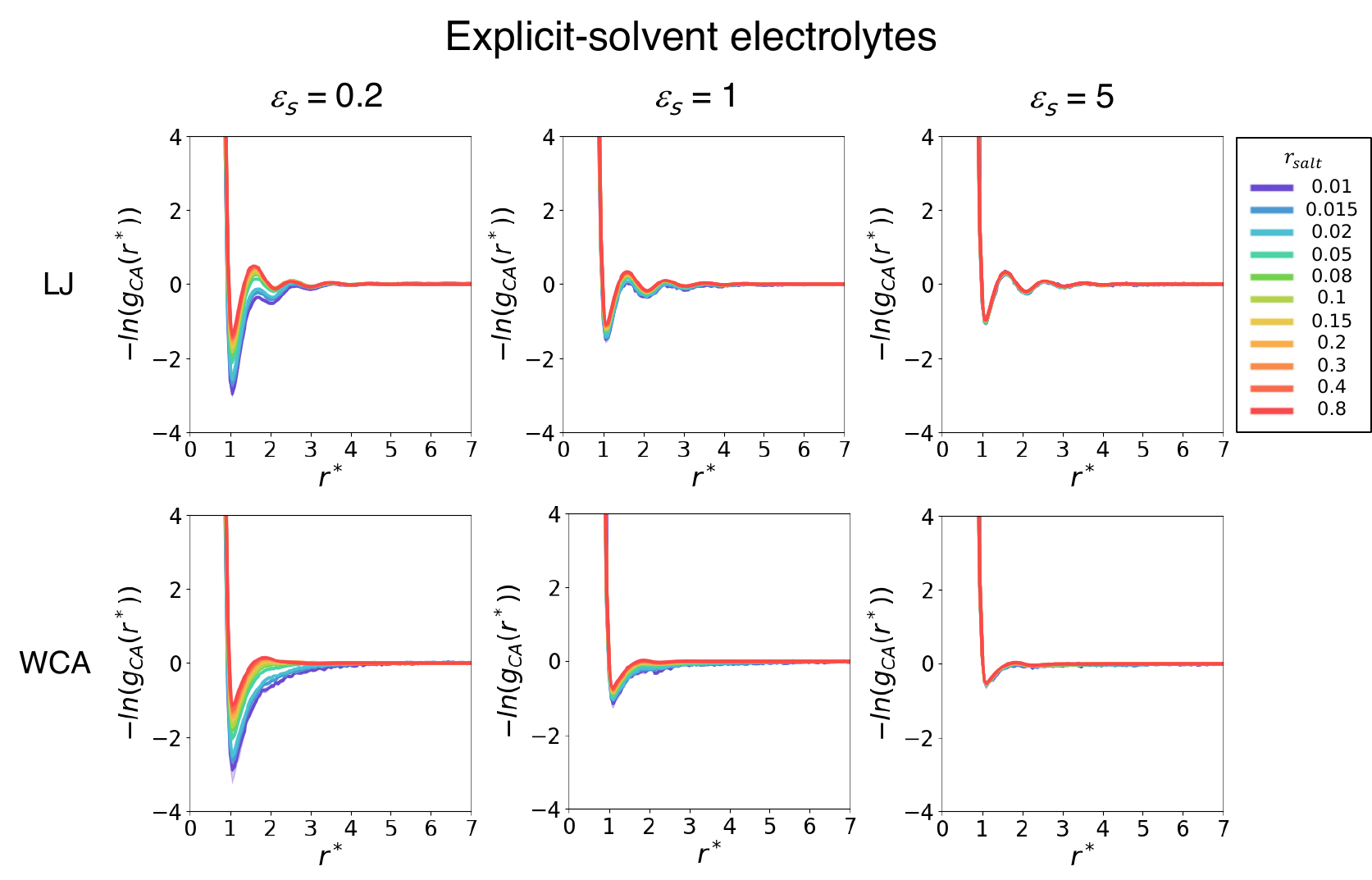}
\caption{Potential of mean force between cations and anions, $\beta w_{+-}(r)=-\ln[g_{+-}(r)]$, for explicit-solvent LJ and WCA electrolytes at various $\varepsilon_s$.} 
\label{si:fig:pmf_MD}
\end{figure}
\clearpage

\begin{figure}[htbp]\centering
\includegraphics[width=\textwidth]{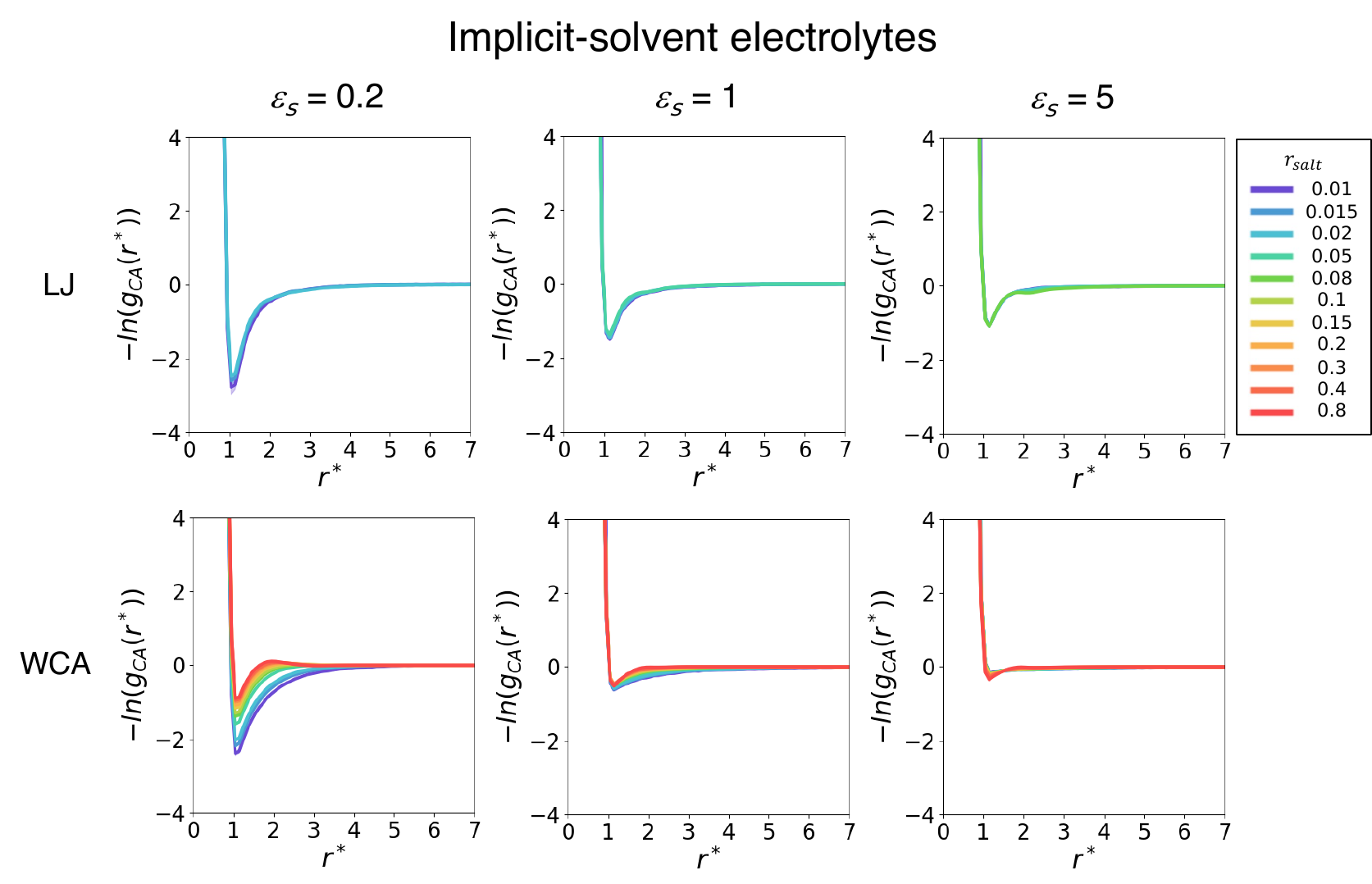}
\caption{Potential of mean force between cations and anions, $\beta w_{+-}(r)=-\ln[g_{+-}(r)]$, for implicit-solvent LJ and WCA electrolytes at various $\varepsilon_s$. The implicit-solvent electrolytes, exhibiting the phase segregation, are excluded.}\label{si:fig:pmf_LD}
\end{figure}
\clearpage

\section{Fitting results of mean-squared displacement for ions and solvent}\label{si:sec:msd}
This section presents the fitting results of the mean-squared displacement (MSD) of both ions and solvent, from which the corresponding diffusion coefficients are obtained. 

\subsection{Explicit-solvent electrolytes}
\begin{figure}[htbp]\centering
\includegraphics[width=\textwidth]{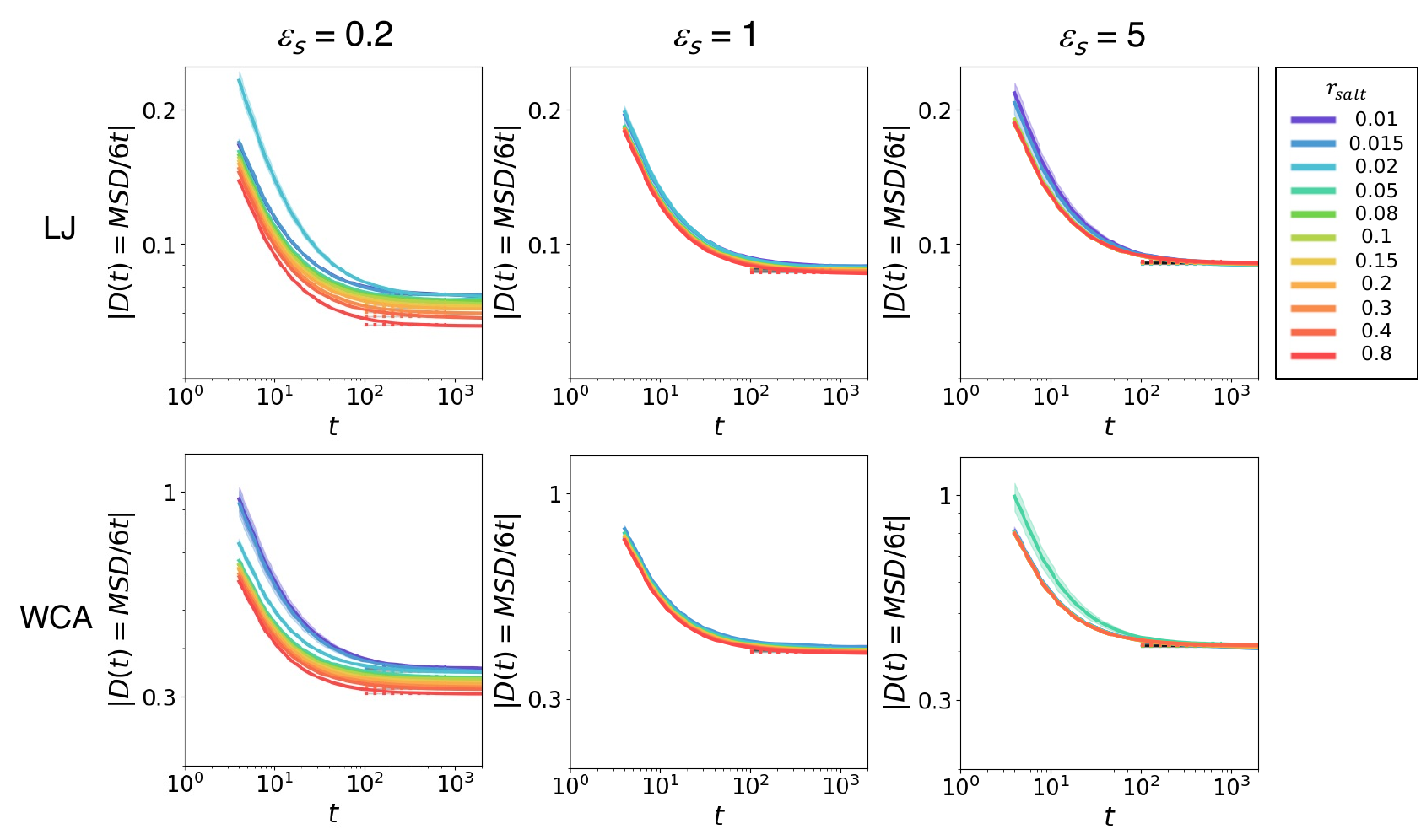}
\caption{Mean-squared displacement of cations in explicit-solvent electrolytes at various $\varepsilon_s$. Top and bottom rows show the results for LJ and WCA systems, respectively. Horizontal lines indicate the diffusion coefficients extracted from the linear-regime fitting within a time interval of [100,800].}
\label{si:fig:MSD_catMD}
\end{figure}
\newpage

\begin{figure}[htbp]\centering
\includegraphics[width=\textwidth]{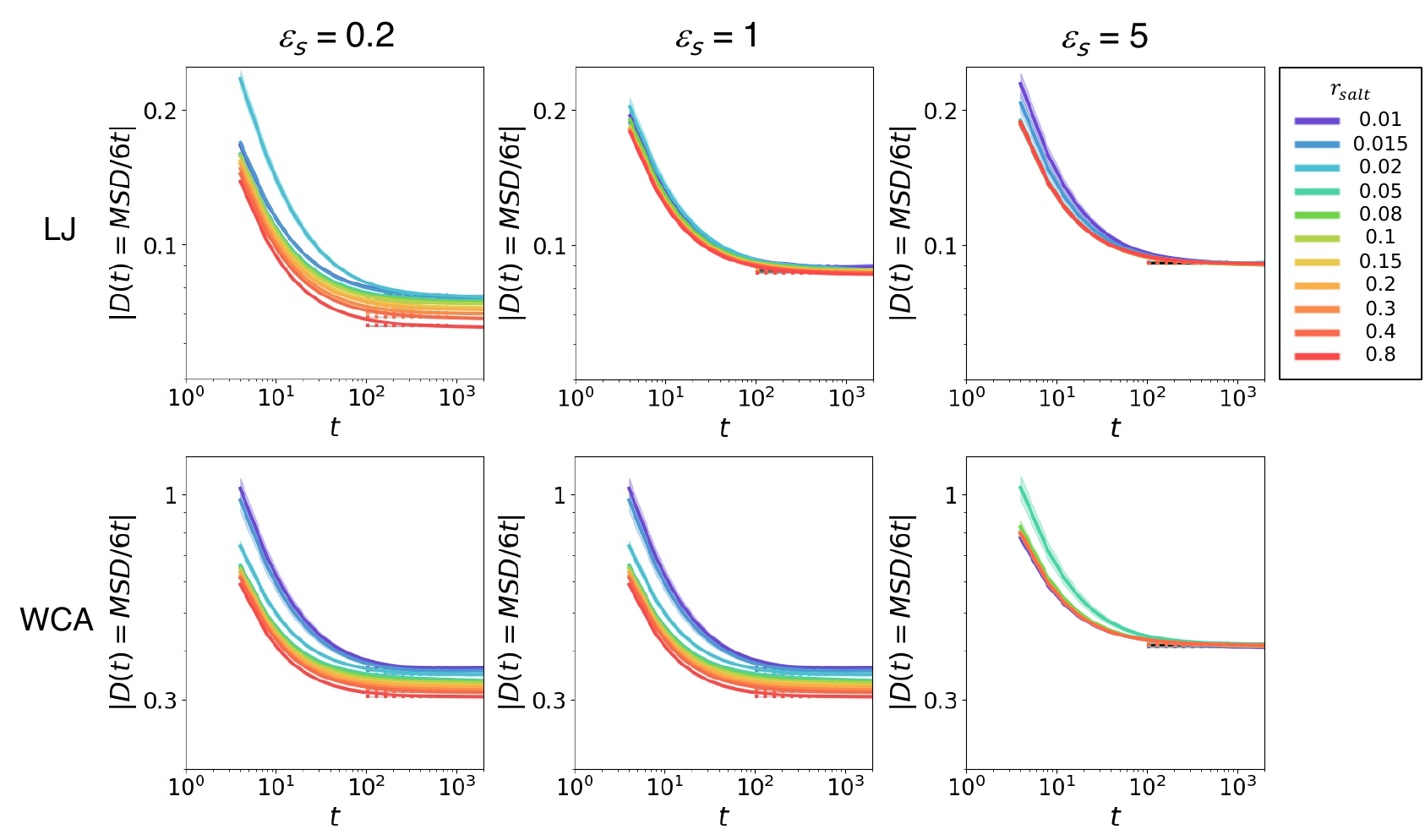}
\caption{Mean-squared displacement of anions in explicit-solvent electrolytes at various $\varepsilon_s$. Top and bottom rows show the results for LJ and WCA systems, respectively. Horizontal lines indicate the diffusion coefficients extracted from the linear-regime fitting within a time interval of [100,800].}
\label{si:fig:MSD_aniMD}
\end{figure}
\newpage

\begin{figure}[htbp]\centering
\includegraphics[width=\textwidth]{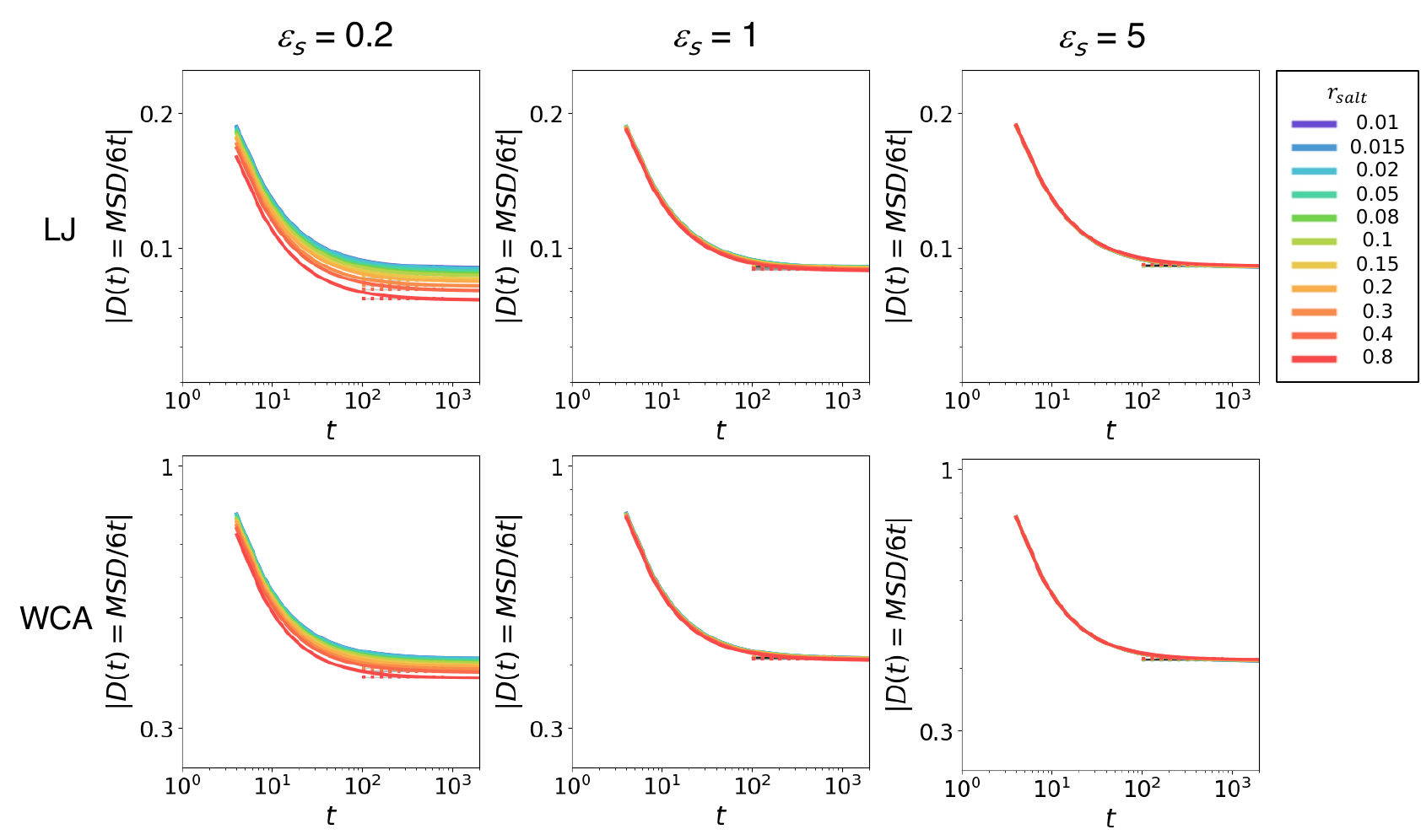}
\caption{Mean-squared displacement of solvent in explicit-solvent electrolytes at various $\varepsilon_s$. Top and bottom rows show the results for LJ and WCA systems, respectively. Horizontal lines indicate the diffusion coefficients extracted from the linear-regime fitting within a time interval of [100,800].}
\label{si:fig:MSD_solvMD}
\end{figure}
\newpage

\begin{figure}[htbp]\centering
\includegraphics[width=\textwidth]{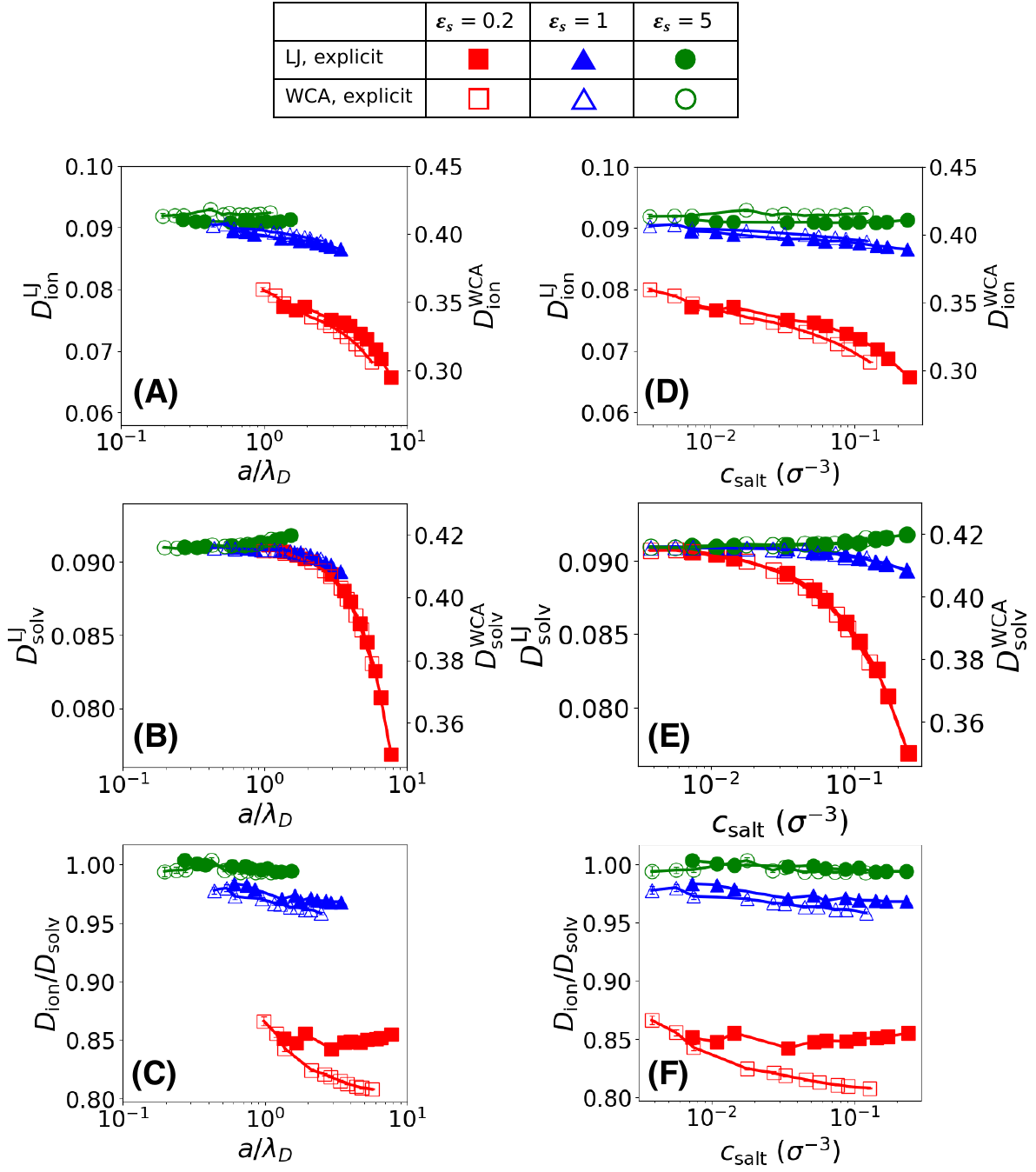}
\caption{Diffusion coefficients in explicit-solvent LJ and WCA electrolytes at various $\varepsilon_s$. (A, D) Ion diffusion coefficient, $D_{\mathrm{ion}}$. (B, E) Solvent diffusion coefficient, $D_{\mathrm{solv}}$. (C, F) The ratio $D_{\mathrm{ion}}/D_{\mathrm{solv}}$. Left and right columns show the result as a function of $a/\lambda_D$ and $c_{salt}$, respectively.}
\label{si:fig:Diff_MD}
\end{figure}
\clearpage

\subsection{Implicit-solvent electrolytes}
\begin{figure}[htbp]\centering
\includegraphics[width=\textwidth]{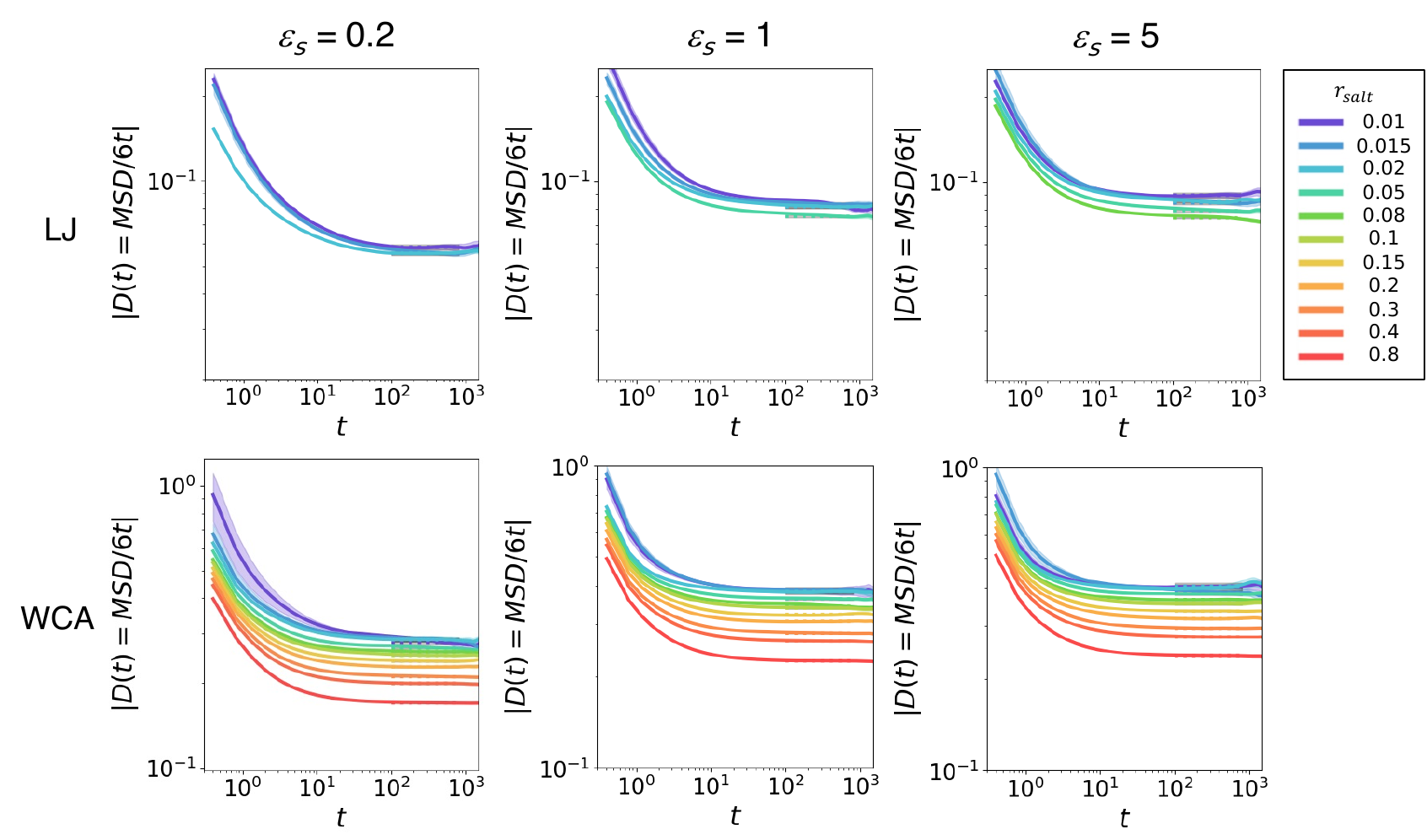}
\caption{Mean-squared displacement of cations in implicit-solvent electrolytes at various $\varepsilon_s$. Top and bottom rows show the results for LJ and WCA systems, respectively. Horizontal lines indicate the diffusion coefficients extracted from the linear-regime fitting within a time interval of [100,800].}
\label{si:fig:MSD_catLD}
\end{figure}
\newpage

\begin{figure}[htbp]\centering
\includegraphics[width=\textwidth]{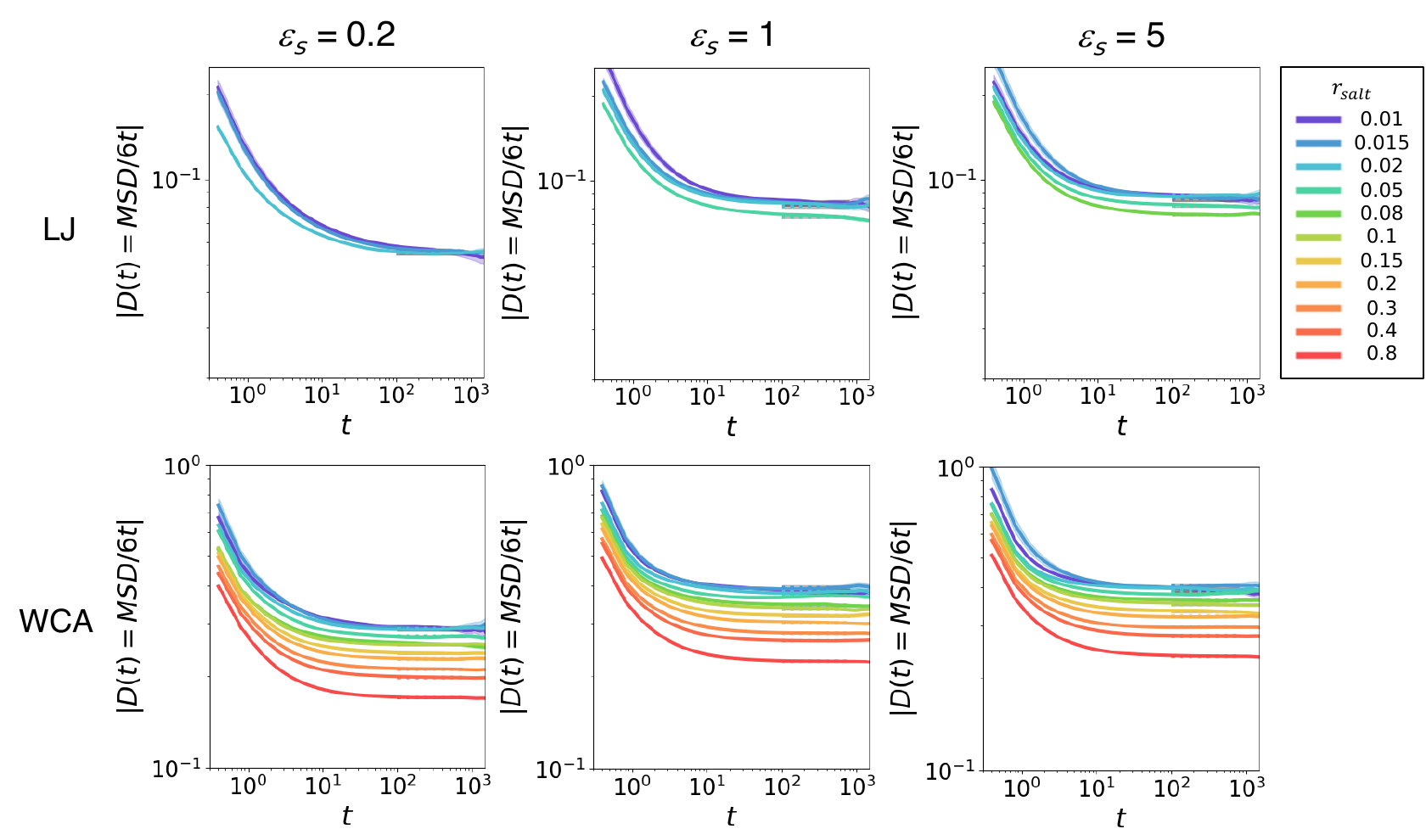}
\caption{Mean-squared displacement of anions in implicit-solvent electrolytes at various $\varepsilon_s$. Top and bottom rows show the results for LJ and WCA systems, respectively. Horizontal lines indicate the diffusion coefficients extracted from the linear-regime fitting within a time interval of [100,800].}
\label{si:fig:MSD_aniLD}
\end{figure}
\newpage

\begin{figure}[htbp]\centering
\includegraphics[width=\textwidth]{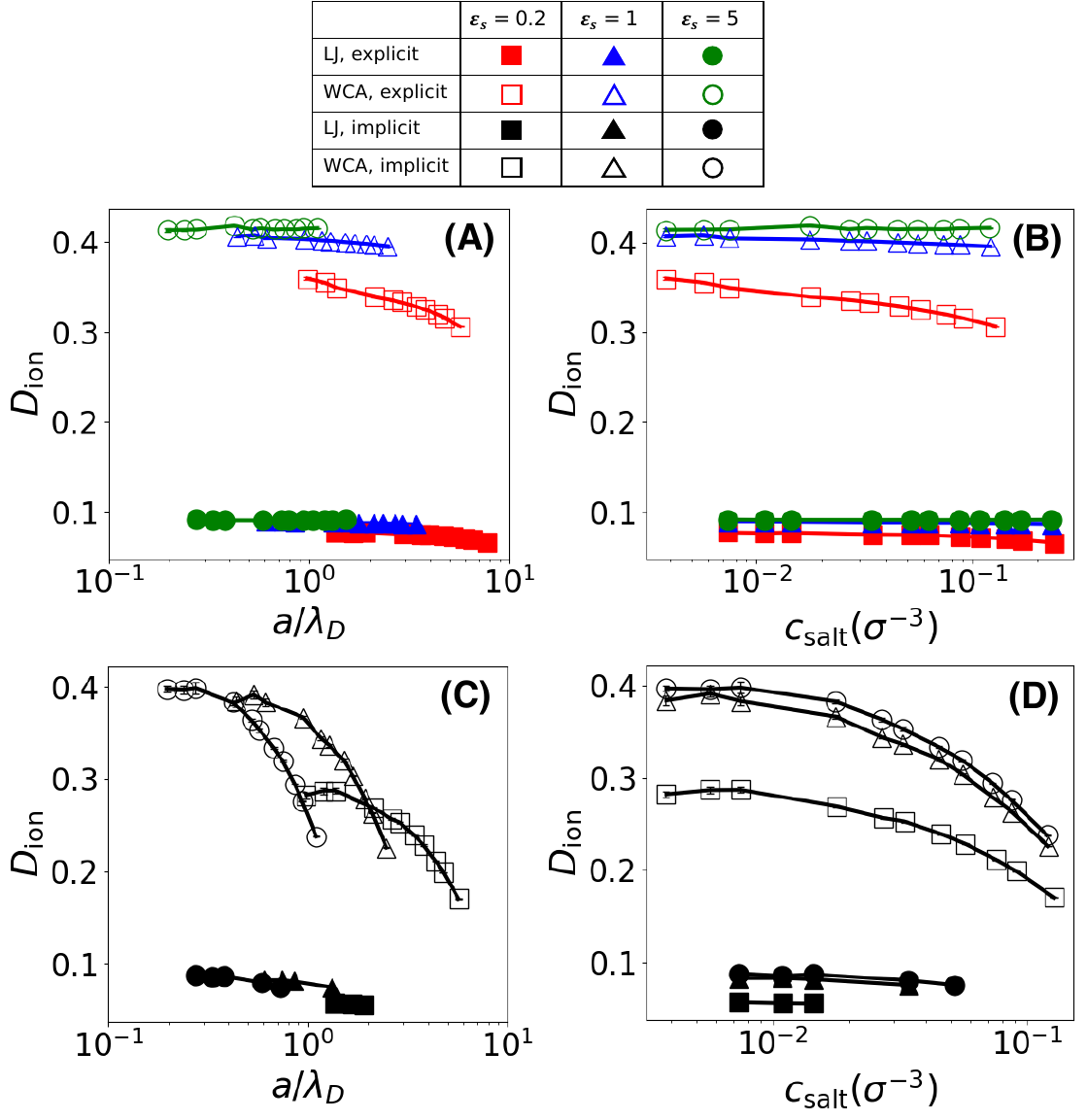}
\caption{Ion diffusion coefficient $D_{\mathrm{ion}}$ as a function of (A, C) $a/\lambda_D$, and (B, D) $c_{salt}$. Top and bottom rows show the results of explicit- and implicit-solvent electrolytes, respectively.}
\label{si:fig:Dion_LD}
\end{figure}
\clearpage

\section{Fitting results of ion-pair survival function}\label{si:sec:relax}
This section presents the fitting results of the ion-pair survival function $H(t)$, from which the ion-pair lifetime $\tau_{\mathrm{pair}}$ is obtained. Unlike other structural and dynamical quantities, $H(t)$ was computed using frequently saved trajectories.
\begin{figure}[htbp]\centering
\includegraphics[width=\textwidth]{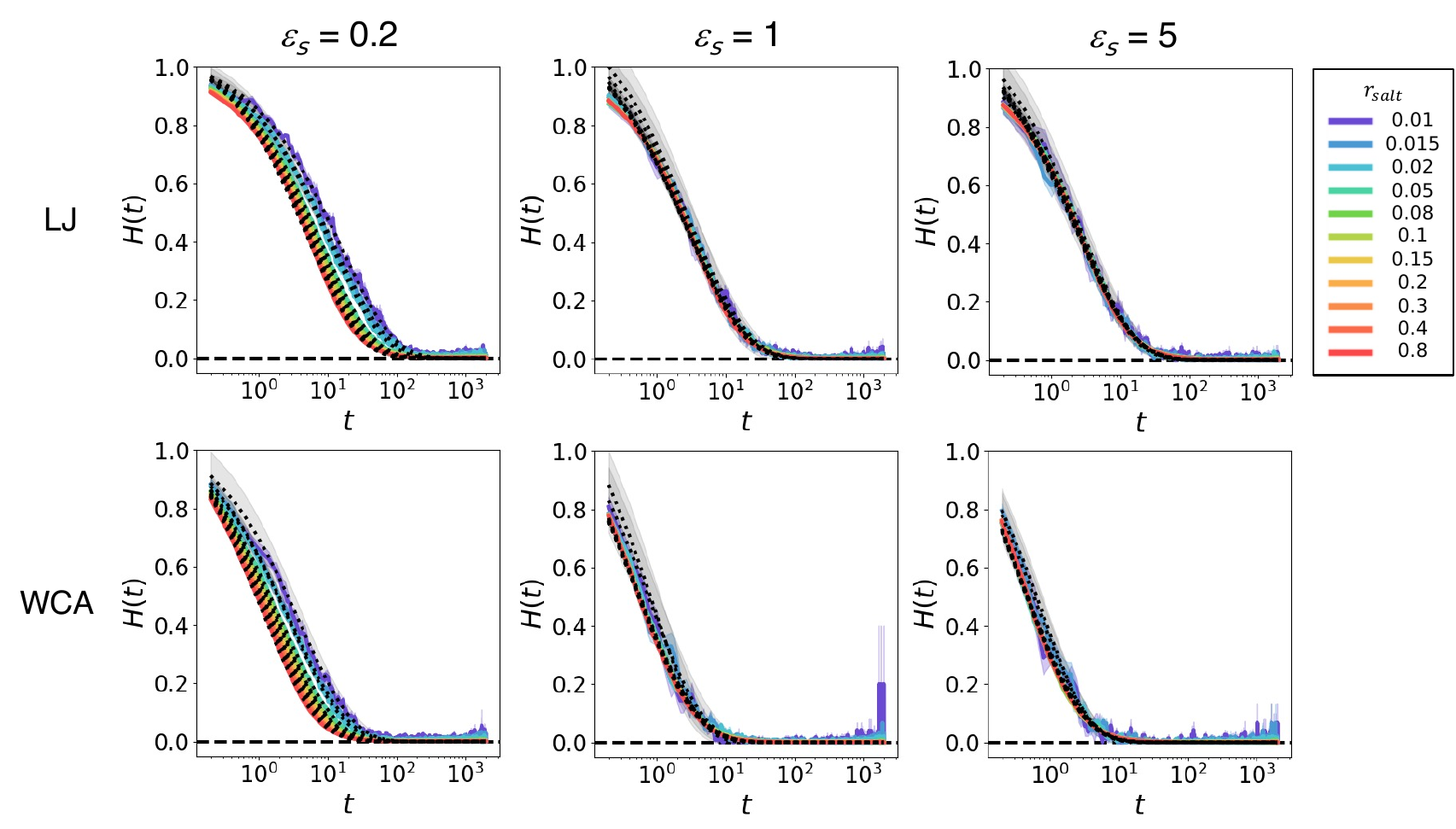}
\caption{Ion-pair survival function $H(t)$ (Eq.~10 in the main text) for explicit-solvent LJ and WCA electrolytes at various $\varepsilon_s$.}
\label{si:fig:ionpair_MD}
\end{figure}
\newpage

\begin{figure}[htbp]\centering
\includegraphics[width=\textwidth]{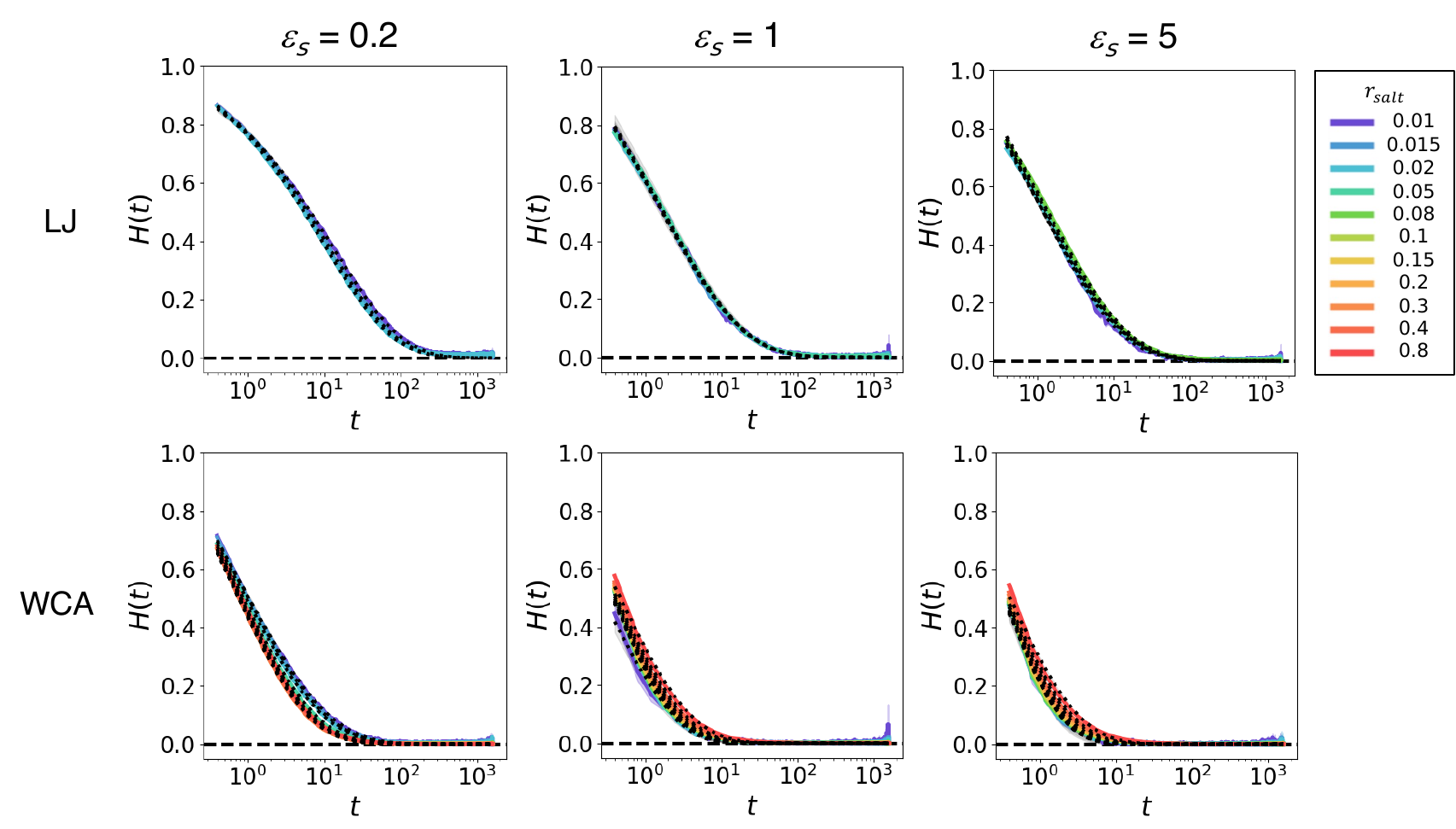}
\caption{Ion-pair survival function $H(t)$ (Eq.~10 in the main text) for implicit-solvent LJ and WCA electrolytes at various $\varepsilon_s$.}
\label{si:fig:ionpair_LD}
\end{figure}
\newpage

\begin{figure*}[htbp]\centering
\includegraphics[width=\textwidth]{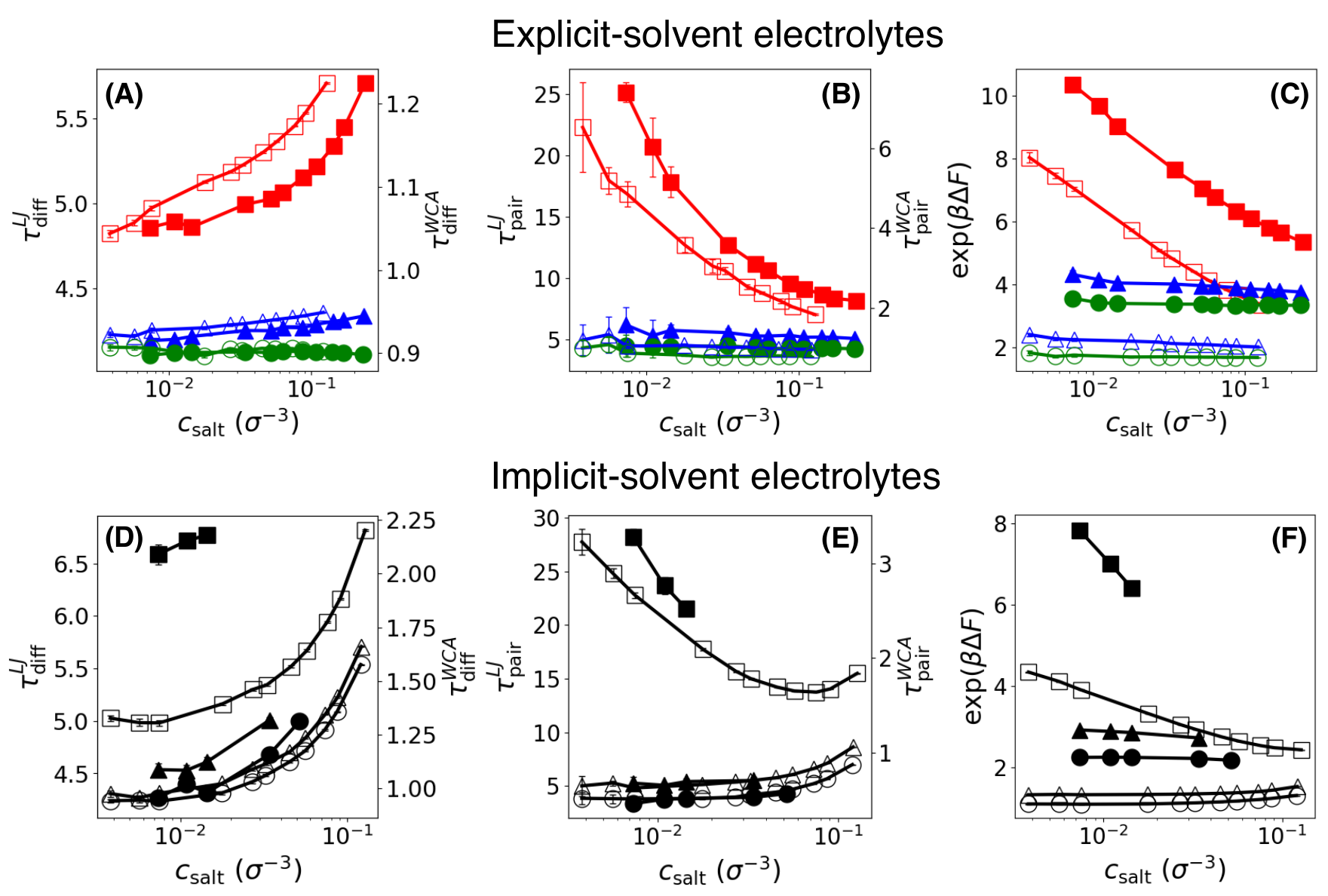}
\caption{Crossovers in ion dynamics and their structural origins as a function of the ratio $c_{salt}$ across all electrolytes examined. 
(A, D) Diffusion relaxation time, $\tau_{\mathrm{Diff}}$. 
(B, E) Mean ion-pair lifetime, $\tau_{\mathrm{pair}}$. 
(C, F) Free-energy barrier for ion-counterion dissociation, $\beta\Delta F$. 
Panels (A–C) on the top row correspond to explicit-solvent models, while panels (D–F) on the bottom row correspond to implicit-solvent models.
In panels (A-B) and (D-E), the left and right axes represent results for LJ (filled markers) and WCA (open markers), respectively.}\label{si:fig:msd}
\end{figure*}
\clearpage

\section{Ionic conductivity}\label{si:sec:cond}
We computed ion conductivity $\sigma$, which includes all dynamic correlations using the Onsager coefficients $\Delta\vec{r}_{\alpha\beta}$~\cite{fong2021ion,fong2020onsager}. The mean-squared charge displacement $\Sigma$ can be computed as follows:
\begin{eqnarray}
\Sigma(t) &=& \sum_{\alpha}\sum_{\beta} \frac{q_\alpha q_\beta}{\langle V\rangle k_BT}  \Delta\vec{r}_{\alpha\beta}(t)\nonumber \\
&=& \frac{e^2}{\langle V\rangle k_BT}(\Delta\vec{r}_{++}(t)+\Delta\vec{r}_{--}(t)-2\Delta\vec{r}_{+-}(t))\nonumber\\
&=& \Sigma_{++}(t)+\Sigma_{--}(t)+\Sigma_{+-}(t) \nonumber\\
&=& \Sigma^{s}_{+}(t)+\Sigma^{d}_{+}(t)+\Sigma^{s}_{-}(t)+\Sigma^{d}_{-}(t)+\Sigma_{+-}(t).
\label{eq:cond_time}
\end{eqnarray}
Here, the superscripts $s$ and $d$ denote the self and distinct part, respectively. 
In the long-time limit, $\Sigma(t)$ grow linearly with time $t$ and its slope is the ionic conductivity, $\sigma$:
\begin{eqnarray}\label{eq:cond}
\sigma= \lim_{t\to\infty} \frac{\Sigma(t)}{6t},
\end{eqnarray}
which includes all the contributions of the correlated motions. The ionic conductivity ($\sigma$) has its five contributions, as do the $\Sigma(t)$, $\Delta\vec{r}_{\alpha\beta}$ and $D_{\alpha\beta}$: $\sigma = \sigma_{+}^s + \sigma_{-}^s + \sigma_{+}^d + \sigma_{-}^d + 2\sigma_{+-}$.
\begin{eqnarray}
\sigma &=& \sigma_{+}^s + \sigma_{-}^s + \sigma_{+}^d + \sigma_{-}^d + 2\sigma_{+-}.
\label{eq:cond_comp}
\end{eqnarray}
That is, the delicate balance between those five components determine the overall conductivity $\sigma$.
\textbf{Nernst-Einstein approximation.} By neglecting all the correlations, only self-terms of Onsager coefficients contribute to the ionic conductivity under the Nernst-Einstein relation. 
 \begin{eqnarray}\label{eq:ne_conductivity}
\sigma_{NE}&=&\sigma_{+}^s + \sigma_{-}^s 
=\frac{e^2\rho_{\text{ion}}}{k_BT}\bigg(\chi_{+}D_{+}+\chi_{-}D_{-}\bigg)\nonumber\\
&=&\frac{e^2\rho_{\text{ion}}}{k_BT}D_{\text{ion}}. 
\end{eqnarray}
Here, $\rho_{\text{ion}}=(N_{+}+N_{-})/\langle V\rangle=2N_{salt}/\langle V\rangle$, and $D_{\text{ion}}=\frac{1}{2}(D_{+}+D_{-})$. 
The contribution $\sigma_{corr}$ of correlated motions to the ionic conductivity is $\sigma_{corr}=\sigma-\sigma_{NE}$ and the \textit{ionicity} $\sigma/\sigma_{NE}=1+\sigma_{corr}/\sigma_{NE}$. $\sigma_{corr}$ is usually negative as the dynamic correlations decrease $\sigma$. 

\newpage
\begin {figure}[htbp]\centering
\includegraphics [width=\textwidth] {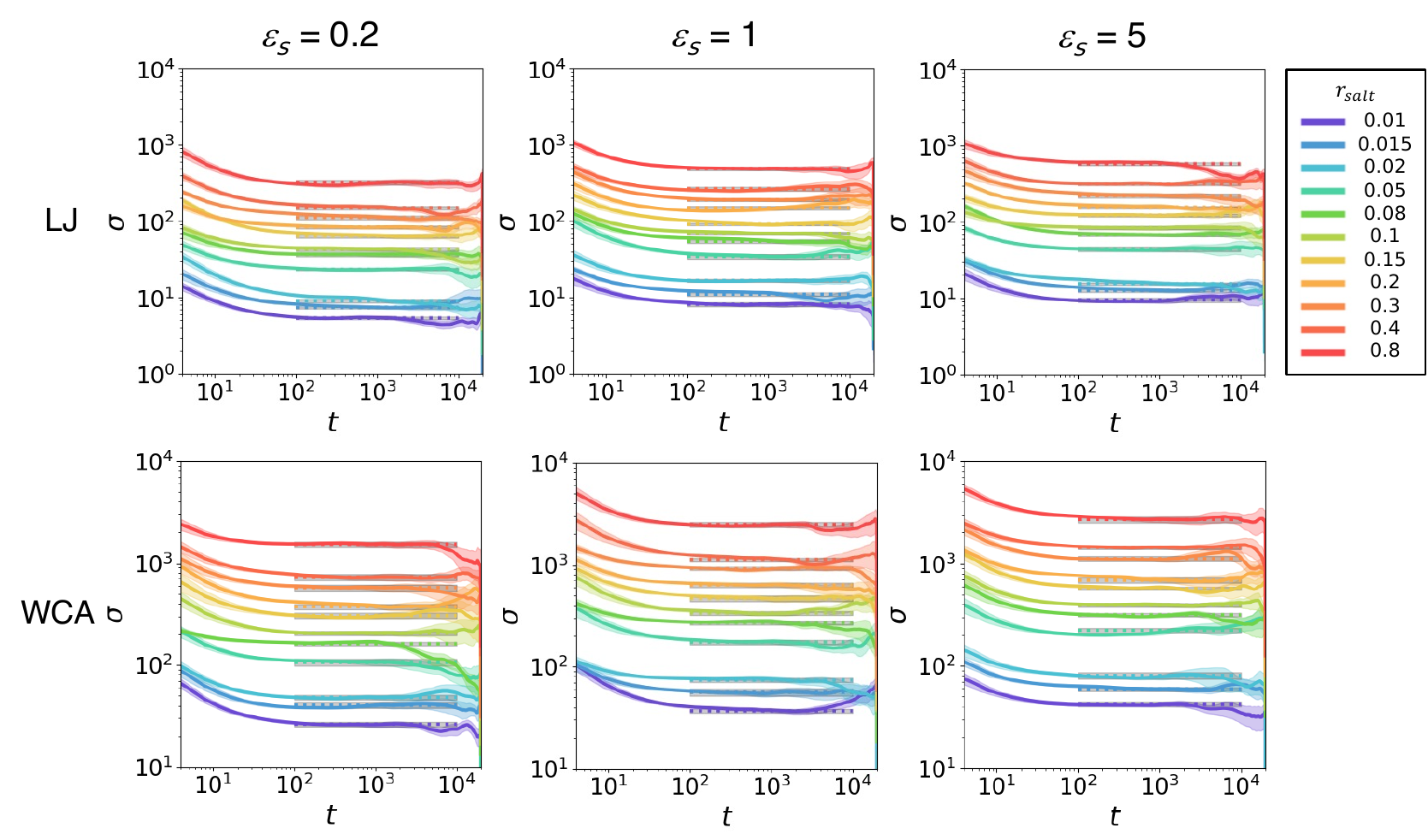}
\caption{Time-dependent ionic conductivity ($\sigma(t)$, Eq.~\ref{eq:cond_comp}) for both explicit-solvent LJ and WCA electrolytes with diverse $\varepsilon_s$. Horizontal lines indicate the ionic conductivity, extracted from the linear-regime fitting within a time interval of [300, 2000].}\label{si:fig:condMD}
\end{figure}
\newpage

\begin {figure}[htbp]\centering
\includegraphics [width=\textwidth] {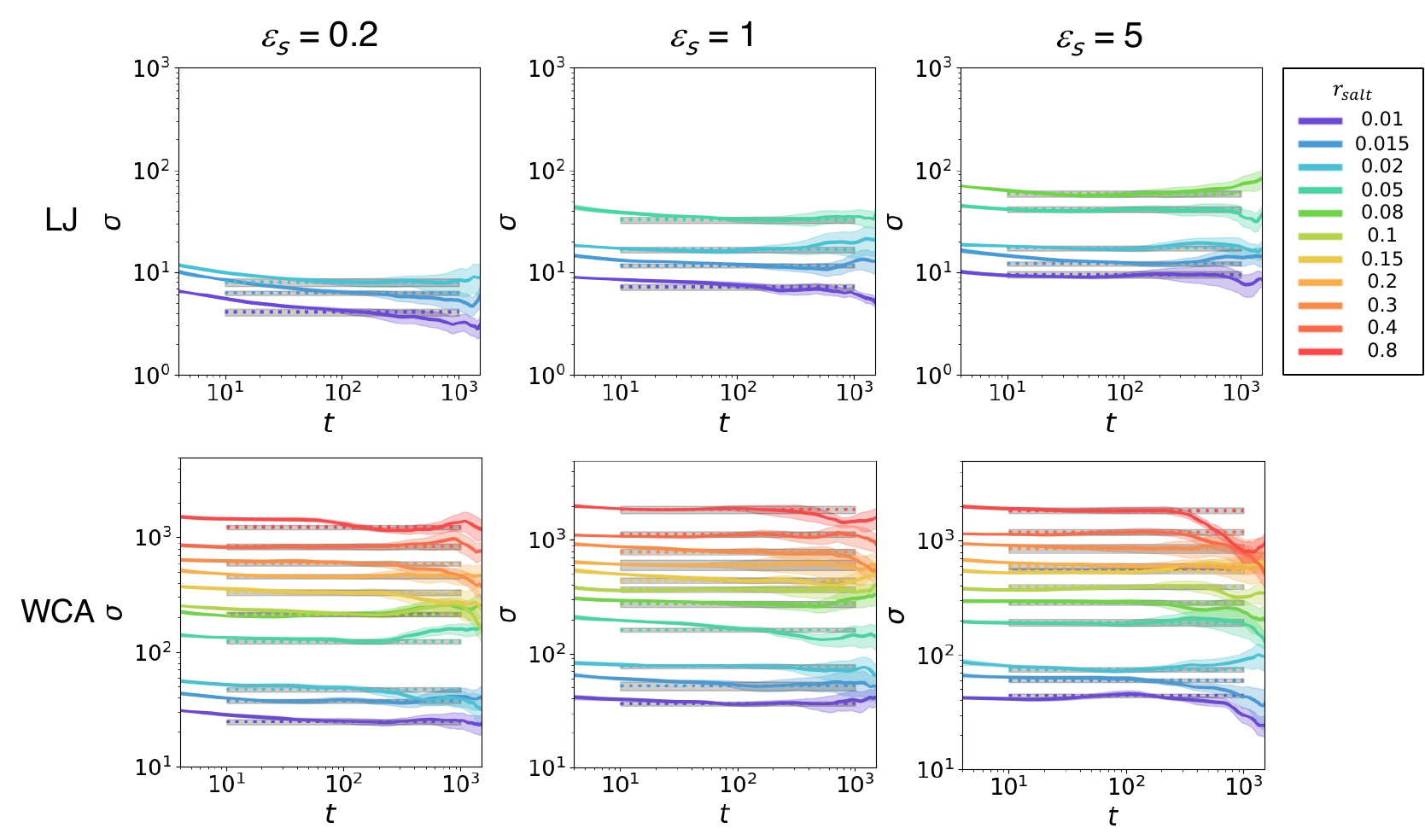}
\caption{Time-dependent ionic conductivity ($\sigma(t)$, Eq.~\ref{eq:cond_comp}) for both implicit-solvent LJ and WCA electrolytes with diverse $\varepsilon_s$. Horizontal lines indicate the ionic conductivity, extracted from the linear-regime fitting within a time interval of [20, 200].}\label{si:fig:condLD}
\end{figure}
\newpage

\clearpage

\begin {figure}[htbp]\centering
\includegraphics [width=5in] {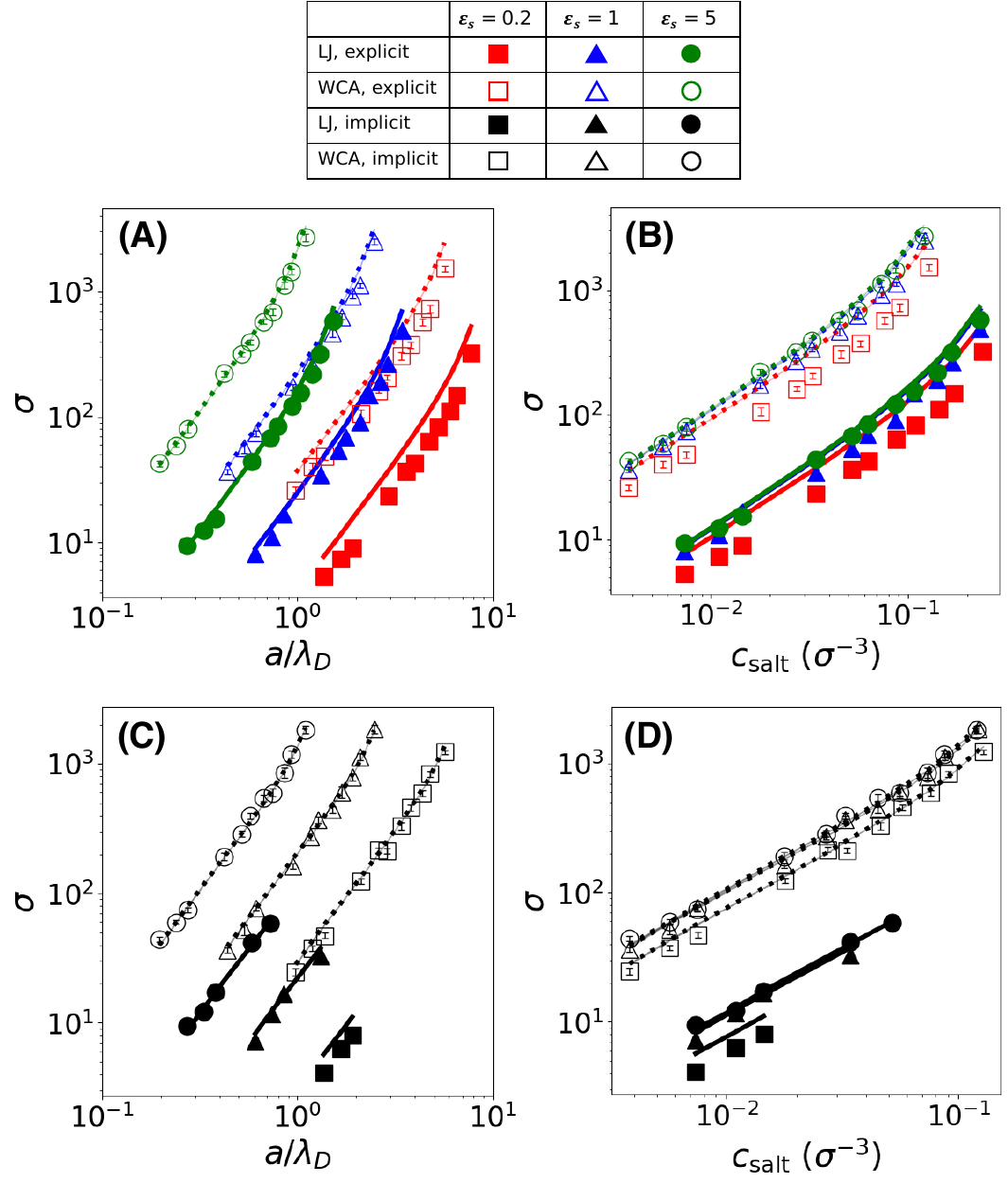}
\caption{Ionic conductivity ($\sigma$, markers) and its Nernst-Einstein (NE) approximation $\sigma_{NE}$ (Eq.~\ref{eq:ne_conductivity}, dashed and solid lines) as a function of (A, C) $a/\lambda_D$ and (B, D) $c_{\mathrm{salt}}$. Top and bottom rows show the results of explicit- and implicit-solvent electrolytes, respectively.}\label{si:fig:condapproximation}
\end{figure}
\newpage

\begin {figure}[htbp]\centering
\includegraphics [width=5in] {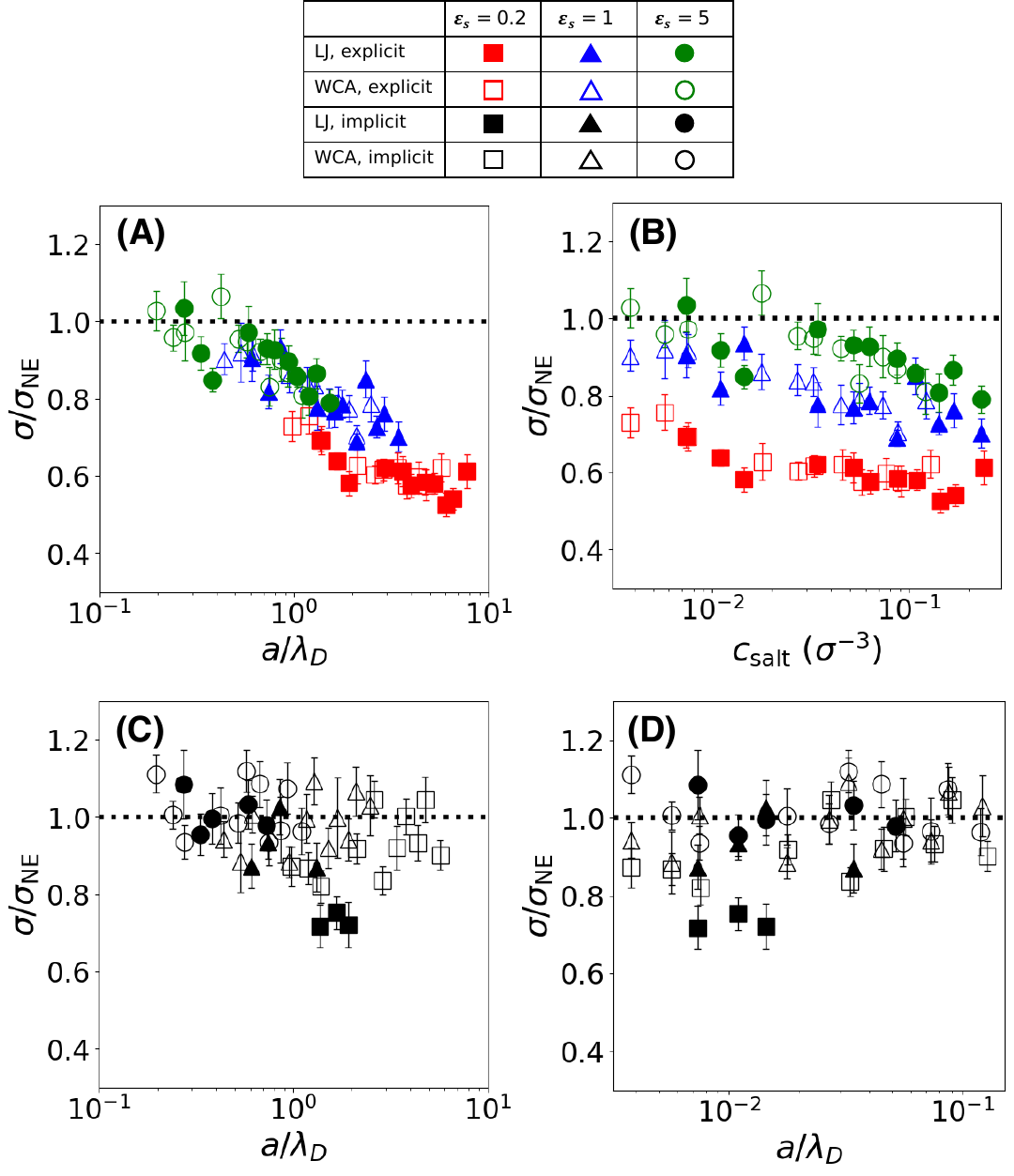}
\caption{Ionicity ($\sigma/\sigma_{\mathrm{NE}}$) as a function of (A, C) $a/\lambda_D$ and (B, D) $c_{\mathrm{salt}}$. Top and bottom rows show the results of explicit- and implicit-solvent electrolytes, respectively.}\label{si:fig:ionicity}
\end{figure}

\clearpage
\newpage

\section{Ionic cluster analysis: comparison between Cayley-tree-based theory and simulation}\label{si:sec:cluster_analysis}
In this section, we provide a more detailed analysis of ionic clusters, focusing on ion association patterns and their comparison with the Cayley-tree-based mean-field theory proposed by McEldrew \textit{et al.}~\cite{mceldrew2020theory, mceldrew2021correlated, mceldrew2021salt, goodwin2023theory, zhang2024long}. Here, we focus in particular on the LJ and WCA electrolytes with $\varepsilon_s=0.2$, where the degree of ionic association exhibits a clear dependence on the salt concentration $c_{salt}$. In contrast, for systems with $\varepsilon_s=1$ and 5, we observe ion association that is independent of $c_{salt}$ (Fig.~\ref{si:fig:associationnumber}). This behavior clearly violates the assumptions of speciation and strong dynamical correlations among ions within aggregates, thereby limiting the applicability of such Cayley-tree-based mean-field theories in these cases. We therefore exclude these systems from the analysis below. 

\subsection{Ion association patterns} 
\begin {figure}[htbp]\centering
\includegraphics [width=\textwidth] {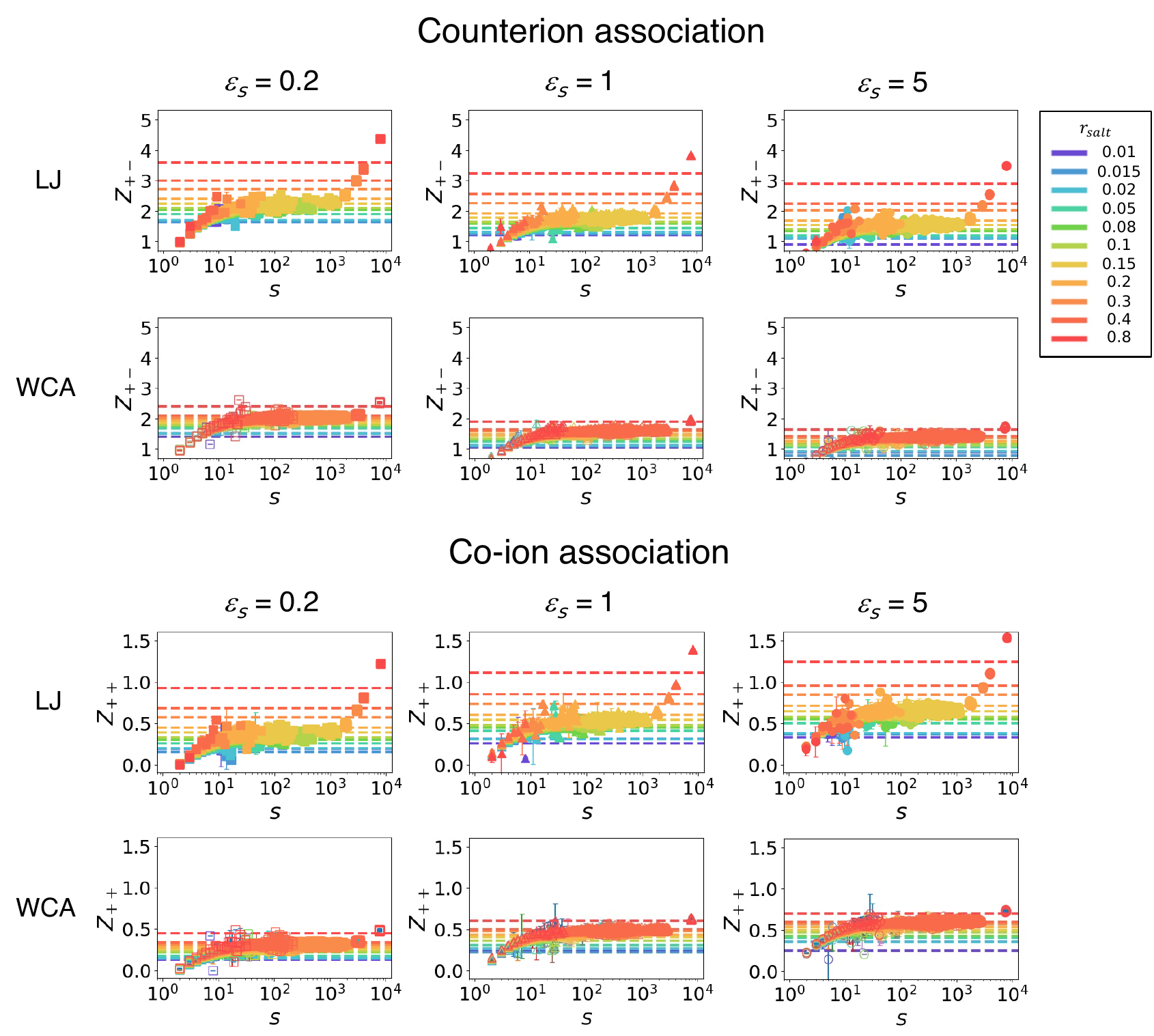}
\caption{The counterion association number $Z_{+-}$ for individual ionic clusters in the LJ and WCA electrolytes with explicit solvent. Top and bottom rows show the simulation results accounting for only counterion associations and for co-ion associations, respectively. From left to right, panels correspond to different values of $\varepsilon_s = $ 0.2, 1, and 5. Dotted horizontal lines represent the averaged association numbers at each salt concentration.}\label{si:fig:associationnumber}
\end{figure}

The counterion association patterns suggest that the ionic clusters are initially tree-like and evolve into more highly branched networks with weak loop formation. Figure~\ref{si:fig:associationnumber} shows the number $Z_{+-}$ of cation-anion associations for individual ionic clusters. The ions are considered associated if their ionic distance is less than a cutoff distance of $1.5 \sigma$, as was used in defining ionic clusters in the original manuscript. We find that $Z_{+-}$ depends on the cluster size $s$, taking values approximately between 1 and 4. In small clusters, $Z_{+-}$ increases with $s$, while it exhibits a plateau at $Z_{+-}\approx2$ over a broad intermediate range of cluster size ($10\lesssim s\lesssim{10}^3$). For the largest clusters, $Z_{+-}$ increases again, which implies the order formation as for NaCl electrolytes~\cite{choi2017ion}. 

\begin {figure}[htbp]\centering
\includegraphics [width=\textwidth] {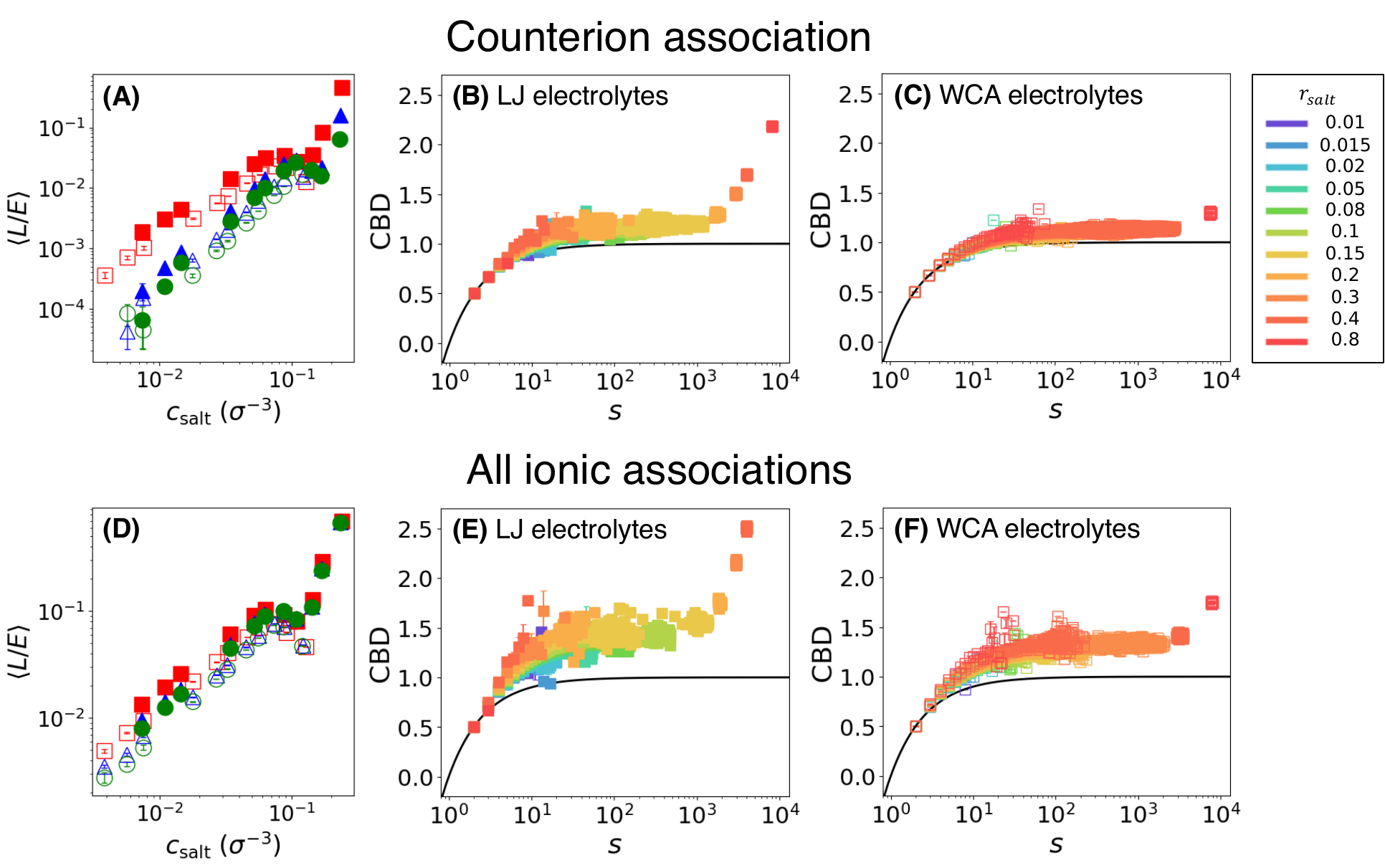}
\caption{Quantitative analysis of loop formation in ionic clusters for the LJ and WCA electrolytes with explicit solvent. Top and bottom rows show the simulation results accounting for only counterion associations and for all ionic associations, respectively. (A, D) The relative number $\langle L/E\rangle$ of intra-cluster loops normalized by the total number of edges, averaged over all ionic clusters. (B, C) The cluster bond density (CBD) as a function of cluster size $s$, accounting for only counterion associations. (E, F) The CBD as a function of cluster size $s$, accounting for all ionic associations, including co-ion associations. The CBD carries information equivalent to $\langle L/E\rangle$; results for $\varepsilon_s =0.2$ are shown here. In panels (B-C) and (E-F), black solid lines represent the prediction for ideal Cayley-tree clusters (i.e., CBD $= 1-1/s$). Filled and open markers correspond to the LJ and WCA electrolytes, respectively. In panels (A, D), colors indicate different values of $\varepsilon_s$: red squares for 0.2, blue triangles for 1, and green squares for 5.}\label{si:fig:lnorm}
\end{figure}

In addition, we find clear deviations from the Cayley-tree assumption in the ionic gelation theory: (i) the presence of co-ion association and (ii) the formation of intra-cluster loops. We observe that a comparable number of co-ion associations emerges within each ionic cluster (Fig.~\ref{si:fig:associationnumber}), highlighting the non-counterion associations in our charge-symmetric simple electrolytes. We also observe that the ionic clusters form intra-cluster loops, whose number increases with increasing $c_{salt}$ (Fig.~\ref{si:fig:lnorm}). For each cluster, the number $L$ of intra-cluster loops was computed from simulations: $L=E-s+1$, where $s$ is the number of constituent ions within a cluster, and $E$ is the number of edges ($Z=2E/s$). The increasing $\langle L/E\rangle$ indicates the increasing contribution of co-ion associations in an ionic cluster (Fig.~\ref{si:fig:lnorm}(A,D)). Equivalently, we also computed the cluster bond density (CBD) $\langle E/s\rangle$ for all ionic clusters (Fig.~\ref{si:fig:lnorm}(B-C,E-F)). For ideal Cayley trees, $E/s=1-1/s$ with $L=0$, and it converges to 1 for large ionic clusters. However, our computation reveals that $\langle E/s\rangle\approx1.3-1.5$, exceeding $1-1/s$ even for moderate-size ionic clusters, indicating that they contain intra-cluster loops, being more ordered than ideal Cayley trees. The computed CBD values below the gel point are comparable to those previously reported for NaTFSI electrolytes~\cite{zhang2024long}. 

Thus, these discrepancies clearly indicate the emergence of nontrivial association patterns ignored in the theory, such as intra-cluster loops~\cite{mceldrew2020theory, goodwin2023theory, zhang2024long} and more complex connectivity within ionic aggregates. Similar results were obtained using the counter-ion associations (Fig.~\ref{si:fig:lnorm}(B-C)), despite slightly lower gelation thresholds with decreased CBD values (Tab.~\ref{si:tab:cluster_compare}). Consequently, the ionic aggregates are better described as branched, weakly looped ionic networks rather than ideal loop-free Cayley-tree structures. 

\begin{table}[h]
\centering
\begin{tabular}{l|c||cc}
\hline
System & $\varepsilon_s$ & $c_{\text{gel}}$ ($\sigma^{-3}$) & $\gamma_s(c_{\text{gel}})$ ($\sigma$) \\
\hline
LJ  & 0.2 & 0.088 & 1.85$\pm$0.01 \\
\hline
WCA & 0.2 & 0.076 & 1.97$\pm$0.01 \\
\hline
\end{tabular}
\caption{Formation of a percolating ionic network near the gelation point $c_{\text{gel}}$ in the explicit-solvent electrolytes, accounting only for counterion associations. The exponent $\gamma_s(c_{\text{gel}})$ quantifies the power-law tail of the size distribution, $P(s)$ (see Eq. 9 in the main text).}\label{si:tab:cluster_compare}
\end{table}

\subsection{Comparison of ionic association between theory and simulations} Owing to the discrepancies found in the ionic associations, our charge-symmetric solvent primitive models, unlike bulky and asymmetric ionic compounds such as LiTFSI, are not expected to quantitatively follow the ionic gelation theory proposed by McEldrew \textit{et al.}~\cite{mceldrew2020theory, mceldrew2021salt}, which is based on Cayley-tree-like clusters. In addition, the simplicity of our model, which employs isotropic interaction potentials between all ions and solvent particles, further raises ambiguities in defining the ion-association functionality. Indeed, it has also been shown that simple electrolytes such as NaCl clearly deviate from Cayley-tree-based ionic gelation predictions due to the formation of ordered, loop-rich aggregates~\cite{mceldrew2020theory, mceldrew2021salt}. In our case, as shown below, the theory captures the counterion association constant below the gelation point reasonably well, yielding values of comparable order of magnitude to those observed in simulations when an appropriate ion-association functionality is chosen. However, it does not accurately locate the gelation threshold over the estimated range of ion-association functionality. 

For comparison with the theory, it is necessary to estimate a single ion-association functionality $f$ that characterizes tree-like, loop-free aggregates at the mean-field level. In this analysis, we focus on the counterion-association functionality $f_{+-}$; due to the charge symmetry, the cations and anions share the same functionality. As mentioned above, there are ambiguities in defining $f_{+-}$ due to the simplicity of our model, which employs isotropic interaction potentials between all ions and solvent particles. Here, we estimate an appropriate range of $f_{+-}$ from the simulations using two independent approaches, both yielding a consistent range of values: $f_{+-}\approx2 - 5$. 

\begin {figure}[htbp]\centering
\includegraphics [width=\textwidth] {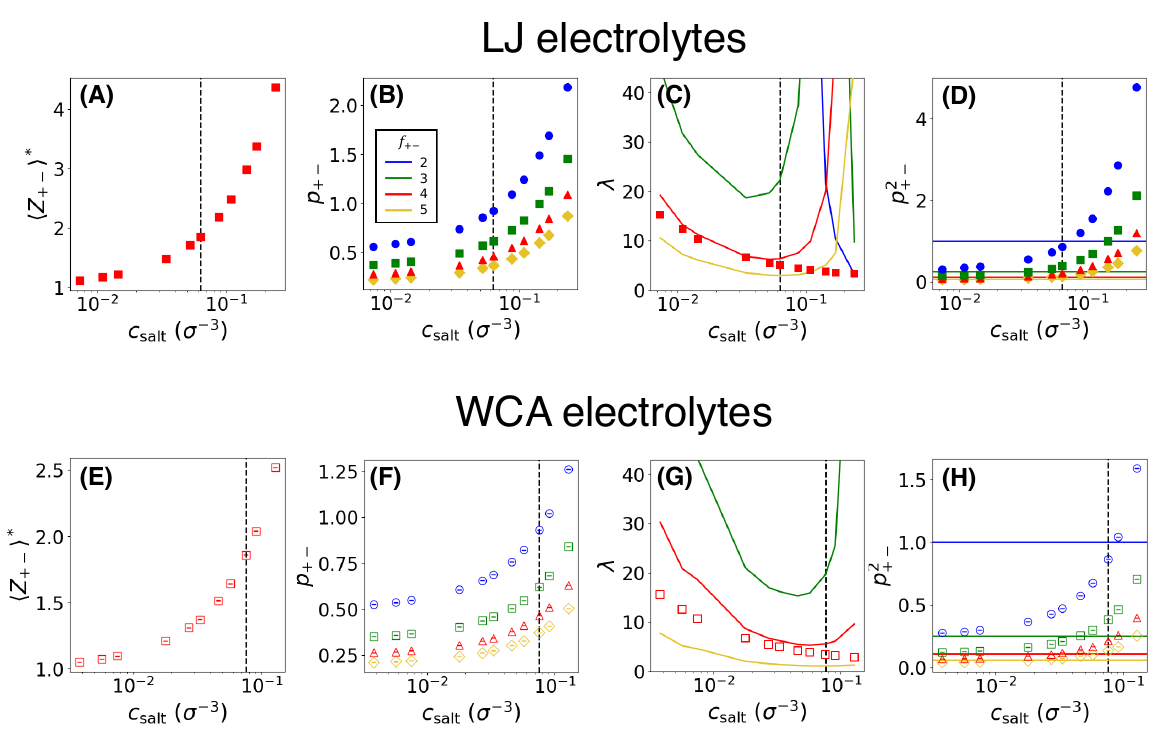}
\caption{Comparison of ionic association between theory~\cite{mceldrew2020theory, mceldrew2021salt} and simulations for both LJ and WCA electrolytes with $\varepsilon_s = 0.2$. The top and bottom rows show the simulation results for the LJ and WCA electrolytes, respectively. (A, E) The counterion association number $Z_{+-}$. (B, F) The corresponding association probability $p_{+-}$ for different values of $f_{+-}$. (C, G) The ionic association constant $\lambda$. (D, H) The squared counterion association probability $p_{+-}^2$ with the theoretical predictions at the gel point, indicated by the horizontal solid lines. In panels (B-D, F-H), symbols represent simulation results, and different colors denote different values of $f_{+-}$. In all panels, vertical black dotted lines indicate the gel point estimated from simulations (see Tab. I in the main text).}\label{si:fig:ionicassociation}
\end{figure}

First, according to the mean-field theory, the average counterion-association number $\langle Z_{+-}\rangle$ determines $f_{+-}$ via
\begin{equation}\label{si:eq:z+-}
\begin{split}
\langle Z_{+-}\rangle^* =\frac{f_{+-}}{f_{+-} - 1},
\end{split}
\end{equation}
where the asterisk denotes the gel point. For both LJ and WCA electrolytes with $\varepsilon_s=0.2$, $\langle Z_{+-}\rangle^{*} \approx 1.85$ (Fig.~\ref{si:fig:ionicassociation}(A,E)). This value suggests an estimated functionality in the range $f_{+-}\approx$ 2 to 5, corresponding to $\langle Z_{+-}\rangle^{*}\approx2-1.5$. Second, we estimated $f_{+-}$ using the number of ionic neighbors within the first solvation shell. The cumulative counter ion-pair distributions $c_{CA}(r)$, evaluated at the cutoff distance $r_{cut}=1.5\sigma$, indicate that the number of counterions is $N_{+-}\approx4$ for the LJ electrolytes and $N_{+-}\approx2$ for the WCA electrolytes with $\varepsilon_s=0.2$ at the highest $c_{salt}$ investigated. 
Another parameter used to estimate the ionic association constant $\lambda$ is the counterion-association probability $p_{+-}$. This probability is concentration-dependent and can be estimated from the simulations as $p_{+-}=\langle Z_{+-}\rangle/f_{+-}$. For the charge-symmetric electrolytes considered in this work, the theoretical relation reads:
\begin{equation}\label{si:eq:lambda_zeta}
\begin{split}
\lambda_{th}\zeta=\frac{{p_{+-}}^2\ }{(1-p_{+-})(1-p_{+-})},
\end{split}
\end{equation}
where $\zeta=f_{+-}p_{+-}c_{salt}$. We note that the definitions of $f_{+-}$ and $p_{+-}$ used in our analysis differ from those of McEldrew \textit{et al.}~\cite{mceldrew2021salt}; in particular, the subscripts here do not imply directionality or asymmetry. From the simulations, the computed association constant $\lambda_{sim}$ is estimated from the minimum value $F_{min}$ of the cation-anion potential of mean force (see Fig. 4 in the main text) via $\lambda_{sim}=e^{-\beta F_{min}}$. The gelation point $c_{gel}$ is then estimated using the relation:
\begin{equation}\label{si:eq:P+-cgel}
\begin{split}
p_{+-}(c_{gel})=\frac{\langle Z_{+-}\rangle^*}{f_{+-}}=\frac{1}{f_{+-} - 1}.
\end{split}
\end{equation}

We find reasonably good agreement between the theoretical and simulated ionic association constants below the gelation threshold for appropriate choices of the functionality, namely $f_{+-} = 4$ (LJ electrolytes) and $f_{+-}= 5$ (WCA electrolytes) (Fig.~\ref{si:fig:ionicassociation}(C, G)). However, the theory cannot predict the gelation threshold, which is expected given that discrepancies in the aggregation patterns emerge with increasing salt concentration. 

Similar limitations of the Cayley-tree-based ionic gelation theory have been reported previously. For example, the gelation points predicted by the Cayley-tree-based ionic gelation theory were found to be systematically lower than those estimated from simulations for nonaqueous LiPF$_6$ and LiBF$_4$ electrolytes~\cite{mceldrew2021salt}. This discrepancy was attributed to the fact that the ionic clusters in these systems do not strictly satisfy the Cayley tree assumption, owing to the formation of intra-cluster loops that suppress ideal tree-like associations. In our primitive solvent models, such deviations are even more pronounced, due to the isotropic interionic interactions and the enhanced formation of intra-cluster loops. Consequently, the gelation thresholds predicted by the Cayley-tree-based theory exhibit substantially larger deviations from our simulation results.

\begin {figure}[htbp]\centering
\includegraphics [width=\textwidth] {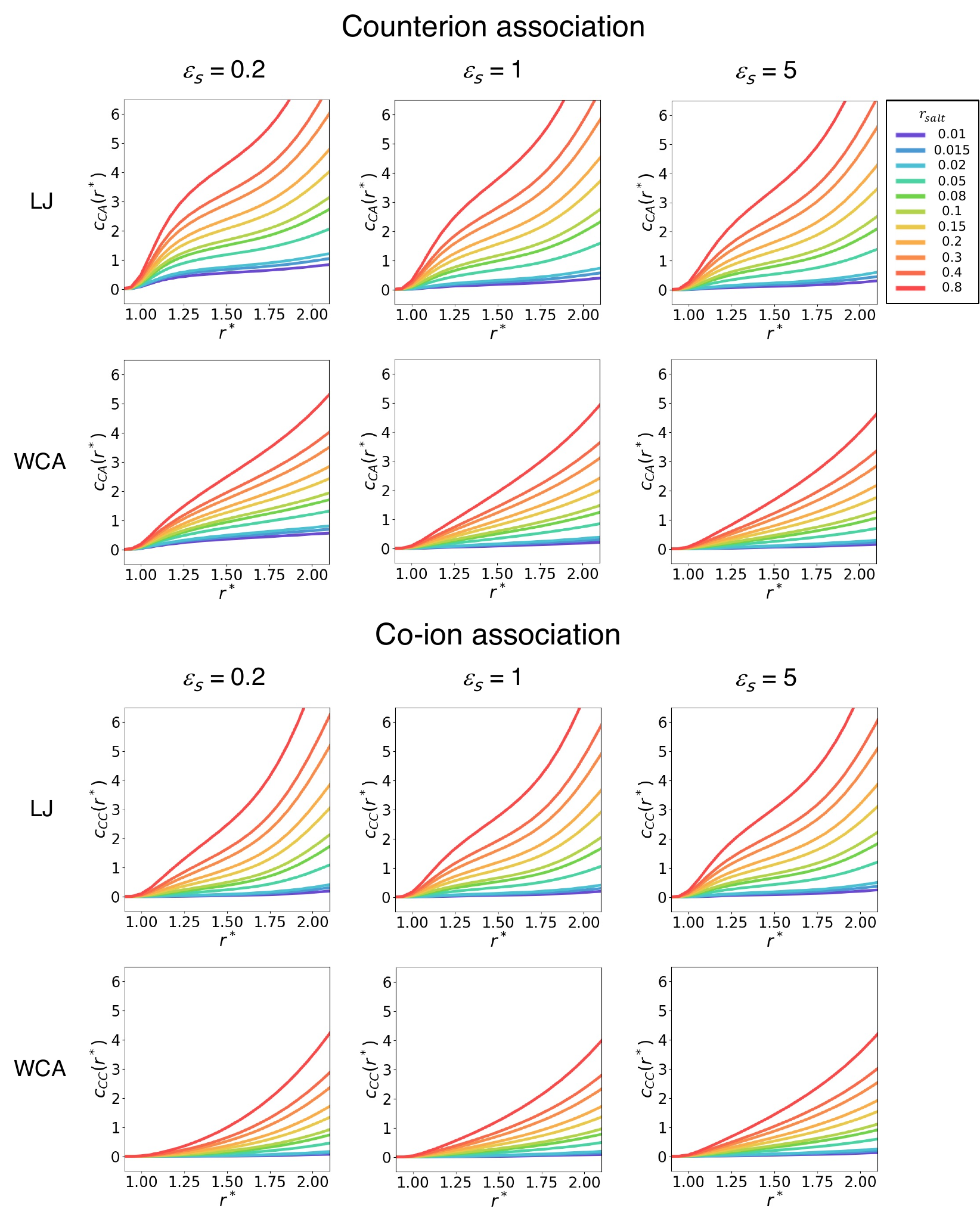}
\caption{The cumulative distribution for counterion $c_{CA}(r)$ and co-ion associations $c_{CC}(r)$. The top and bottom rows show the simulation results for the LJ and WCA electrolytes, respectively. From left to right, panels correspond to different values of $\varepsilon_s$ = 0.2, 1, and 5. Owing to the charge symmetry, cation-cation associations are only considered here.}\label{si:fig:cr}
\end{figure}

\clearpage
\newpage